\documentclass[aps,pra,twocolumn,superscriptaddress,floatfix, 10pt, showpacs,final]{revtex4}
\usepackage[usenames]{color}
\usepackage{amsmath}
\usepackage{euscript}
\usepackage{graphicx}

\usepackage{dcolumn}
\usepackage{amssymb}
\usepackage{mathrsfs}

\begin{document}
\title{Dissociative electron attachment to the H$_2$O molecule. II. Nuclear dynamics on coupled electronic surfaces within the local complex potential model}
\author{Daniel J. Haxton}
\altaffiliation{Present address: Department of Physics and JILA, University of Colorado, Boulder, CO 80309, USA.}
\affiliation{Department of Chemistry, University of California, Berkeley, California 94720}
\affiliation{Lawrence Berkeley National Laboratory, Chemical Sciences, Berkeley, California 94720}

\author{T. N. Rescigno}
\affiliation{Lawrence Berkeley National Laboratory, Chemical Sciences, Berkeley, California 94720}
\author{C. W. McCurdy}
\affiliation{Lawrence Berkeley National Laboratory, Chemical Sciences, Berkeley, California 94720}
\affiliation{Departments of Applied Science and Chemistry, University of California, Davis, California 95616}


\begin{abstract}

We report the results of a first-principles study of dissociative electron attachment to H$_2$O. The cross sections are obtained from nuclear dynamics calculations carried out in full dimensionality within the local complex potential model by using the multi-configuration time-dependent Hartree method. The calculations employ our previously  obtained  global, complex-valued, potential-energy surfaces for the three ($^2B_1$, $^2A_1$, and $^2B_2$) electronic Feshbach resonances involved in this process. These three metastable states of H$_2$O$^-$ undergo several degeneracies, and we incorporate both the Renner-Teller coupling between the $^2B_1$ and $^2A_1$ states as well as the conical intersection between the $^2A_1$ and $^2B_2$ states into our treatment.  The nuclear dynamics are inherently multidimensional and involve branching between different final product arrangements as well as extensive excitation of the diatomic fragment. Our results successfully mirror the qualitative features of the major fragment channels observed, but are less successful in  reproducing the available results for some of the minor channels.  We comment on the applicability of the local complex potential model to such a complicated resonant system.
\end{abstract}

\maketitle

\section{Introduction} \label{introsect}

In the preceeding paper~\cite{haxton5}, referred to hereafter as paper I, we presented global representations of the three ($^2B_1$, $^2A_1$, and $^2B_2$) complex-valued potential-energy surfaces of the   metastable  states of H$_2$O$^-$ that underlie dissociative electron attachment to water. This paper is concerned with the calculation of the cross sections for that physical process.
Prior experimental and theoretical results 
\cite{Lozier,Buchel,Schultz,compton,Melton,sancheschultz,tjhall,belic,curtiswalker,
Claydon,Jungen,Gil,Morgan,Gorfinkiel,haxton1,haxton2,haxton3,haxton4,fedor} have 
characterized the various breakup channels and determined
the spatial symmetries of the three metastable electronic states of H$_2$O$^-$,
the $^2B_1$, $^2A_1$, and $^2B_2$ electronic Feshbach resonances, which are
responsible for production of H$^-$ and O$^-$. As explained in Ref.~\cite{haxton3} and paper I, the energetically lowest H+OH$^-$ channel does not directly correlate with any of the three Feshbach states.  We therefore conclude that OH$^-$ production is due to nonadiabatic effects.

We pursue this problem theoretically using a coupled Born-Oppenheimer treatment of the nuclear
motion.  The first task, which was described in paper I, is the construction of three-dimensional, complex-valued potential-energy surfaces for these three states, which have a negative imaginary component due to the
finite probability of electron autodetachment back to H$_2$O+$e^-$. These complex-valued potential-energy surfaces, which are functions of the nuclear geometry $\vec{q}$, are defined as
\begin{equation}
V(\vec{q}) = E_R(\vec{q}) - i\frac{\Gamma(\vec{q})}{2} ,
\label{lcppot}
\end{equation}
where $E_R$ is the resonance position and $\Gamma$ is the width
of the resonance, which is related to the lifetime by $\tau=1/\Gamma$.
(We use atomic units throughout this paper.)  
The present article, which we label paper II, is concerned with the
use of these potential curves within the local
complex potential (LCP) model \cite{BirtwistleHerzenberg,DubeHerzenberg,BardsleyWadehra,OmalleyTaylor,Omalley}
to calculate the nuclear dynamics leading to dissociation.  The analysis of
the dynamics yields the DEA cross section
as a function of incident electron energy.

We must account for two major nonadiabatic physical effects in calculating the
quantum dynamics of the nuclei.  
As described in paper I, the three potential-energy
surfaces have several degeneracies that lead to coupling among them.  First,
the $^2B_1$ and $^2A_1$ states become members of a degenerate $^2\Pi$ pair
in linear geometry, and for this reason there will be Renner-Teller coupling
between them.  We expect this coupling to be relevant for 
DEA via the  $^2A_1$ state, because the gradient of its potential-energy surface will cause the system to move toward linear geometry
after the electron attaches.  Second, there is a conical 
intersection~\cite{haxton3}
between the $^2B_2$ and $^2A_1$ states which leads to coupling between
them.  For this reason, as described in paper I, we constructed a set of diabatic $^2B_2$ and $^2A_1$
surfaces, along with a coupling term, which we use in the calculations
presented in this paper.

\begin{figure}
\begin{center}
\resizebox{0.95\columnwidth}{!}{\includegraphics*[0.95in,0.60in][4.35in,3.0in]{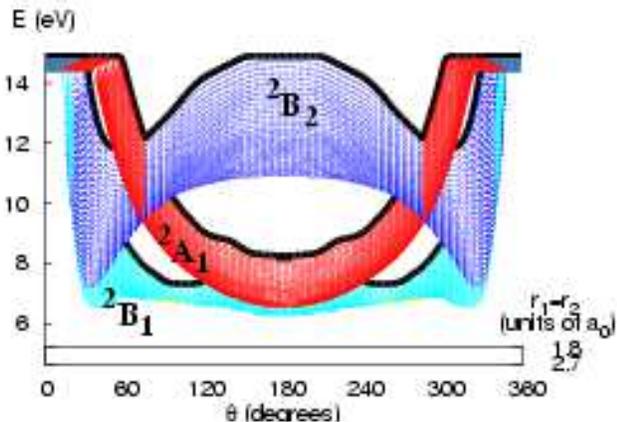}}
%
%
%
\end{center}
\caption[Dependence of real parts of resonance energies upon bending
coordinate]{Real parts of resonance energies $E_R$ as constructed
in paper I within
$C_{2v}$ geometry ($r_1=r_2$), plotted with respect
to bending angle and symmetric stretch distance.}
\label{c2vfig} 
\end{figure}

In Fig.~\ref{c2vfig}, we show the real parts $E_R$ of the constructed potential-energy surfaces
along a two-dimensional cut which includes the equilibrium geometry of the
neutral ($r_1=r_2=1.81a_0$; $\theta_{HOH}=104.5^\circ$).  The degeneracies that lead to the
nonadiabatic effects listed above can be seen in this figure.
The two-dimensional cut depicted is that
for which the two OH bond lengths are equal ($r_1=r_2$), corresponding
to $C_{2v}$ symmetry.  (In $C_{2v}$ symmetry, the adiabatic and diabatic $^2A_1$
and $^2B_2$ surfaces coincide.)
The backside
of this cut lies at $r_1=r_2=1.81a_0$, which is the equilibrium value of the
bond lengths in neutral H$_2$O, and is marked with solid lines.
The surfaces extend forward in Fig.~\ref{c2vfig}
along the symmetric stretch direction to geometries at which $r_1=r_2=
2.7a_0$.  The conical intersection comprises the set of points along
which the $^2A_1$ and $^2B_2$ surfaces intersect.  The Renner-Teller
degeneracy between the $^2B_1$ and $^2A_1$ states 
occurs at $\theta$=180$^\circ$.

Although Fig.~\ref{c2vfig} shows only one cut of the potential-energy surfaces, and only their real part, it is useful for introducing
certain features of these surfaces and the dynamics that will result.
Dissociative attachment via the lower $^2B_1$ and $^2A_1$ states leads 
primarily to the product H$^-$+OH (X $^2\Pi$).  The two OH bond lengths
for such an arrangement are unequal, and therefore this product arrangement
cannot be seen in Fig. \ref{c2vfig}.  However, we can see that at
the equilibrium geometry of the neutral, the $^2B_1$ surface is relatively
flat with bend, while the $^2A_1$ surface slopes steeply downward toward
linear geometry ($\theta$=180$^\circ$).  As a result, the dynamics
beginning on the $^2A_1$ surface will lead toward linear geometry,
and we expect that the Renner-Teller coupling between these two states
will be more important for DEA via the $^2A_1$ than via the $^2B_1$ state.

The channel H$_2$+O$^-$ is the minor channel for DEA via the
$^2B_1$ and $^2A_1$ states, but the major channel for the $^2B_2$
state.  We can see why this is the case from Fig.~\ref{c2vfig};
the gradient of the $^2B_2$ (2 $^2A'$) surface leads downward toward
the conical intersection, where the system may make a nonadiabatic
transition to the lower surface and access the clearly visible
H$_2$+O$^-$ well on the 1 $^2A'$ (lower cone) surface.  The 2 $^2A'$ surface
does not have a low energy asymptote in this geometry, instead
correlating to O$^-$+H$_2$ ($\sigma_g^1 \sigma_u^1$).

The outline of this paper is as follows.
In Sec. \ref{priorsect}, we summarize previous experimental and theoretical
work on this problem.  In Sec. \ref{lcpsect}, we present
the local complex potential model, which forms the foundation of our
theoretical implementation. The Hamiltonian for the rovibrational nuclear motion of a triatomic
molecule, and the additional terms which arise when the Renner-Teller effect
is included, are described in Sec. \ref{triatomicsect}.  In Sec. \ref{mctdhsect} we describe the multi-configuration time-dependent
hartree (MCTDH) method, which we use to calculate the nuclear dynamics, and the
formalism for calculating the DEA cross sections.
In Sec. \ref{resultssect} we present the final results of this study: cross
sections, as a function of incident electron energy, resolved into the final
rovibrational product states.

\section{Previous experimental and theoretical results} \label{priorsect}

Dissociative electron attachment to water molecules has been the subject of
previous  experimental investigation, starting as early
as 1930~\cite{Lozier}, and as recently as the past year (2006)~\cite{fedor}.  Early experiments on dissociative electron
attachment to H$_2$O focused mainly on the identification
of the negative ion species formed, the measurement of the total
cross sections, and the energy locations of the structures in the
resonance process~\cite{Melton}.  Buchel'nikova~\cite{Buchel} and
Schultz~\cite{Schultz} established that the main products of 
dissociative electron attachment to water are ${\rm H^-}$ and ${\rm O^-}$,
with the production of ${\rm O^-}$ being almost ten times smaller than
that of ${\rm H^-}$ at lower energies, but with ${\rm O^-}$ dominating
at higher electron-impact energies.

Both Compton and Christophorou~\cite{compton} and Melton~\cite{Melton}
carried out comprehensive studies 
of negative ion formation in water and measured absolute cross sections for
DEA.  Three resonance peaks were observed. ${\rm H^-}$ production was observed
at approximately 6.5 eV and 8.6 eV, with the second peak much
less intense than the first.  The species ${\rm O^-}$ was observed in increasing intensities 
in three peaks at 7.0 eV,
9.0 eV, and 11.8 eV~\cite{compton}.  

The species OH$^-$ is also
observed in the dissociative electron attachment experiments, though at an intensity
one order of magnitude below the minor O$^-$+H$_2$ channel, which is itself
observed at an intensity approximately one order of magnitude lower than the
dominant H$^-$+OH channel.  
Melton~\cite{Melton} argued that OH$^-$+H
was a true channel of dissociative electron attachment to H$_2$O molecules, while in  subseqent studies (e.g., Ref. \cite{Klots}) it was argued that  OH$^-$ is produced by 
DEA to water clusters [H$_2$O]$_n$.  The question of OH$^-$ production has been reexamined
in the recent experimental study of Fedor \textit{et al.}\cite{fedor}. These authors
 have concluded that, indeed, it is a direct product of dissociative electron attachment to water. 
This minor channel is not examined in the present treatment, and no mechanism has, as yet, been advanced.

The effects of isotopic substitution have also been an issue of some
debate.  The replacement of H$_2$O by D$_2$O as the molecular target
has the effect of nearly doubling the reduced masses corresponding
to OH (OD) bond motion.  One would expect, at least in a simple
one-dimensional picture, that the nuclear dynamics may
be substantially altered by such replacement, and in particular, the
time to dissociation is increased. 
A longer dissociation time
allows a greater amount of electron autodetachment to take
place; therefore, performing the same experiment with different isotopic
variants provides information on the lifetime of the electronic
state involved. 
The cross sections for DEA via both H$_2$O and D$_2$O 
were measured, compared, and discussed in detail 
by Compton and Christophoreau~\cite{compton}.  
A smaller peak cross section for D$^-$ 
production than for H$^-$ production via the lowest-energy $^2B_1$ state 
was observed.  On the basis of these results, these authors
derived an approximate lifetime of 2.1 $\times$ 10$^{-14}$ seconds
for the lowest-energy $^2B_1$ Feshbach resonance.
We have already published an initial study of DEA via this 
state~\cite{haxton1}, which arrived at results and conclusions much different
from those of Ref.~\cite{compton}.
The calculated results yielded a higher peak
cross section for D$^-$ production via the $^2B_1$ resonance than
for H$^-$ production, and a similar energy-integrated cross section,
in stark contrast to the results of Compton\cite{compton}.
The calculations indicated a larger lifetime
of 10.9 $\times$ 10$^{-14}$ seconds for the $^2B_1$ state, and
the nuclear dynamics that we calculated indicated that only
a small portion of the dissociating anion flux is lost to
autodetachment.  

The recent experimental results of Fedor \textit{et al.}\cite{fedor}
have substantially resolved this controversy.  These authors
obtain results different from those of Ref.~\cite{compton}, reversing
the trend in peak heights for H$^-$ versus D$^-$ production via
the $^2B_1$ resonance.  They observe a higher peak for D$^-$ production
than for H$^-$ production, which brings the current experimental
and theoretical results into qualitative agreement.

Although the peak heights provide considerable information about
the physical process of dissociative electronic attachment to water,
further information is gained by resolving the angular dependence
of the fragments produced, and the final (ro)vibrational state of the
diatomic fragment.
A series of measurements by Trajmar and Hall \cite{tjhall} and Belic,
Laudau, and Hall \cite{belic} revealed the energy and angular 
dependence of ${\rm H^-}$ in dissociative electron 
attachment to ${\rm H_2O}$. 
The determination of the angular dependence aided the assignment
of the spatial symmetries of the three resonant states,
$B_1$, $A_1$, and $B_2$, which had previously been misassigned.
By resolving the kinetic energy of the H$^-$ fragment, this
experiment yielded information about the
vibrational and rotational state distribution of the OH fragments.

Curtis and Walker~\cite{curtiswalker} measured cross sections
for dissociative electron attachment to D$_2$O and obtained two important
results.  By measuring the kinetic energy of recoil of the D$^-$
fragments produced, these researchers established that 
both ground state OD ($^2\Pi$)  and excited OD ($^2\Sigma$) accompany
the D$^-$ anions produced within the third resonance peak, and that
the three-body breakup channel D$^-$+D+O is observed toward
the high-energy tail of the second peak.

The experimental studies determined that there are three 
metastable electronic resonance states of the H$_2$O$^-$ anion, 
the $^2B_1$, $^2A_1$, and $^2B_2$, which are primarily responsible for dissociative electron
attachment to water.  These three electronic states correspond to the three
peaks seen in the experimental cross sections.  Although the third
peak is not obvious in the H$^-$ cross sections,
it is present, though much smaller than the first and second peaks.

Several salient features of the early experiments suggest that the nuclear
dynamics of this process may hold some surprises.  For dissociative
attachment through the $^2B_1$ resonance, the cross section for
producing ${\rm H^-+OH}$ is roughly 40 times larger at its peak than
the cross section for producing the energetically favored products,
${\rm O^-+H_2}$ \cite{compton,Melton}.
The lowest-energy atom/diatom arrangement, ${\rm H+OH^-}$, is produced
in even smaller quantities.
In addition, the branching ratios for the different product states vary
greatly depending on which Feshbach resonance is formed by the attachment.
These observations indicate that the
products of this reaction are determined by the dynamics of the process
itself rather than by the energetics of the possible product channels, and
that moreover those dynamics are different for each of the resonance states
of the water anion.
The detailed experiments of Beli\`c, Landau and Hall \cite{belic}
in 1981 indicated that the dissociation dynamics involve correlated 
motion among multiple degrees of freedom.
For instance, the channel producing ${\rm H^-+OH}$ through the $^2B_1$ 
resonance state is accompanied by extensive vibrational excitation
of the OH fragment.   

Therefore, given the competition between dissociation channels
and the observed product vibrational excitation, one expects that the
dynamics of dissociative attachment to this molecule are intrinsically
polyatomic, and can only be described theoretically by a treatment using
the full dimensionality of nuclear motion.

Compared with the large number of experimental measurements, detailed
theoretical work on dissociative electron-water collisions  has been
relatively scarce.
The paucity of 
theoretical work on DA stems from the fact that, in water,
DA proceeds, not through tunneling shape resonances, but through
Feshbach resonances that involve changes in the electronic structure of the target.
Early theoretical work 
focused on the electronic structure~\cite{Claydon} and 
configuration-interaction~\cite{Jungen}
calculations on various states of ${\rm H_2O^-}$ that are possible
resonances. These calculations, together with experimental observations,
formed the basis of the assignment of the three Feshbach resonances
that are responsible for electron-impact dissociation of water
in the gas phase. 

Contemporary theoretical work has included {\em ab initio} 
complex Kohn~\cite{Gil} and $R$-matrix~\cite{Morgan} calculations,
at the equilibrium nuclear geometry, of the resonances and excitation 
cross sections into low-lying dissociative electronic states. More recently,
Gorfinkiel, Morgan, and Tennyson~\cite{Gorfinkiel} carried out $R$-matrix 
calculations of dissociative excitation of water through the four lowest excited
states (the $^{1,3}B_1$ and $^{3,1}A_1$ states). A limited study of
the effects of nuclear motion were included in that work by increasing
one of the OH bonds while keeping the equilibrium HOH bond
angle and the other OH bond length constant.  The only theoretical work on
the dynamical aspects of dissociative electron attachment to water are earlier
classical trajectory analyses based on either repulsive~\cite{Goursaud1}
or attractive~\cite{Goursaud2} model resonace surfaces.

We previously reported calculations of the cross sections for dissociative
attachment through the lowest-energy $^2B_1$ resonance~\cite{haxton1,haxton2}
that incorporated a full quantum treatment
of the nuclear motion of the resonant state.  That study found good
agreement with experiment for dissociative attachment through the lowest
resonance state ($^2B_1$) of the water anion to produce H$^-$, and it
established that the associated dynamics are intrinsically polyatomic
and thus cannot be described successfully by one-dimensional models.
The present treatment supersedes our earlier study and extends the treatment to include the 
higher resonance states as well.

We have recently presented a qualitative study~\cite{haxton3} 
of the potential-energy surfaces for the three Feshbach resonances,
which demonstrated that for these metastable, anion states, there
exist numerous intersections and degeneracies within the adiabatic
manifold.  This study identified the conical intersection between
the $^2A_1$ and $^2B_2$ states, as well as a novel degeneracy between
the $^2B_2$ Feshbach resonance and a $^2B_2$ shape resonance.
This degeneracy defines a branch seam, and the two resonance energies
are seen to comprise two components of a double-valued adiabatic
potential-energy surface.  This seam and the resulting dynamics
may have an effect upon the three-body, H+H+O$^-$
cross section, although we do not include it in the present treatment.
Finally, in a separate publication~\cite{haxton4}, we derived a 
``constant-eigenmode approximation'' and used it to 
calculate the angular dependence of the H$^-$
fragment production~\cite{haxton4} via the $^2B_1$ resonance.
We found excellent agreement with the results of Belic,
Landau and Hall~\cite{belic}, and demonstrated that the observed
angular dependence is a result of partial-wave mixing in the
resonance-background coupling.

\section{Local Complex Potential model} \label{lcpsect}

We treat the nuclear dynamics of dissociative electron attachment within
the local complex potential  model.
This model is concerned with the proper accounting 
for the decay of the resonant state, and its effect upon the nuclear dynamics.
The LCP model includes the simplest such accounting, in which the decay
rate is a local function of the nuclear geometry.  

\subsection{Feshbach partitioning and the nuclear wave equation}

The local complex potential model~\cite{BirtwistleHerzenberg,DubeHerzenberg,BardsleyWadehra}
, also known as the  ``Boomerang'' model when applied to vibrational excitation,
describes resonance nuclear motion by an inhomogeneous Schr\"odinger equation
and a complex, but purely local potential. It is perhaps easiest to derive by applying Feshbach partitioning~\cite{Feshbach} within the Born-Oppenheimer framework to derive a nuclear wave equation~\cite{OmalleyTaylor,Omalley}. The derivation  begins by defining
a discrete (square-integrable) 
approximation to the resonant electronic state, $\psi_Q(\vec{r_e};\vec{q})$ which
depends parametrically on the nuclear coordinates $\vec{q}$ and which is 
unit-normalized with respect to integration over the
electronic coordinates $\vec{r_e}$.  One then defines 
the geometry-dependent Feshbach projection operator $Q$, 
which operates on the electronic degrees
of freedom,
\begin{equation}
\label{q_def}
Q(\vec{q}) = \big\vert \psi_Q(\vec{q}) \big] \big[ \psi_Q(\vec{q}) \big\vert ,
\end{equation}
and its complement $P$:
\begin{equation}
P(\vec{q}) = \mathbf{1} - Q(\vec{q}) ,
\end{equation}
with $P^2=P$, $Q^2=Q$ and $PQ=QP=0$.
(Brackets denote integration over the electronic degrees of freedom only.)
Partitioning the full wave function for total energy $E$ as $\Psi^+=P\Psi^+ + Q\Psi^+$,
we can formally derive  the following inhomogenous equation for $Q\Psi^+$:
\begin{equation}
\begin{split}
&  \left(E - Q\mathbf{H}Q - Q\mathbf{H}P\frac{1}{E-P\mathbf{H}P+i\epsilon}P\mathbf{H}Q \right) Q \Psi^+ \\
& \qquad \qquad = Q\mathbf{H}P \Psi^+,
\end{split}
\label{transcendental2}
\end{equation}
where $H$ is the sum of the electronic Hamiltonian and nuclear kinetic energy, 
$\mathbf{H} = H_{el} + T_{\vec{q}}$.

In view of Eq.~(\ref{q_def}), we can write
\begin{equation}
Q\Psi^+(\vec{r_e}; \vec{q}) = \psi_Q(\vec{r_e}; \vec{q}) \ \xi(\vec{q}).
\end{equation}
The function $\xi(\vec{q})$ describes the relative motion of the nuclei in the negative-ion resonance state. To derive an equation for $\xi(\vec{q})$, the first approximation that is made is the Born-Oppenheimer approximation:  we neglect all non-adiabatic
couplings arising from the operation of the nuclear kinetic energy upon the adiabatic
basis.  Then multiplying Eq. (\ref{transcendental2}) from the left by $\psi_Q(\vec{r_e};\vec{q})$
and integrating oven the electronic coordinates gives the nuclear wave equation,
\begin{equation}
\left(
E - V_Q(\vec{q})
 - \Delta(E) 
 - T_{\vec{q}} 
\right) \xi(\vec{q}) = QH_{el} P \Psi^+,
\label{transcendental3}
\end{equation}
where
\begin{subequations}
\begin{equation}
V_Q(\vec{q}) \equiv \left[ \psi_Q \left\vert H_{el} \right\vert \psi_Q \right]
\end{equation}
and
\begin{equation}
\Delta(E) \equiv Q H_{el} P \frac{1}{E-P H_{el} P-T_{\vec{q}}+i\epsilon}P H_{el} Q.
\end{equation}
\end{subequations}
The real-valued potential $V_Q(\vec{q})$ is the expectation value of the
electronic Hamiltonian with respect to the discrete state $\psi_Q$; the
additional, energy-dependent term $\Delta(E)$ is called the ``level-shift operator''
and is nonlocal in the nuclear degrees of freedom $\vec{q}$,
owing to the presence of the nuclear Green's function.  The residue of this Green's
function gives the level-shift operator $\Delta(E)$ a negative-definite
imaginary component.

In order to bring Eq.(\ref{transcendental3}) into the form of the local
complex potential model, it is necessary to make a local approximation
to the level-shift operator $\Delta(E)$, and also to approximate the driving
term. The assumptions that underlie these approximations are well
understood~\cite{Bieniek, Kurilla}. A local approximation to the level-shift operator yields
\begin{equation}
V_Q(\vec{q}) + \Delta(E) \approx E_R(\vec{q}) - i\frac{\Gamma(\vec{q})}{2},
\label{lcpapprox}
\end{equation}
where $E_R$ and $\Gamma$ are the location and \textit{total} width of the resonance.
A first-order perturbation treatment (Fermi's golden rule)
of the driving term yields~\cite{haxton4}
\begin{equation}
 QH_{el} P \Psi^+ \approx
\sqrt{\frac{\Gamma_0(\vec{q})}{2\pi}}\chi_{\nu_i}(\vec{q}) \equiv \phi_{\nu_i}(\vec{q},0),
\label{drivdef}
\end{equation} 
where $\Gamma_0$ is the \textit{partial} width for decay to the
ground electronic state of the target, and $\chi_{\nu_i}$ is the 
initial rovibrational
state of the target.  

The final working equation of the LCP model then reads
\begin{equation}
\left(E - E_R(\vec{q}) +\frac{i\Gamma(\vec{q})}{2} - T_{\vec{q}}\right)\xi_{\nu_i}(\vec{q})=\sqrt{\frac{\Gamma_0(\vec{q})}{2\pi}}\chi_{\nu_i}(\vec{q}).
\label{eq:boomerang}
\end{equation}
The location and widths of the various resonance states were obtained from configuration interaction and fixed-nuclei variational electron scattering calculations, respectively, as detailed in paper I. In the case of the $^2B_1$ resonance, which generally lies below its $^3B_1$ neutral parent, the resonance can only decay into the ground electronic channel. In that case, the total and partial widths, $\Gamma$ and  $\Gamma_0$,  coincide and can be obtained by fitting the eigenphase sum to a Breit-Wigner form. For the higher resonances, a more elaborate fitting procedure is required to obtain the partial widths, as outlined in  Ref.~\cite{haxton4} and in paper I.

\subsection{Time-dependent formulation of the LCP model}

A direct solution of the differential equations of the local complex potential model can pose significant difficulties for problems with multiple degrees of freedom and, in such cases, a time-dependent formulation of the problem can offer distinct computational advantages. Such a formulation can be made, as demonstrated by McCurdy and
Turner~\cite{McT}, by formally writing the solution of Eq. (\ref{eq:boomerang}) as 
\begin{equation}
\label{boomsol}
\xi_{\nu_i}(\vec{q}) = (E-H +i\epsilon)^{-1} \phi_{\nu_i}(\vec{q},0)
\end{equation}
and writing the nuclear Green's function as the Fourier transform of the propagator for the time-dependent Schr\"odinger equation:
 \begin{equation}
\begin{split}
\xi_{\nu_i}(\vec{q}) & = 
\lim_{\epsilon \to 0} 
\ i \int^{\infty}_0 e^{i(E +i\epsilon)t}
 e^{-iHt}\phi_{\nu_i}(\vec{q},0) dt \\ 
& = \lim_{\epsilon \to 0} 
\ i \int^{\infty}_0 e^{i(E +i\epsilon)t}
\phi_{\nu_i}(\vec{q},t)dt, 
\end{split}
\label{fourier}
\end{equation}
where we define the time-dependent nuclear wave function as
\begin{equation}
\phi_{\nu_i}(\vec{q},t) = e^{-iHt}\phi_{\nu_i}(\vec{q},0).
\label{timedepwfcn}
\end{equation}

The driving term $\phi_{\nu_i}(\vec{q},0)$ of the LCP equation can thus be viewed as the initial value of a wave packet that subsequently evolves on the complex potential surface of the resonance anion. Since the potential surface is complex, the packet decays as a function of time until it effectively escapes the region of the surface where the width is nonzero.

\section{Triatomic Jacobi coordinate system and Hamiltonian} \label{triatomicsect}

\begin{figure}
\begin{center}
\begin{tabular}{c}
\resizebox{0.95\columnwidth}{!}{\includegraphics*[2.4in,7.4in][6.2in,8.8in]{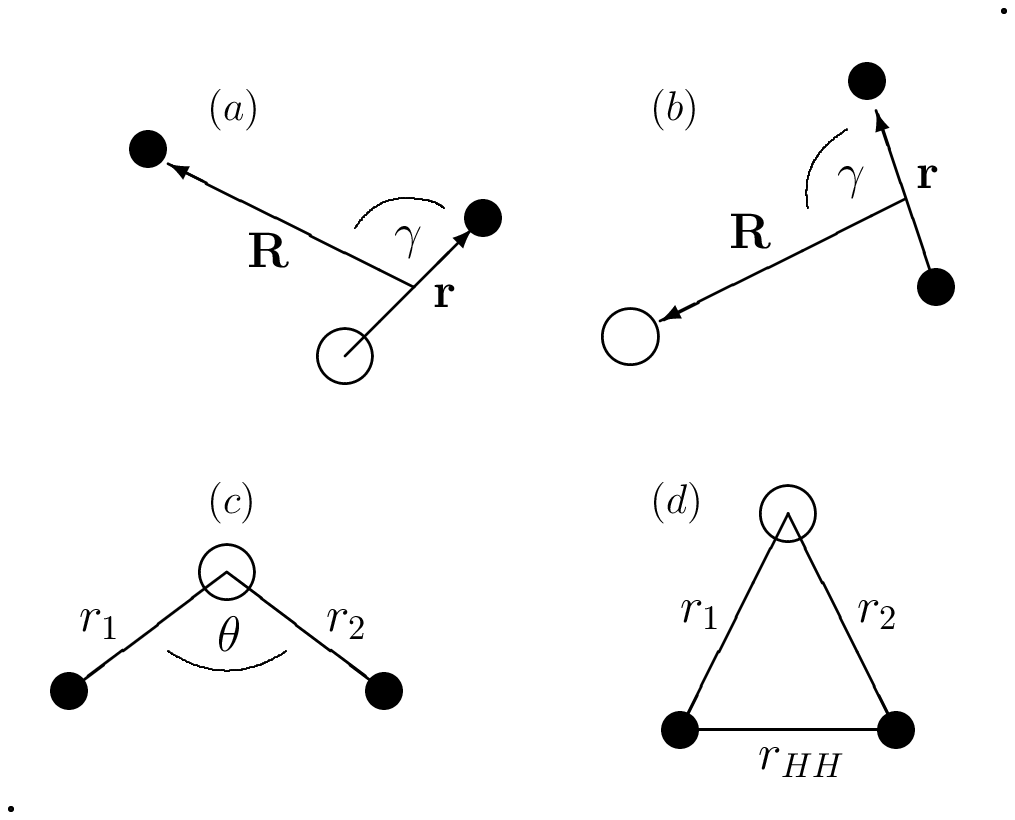}} \\
\resizebox{0.95\columnwidth}{!}{\includegraphics{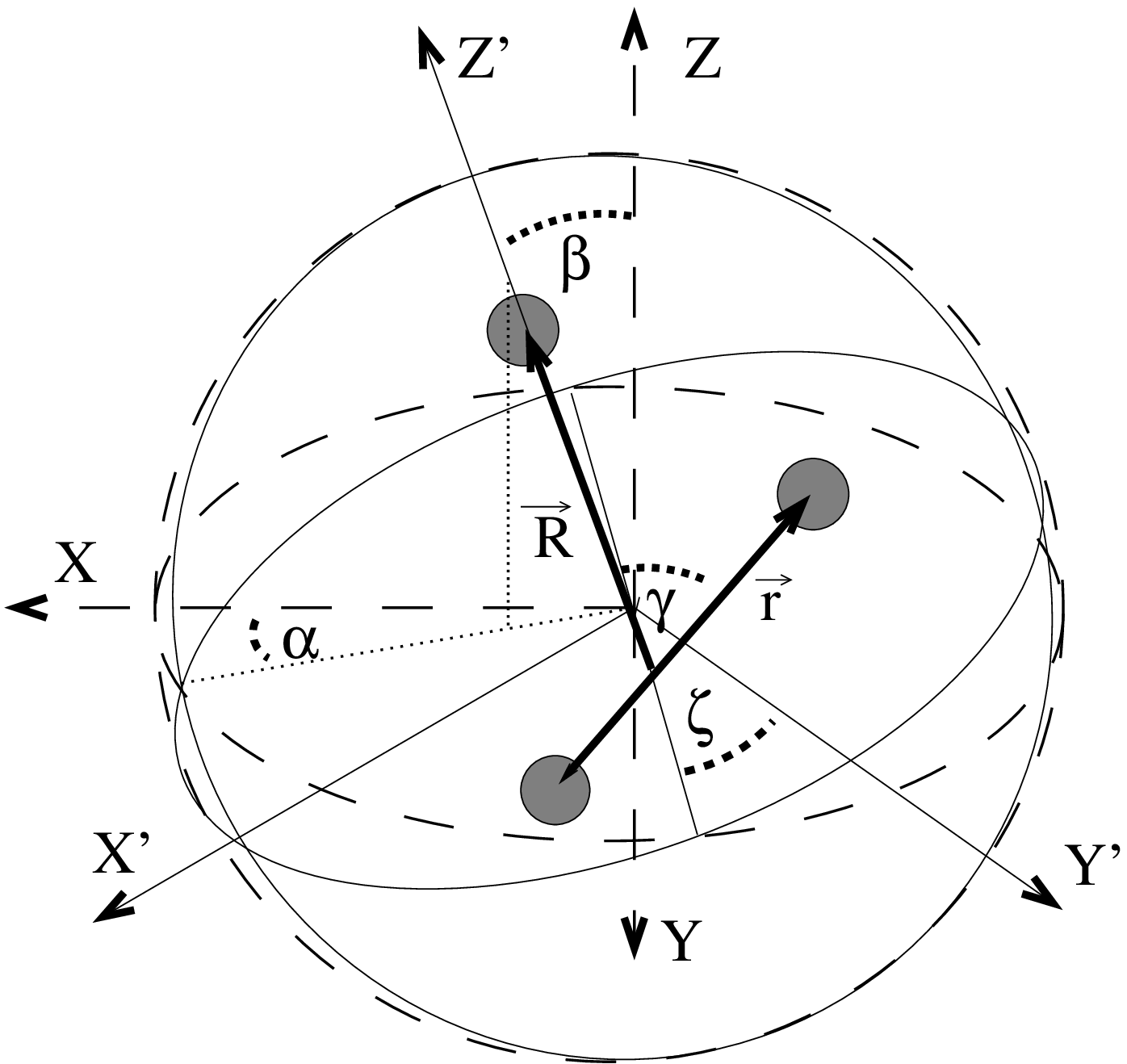}} 
\end{tabular}
\end{center}
\caption{Jacobi coordinate systems used to analyze the OH+H, ($a$) and H$_2$+O ($b$) arrangement channels  and the ``R-embedding''  coordinate system with origin at the center of mass. Primed and unprimed axes refer to BF and SF frames, respectively. The BF $X'Z'$ and $X'Y'$ planes are both
marked with a thin line circle and the SF $XZ$ and $XY$ planes are marked with dashed circles.
The line of nodes is also drawn. The molecule resides in the BF $X'Z'$ plane.  
\label{bigcoords}}
\end{figure}

The LCP model equations were solved in the coordinate systems
depicted in Fig. \ref{bigcoords}.  For the three internal degrees of freedom
of this triatomic molecule, we employ Jacobi coordinate systems, which
are depicted at the top of this figure.  The Jacobi coordinate system on the  left, 
marked ``(a),'' is used to analyze the OH+H arrangement; the one on the right,
marked ``(b),'' is used for the H$_2$+O arrangement.
The vector $\vec{r}$ connects the nuclei
of the diatomic.  The vector $\vec{R}$ connects the center of mass of the
diatomic to the third atom.  $R$
is the length of $\vec{R}$, $r$ is the length of $\vec{r}$, and $\gamma$
is the angle between the $\vec{R}$ and $\vec{r}$ vectors.
For (a), $\gamma=0$ denotes a linear OHH configuration.

In addition to the three internal degrees of freedom there are also the
three Euler angles that orient the internal or body-fixed (BF) frame with respect to the 
lab or space-fixed (SF) frame.  The origin of both frames is the center of mass.
The space-fixed $Z$-axis is always chosen to be parallel with the wavevector of
the incident electron. For calculations with total rotational angular momentum 
$J=0$, the Hamiltonian only operates on the internal
degrees of freedom. For $J\neq 0$ we must take the Euler angles into account, and we denote them by $\alpha,
\beta, \zeta$.  

The total (electronic plus nuclear, ignoring spin) angular momentum, $J$, and its projection
upon the space-fixed $Z$ axis, $M$, are quantum numbers conserved by the Hamiltonian.  We also 
use the quantum number $K$ to specify the projection of the angular momentum on a BF axis. $K$ is not a conserved quantity and there is some flexibility in its definition. We use the ``$R$-embedding'' scheme~\cite{rembed}
in which $\vec{R}$ is taken to be collinear with the BF $Z'$ axis and the angular momentum
number $K$ is quantized around this axis. With this convention, the Euler angles $\alpha$ and $\beta$ are the polar angles which orient the $R$
vector with respect to the SF frame, and $\zeta$ is the third Euler
angle specifying orientation about the BF $Z'$ axis.  A schematic of the
coordinate system is also shown in Fig.~\ref{bigcoords}.

We may write a general expression for
the six-dimensional rovibrational wave 
function for a triatomic with specified $J$ and $M$ value 
as follows:
\begin{equation}
\chi_{\nu_i}(R,r,\gamma,\alpha,\beta,\zeta) = \sum_K
\widetilde{D}^J_{MK}(\alpha,\beta,\zeta) 
\frac{\chi_{\nu_i}^K(R,r,\gamma)}{Rr}, \label{triexpand}
\end{equation}
where the basis of $\widetilde{D}^J_{MK}(\alpha,\beta,\zeta)$ is the set 
of normalized Wigner rotation matrices (and BF angular momentum eigenstates)
\begin{equation}
\label{expansion}
\widetilde{D}^J_{MK}(\alpha,\beta,\zeta) = \sqrt{\frac{2J+1}{8\pi^2}}D^J_{MK}(\alpha,\beta,\zeta) 
\end{equation}
such that 
\begin{equation}
\begin{split}
\int_0^{2\pi} & d\alpha \int_{-1}^{1} d(\mathrm{cos} \ \beta) \int_0^{2\pi}
d\zeta \ 
\widetilde{D}^J_{MK}(\alpha,\beta,\zeta)
\widetilde{D}^{J'*}_{M'K'}(\alpha,\beta,\zeta) \\ & =
\delta_{J,J'} \delta_{M,M'} \delta_{K,K'}. \label{dnormalized}
\end{split}
\end{equation}
In Eqs.(\ref{expansion}) and (\ref{dnormalized})
we follow 
the conventions of Zhang~\cite{jzhzhang}, which for the $D^J_{MK}$ 
is the same as that of Edmonds~\cite{edmonds}.

The standard~\cite{petrongolo, hd2} 
BF Hamiltonian for the 
radial solutions $\chi^K_{\nu_i}$ of this expansion
incorporates coupling among the different $K$ values for
a given total angular momentum $J$.  The neglect of this
coupling is termed the ``coupled states'' or ``centrifugal
sudden'' (CS) approximation\cite{CS1,CS2}, and we employ
this approximation for our calculations, since the kinetic
energy of the recoiling fragments is large compared to their
centrifugal energy.  The resulting Hamiltonian
is thus diagonal in $K$ and can be written
\begin{equation}
\label{hamiltonian}
\begin{split}
H^J_K = & \frac{-1}{2\mu_R} \frac{\partial^2}{\partial R^2} +
\frac{-1}{2\mu_r}
\frac{\partial^2}{\partial r^2} + \left(\frac{\hat{j}^2}{2\mu_r r^2}+\frac{\hat{j}^2}{2\mu_R R^2}\right) \\ & + 
\frac{J(J+1) - 2K^2}{2\mu_R R^2} + V(R,r,\gamma)  \\
\hat{j}^2 = & -\left(\frac{1}{\sin \gamma}\frac{\partial}{\partial \gamma}
\sin 
\gamma \frac{\partial}{\partial \gamma} - \frac{K^2}{\sin^2 \gamma}\right) 
\qquad 
\end{split}
\end{equation}
where $\mu_r$ and $\mu_R$ are the reduced masses in either degree of freedom and
$V$ is the (coupled set of) Born-Oppenheimer potential-energy surface(s) that we calculate.

\subsection{Inclusion of Renner-Teller coupling}

For dynamics beginning on the $^2A_1$ (1 $^2A'$) resonance surface, the gradient
of that surface will force the wave packet toward linear geometry,
at which point this resonance state is degenerate with the $^2B_1$ resonance (see Fig.
\ref{c2vfig}).
The Renner-Teller effect~\cite{Renner,jungenmerer, carterhandyRT, petrongolo, RT1996, bersuker, RT2006_1, RT2006_2} 
will therefore couple these two components of the $^2\Pi$ state, and we modify the
Hamiltonian of Eq.(\ref{hamiltonian}) accordingly.

The quantum numbers
$J(J+1)$ and $K$ in
Eq.(\ref{hamiltonian}) are obtained as eigenvalues of the total angular momentum operators
$\hat{J^2}$ and $\hat{J_{z'}}$:
\begin{equation}
\begin{split}
\hat{J_{z'}} & \left( \widetilde{D}^J_{MK}(\alpha,\beta,\zeta) \frac{\chi_{\nu_i}^K(R,r,\gamma)}{Rr} \right) = \\
& K  \left( \widetilde{D}^J_{MK}(\alpha,\beta,\zeta) \frac{\chi_{\nu_i}^K(R,r,\gamma)}{Rr} \right) ,
\end{split}
\end{equation}
etc., where $\hat{J_{z'}}$ has a simple form in terms of derivative operators in 
($\alpha$, $\beta$, $\zeta$)~\cite{schatz}.  Properly,
the operators  that appear in the Born-Oppenheimer
Hamiltonian for the rovibrational motion of the nuclei, Eq.(\ref{hamiltonian}), 
should be not $\hat{J^2}$ and  $\hat{J_{z'}}$ but
the \textit{nuclear} angular momentum operators $\hat{R^2}$ and $\hat{R_{z'}}$, where
\begin{equation}
\hat{R_i} = \hat{J_i} - \hat{l_i},
\end{equation}
in which expression $\hat{l_i}$ is an electronic angular momentum operator; 
the Hamiltonian~\cite{petrongolo} with this form is exact except for the omission of the mass-polarization
term.

The exact Hamiltonian~\cite{petrongolo} introduces numerous new diagonal and off-diagonal (off-diagonal
in $K$, electronic state, and both) coupling terms to the triatomic Hamiltonian.  The 
term that is most commonly labeled the Renner-Teller coupling 
comes from the $\hat{j^2}$ term in
Eq.(\ref{hamiltonian}):
\begin{equation}
\begin{split}
\left(\frac{1}{2\mu_r r^2} +  \frac{1}{2\mu_R R^2}\right) \frac{K^2}{\sin^2 \gamma} 
& \rightarrow\left(\frac{1}{2\mu_r r^2}+\frac{1}{2\mu_R R^2}\right) \frac{\hat{R_{z'}^2}}{\sin^2 \gamma} \\ 
  = \left(\frac{1}{2\mu_r r^2}+\frac{1}{2\mu_R R^2}\right) & \frac{K^2-2K\hat{l_{z'}}+\hat{l_{z'}^2}}{\sin^2 \gamma}. \\ 
\label{angcentrifugal}
\end{split}
\end{equation}
It is the $2K\hat{l_{z'}}$ term which couples the two components (sine and cosine, $^2B_1$ and 1 $^2A'$ ) of the $^2\Pi$ state
at linear geometry.  At such geometries the operator $\hat{l_{z'}}$ is diagonalized by
\begin{equation}
\hat{l_{z'}} \left(\psi_{A'} \pm i\psi_{B1}\right) = \pm 1 \times \left(\psi_{A'}  \pm i\psi_{B1}\right) 
\label{lsubz}
\end{equation}

The matrix elements of $l_{z'}$ may either be computed~\cite{jungenmerer, RT2006_1, RT2006_2}, or approximated by their values
at linear geometry~\cite{carterhandyRT, RT1996}.  We take the latter route, 
i.e., we assume that Eq.(\ref{lsubz}) holds everywhere.
This approximation has little effect on the dynamics because only near linear geometry 
does the coupling become large.
We perform our Renner-Teller calculations in the ($l_{z'}=\pm 1$) diabatic basis
because it allows us to incorporate the boundary condition in $\gamma$ using
the ``\textit{K}-legendre'' discrete variable representation~\cite{hd2}.  
With this assumption, for a given value of $K$, the ($l_{z'}=\pm 1$) diabatic states have 
$R_{z'}=K \pm 1$.
The kinetic energy operator in Eq.(\ref{angcentrifugal}) is diagonal in this diabatic
basis.  The coupling then arises from the electronic Hamiltonian, which is not
diagonal in this basis.
The electronic Hamiltonian in this basis takes the form
\begin{equation}
V = \frac{1}{2}\left(\begin{array}{cc} 
V_{A'}+V_{B_1} \ & \ V_{A'}-V_{B_1} \\
V_{A'}-V_{B_1} \ & \ V_{A'}+V_{B_1} \\ 
\end{array}\right) ,
\end{equation}
i.e., the diabatic states are degenerate.  When $K=0$,
there is no Renner-Teller effect, since the coupling term in Eq.\ref{angcentrifugal} vanishes.

\section{The Multiconfiguration Time-Dependent Hartree  Method} \label{mctdhsect}

The Multiconfiguration Time-Dependent Hartree or MCTDH~\cite{mey90:73,man92:3199,Becketal,mey03:251} 
method is an efficient adaptive scheme for propagating quantum-mechanical
wave packets for systems with multiple degrees of freedom.
We use this method to perform the propagation in Eq. (\ref{timedepwfcn}).
We use the implementation within the MCTDH package\cite{mctdh82:i}, 
a freely available suite of codes built at the University of Heidelberg,
Germany.

In the MCTDH method, as in other methods developed for solving the
time-dependent Schr\"odinger equation, we start with a time-independent
orthonormal product basis set,
\begin{equation}
\{\chi_{j_1}^{(1)}(q_1)...\chi_{j_f}^{(f)}(q_f)\},\hspace{.25in}
j_\kappa=1 \cdots N_\kappa
\label{MCTDH1}
\end{equation}
for a problem with $f$ degrees of freedom and nuclear coordinates 
labeled $q_1,...q_f$. For
computational efficiency, the basis functions 
$\chi_{j_\kappa}^{(\kappa)}$ are chosen as the basis functions of a 
discrete variable representation (DVR)~\cite{Light}.

The central idea of the MCTDH technique is the representation of the
nuclear wave packet as a sum of separable terms,
\begin{equation}
\phi_{\nu_i}(\vec{q},t)=\sum_{j_1=1}^{n_1}...\sum_{j_f=1}^{n_f}A_{j_1...j_f}(t)
\prod_{\kappa=1}^f\varphi_{j_\kappa}^{(\kappa)}(q_\kappa,t),
\label{MCTDH2}
\end{equation}
with $n_\kappa\ll N_\kappa$.  Each ``single particle function'' (or SPF) 
$\varphi_{j_\kappa}^{(\kappa)}(q_\kappa,t)$ is itself represented in terms 
of the primitive basis,
\begin{equation}
\varphi_{j_\kappa}^{(\kappa)}(q_\kappa,t)=\sum_{i_\kappa=1}^{N_\kappa}
c_{i_\kappa j_\kappa}^{(\kappa)}(t) \chi_{i_\kappa}^{(\kappa)}(q_\kappa). 
\label{MCTDH3}
\end{equation}

One can determine equations of motion for the parameters $c_{i_\kappa j_\kappa}^{(\kappa)}(t)$
and $A_{j_1...j_f}(t)$.
Since both the coefficients $A_{j_1...j_f}$ and the single-particle
functions $\varphi_{j_\kappa}^{(\kappa)}$ are time-dependent, 
the wave function representation is made
unique by imposing  additional constraints
on the single-particle functions which keep them orthonormal for all
times~\cite{Becketal}.  

The evaluation of the Hamiltonian matrix, which must be carried out at
every time step, may be expedited~\cite{man92:3199,Becketal} if the Hamiltonian
can be written as a sum of products of single-coordinate operators.
The MCTDH package~\cite{mctdh82:i} includes a utility which performs a fit
of a given potential to a separable representation of this form.  Details
can be found in Beck {\it{et al.}}\cite{Becketal}.  All potential-energy
surfaces used in the current calculation were represented in this manner,
using this utility to fit them specifically for each choice of the
DVR grids.

For calculations on the electronically coupled $^2A_1$ and $^2B_2$ states,
the underlying DVR is the same for each electronic state, 
but each electronic state has its own set of
single-particle functions $\varphi_{j_\kappa}^{(\kappa)}$.  This 
is referred to as the ``multi-set'' formalism, as opposed to ``single-set.''
The Renner-Teller coupled 1 $^2A'$ - $^2B_1$ calculation is performed under
the single-set formalism.

\subsection{Complex absorbing potentials}

The sine DVR bases in the $r$ and $R$ degrees of freedom incorporate standing
wave boundary conditions at their edges.  
Therefore, when the dissociating wave packet reaches the end of the DVR grid, it must be absorbed to prevent unphysical backward reflections.  To this end we include an artificial
negative imaginary component to the surface called a 
``complex absorbing potential'' or CAP\cite{CAPref1, CAPref2}:
\begin{equation}
V_{CAP} \ = \ \begin{cases} \begin{split} 
0 & \quad \quad (R \le R_c) \\
i \eta (R-R_c)^2 & \quad \quad (R \ge R_c) ;
\end{split}
\end{cases}
\label{eq:cap}
\end{equation}
a similar expression for the CAP in the $r$ degree of freedom
also applies.
Formally, the CAP's provide the $+i\epsilon$ limit
in Eq.(\ref{fourier}).

We use a value for $\eta$ equal
to 0.007 hartree, and place $R_c$ three bohr before the end of our grid,
except for the 1 $^2A'$ calculations for $H_2$ and $D_2$, for which
we use a strength of 0.0018 hartree and a value of $R_c$ five bohr before
the end of the grid.

\subsection{Dissociative attachment cross sections from outgoing projected flux}

The cross sections for dissociative attachment
can be calculated directly from the time-propagated wave packet by
computing the energy-resolved, outgoing projected flux. The energy resolution 
is achieved by Fourier transform and a final state resolution is achieved by
the introduction of appropriate projection operators. For DEA to a specific
final rovibrational state labeled by rotational ($j$) and vibrational ($\nu$)
indices, we
use the projection operator
\begin{equation}
P_{j \nu} = \left\vert \frac{\chi_{j \nu}}{r} 
 \right\rangle \left\langle \frac{\chi_{j \nu}}{r} \right\vert.
\label{projector}
\end{equation}

The flux operator, which measures the flux passing through a surface 
defined by $R=R_c$, is defined as
\begin{equation}
\hat{F} =  i[H,h(R-R_c)], 
\end{equation}
where $h$ is a heaviside function.
The energy-resolved  projected flux is
then given by
\begin{equation}
\begin{split}
F_{j \nu}(E) \  & = \ 
  \frac{1}{2\pi} \int^\infty_0 dt \int^\infty_0 dt' \\ \times
& \langle \phi_{\nu_i} \vert e^{i(H-E)t} P_{j \nu} \hat{F} P_{j \nu} e^{-i(H-E)t'}
\vert \phi_{\nu_i}\rangle .
\end{split}
\end{equation}
The MCTDH package \cite{mctdh82:i} includes a utility which computes 
the outgoing projected flux.
In the actual calculations, the flux operator appearing in
the equation above is replaced by an expression involving the complex
absorbing potential, Eq.(\ref{eq:cap}).
This formulation of the flux operator is very convenient numerically
and entirely equivalent to the traditional formal definition of the
operator in this context, in the limit that the CAP does not perturb
the propagating wave packet beyond first order, which in the present case, 
given the nuclear
masses, holds as a good approximation.  
For more details on this CAP flux
formalism see Refs.~\cite{jae96:6778, Becketal, mey03:251}.

The resulting energy-resolved projected flux is that associated with 
the time-independent solution of the driven Schr\"odinger equation of 
the LCP model in Eq. (\ref{eq:boomerang}),
\begin{equation}
F_{j \nu}(E)  =  
\frac{1}{2\pi} \langle \xi_{\nu_i} \vert P_{j \nu} \hat{F} P_{j \nu} \vert 
\xi_{\nu_i}\rangle 
\label{withxi}.
\end{equation}
In terms of $F_{j \nu}$, the DEA cross section
is~\cite{haxton2}
\begin{equation}
\sigma_{DEA}^{j \nu} = 
    \frac{4\pi^3}{k^2} \ F_{j \nu}\left(E_{\nu_i}+\frac{k^2}{2}\right).
\label{cross}
\end{equation}
For the H$^-$+OH channel, an additional factor of
two is included in Eq.~(\ref{cross}) to account for the fact that
in a given calculation we perform the flux analysis for only one of the
two ${\rm H^-}$+OH arrangements, namely the one for which the Jacobi
coordinates are appropriate.

The definition of the rovibrational states $\chi_{j \nu}$ is complicated by 
the ion-dipole interaction of the fragments.
In our earlier study DEA to water via the 
$^2B_1$ Feshbach resonance\cite{haxton2},
we attempted a complete final state analysis, and projected upon
pendular (restricted rotor) states~\cite{pendular},
not free rotational states, and assumed that these pendular states evolve
adiabatically to their free rotational state asymptotes.  This analysis did
not yield any major insight, and so for the present calculations
we simply project upon free rotational
states.  As a consequence, there is a small error in our final state resolution,
but the magnitude of this error will span a range of states and
range of energies approximately equal to the magnitude of the ion-dipole
interaction at the edge of our grid, which is small compared to the kinetic
energy spread of the fragments.

\subsection{The DVR bases and other MCTDH parameters}

In most of the calculations reported here, we used DVR primitive basis sets for
all internal degrees of freedom, choosing the standard
sine DVR\cite{Becketal} for the $r$ and $R$ degrees of freedom and, for $J=0$, 
the Legendre DVR\cite{Becketal}
for $\gamma$.  For $J>0$, as previously discussed, the DVR for $\gamma$ must be modified to 
account for singularities in the 
Hamiltonian [see Eq.(\ref{hamiltonian})] due to the term $K^2/{{\mathrm{sin}}^2 
(\gamma)}$. This
is done by using an extended Legendre DVR~\cite{cor92:4115, cor93:1,hd2}, 
which is implemented in the Heidelberg MCTDH package~\cite{mctdh82:i}.

\subsection{Initial states}

The initial rovibrational states $\chi_{\nu_i}$ of Eq.~(\ref{drivdef})
were obtained via relaxation and improved relaxation~\cite{mey03:251}
as implemented within the MCTDH package~\cite{mctdh82:i}.
In relaxation runs, an initial guess $\chi_{g}(\vec{q},0)$ for the ground state is propagated in
imaginary time, which yields the ground state $\chi_0(\vec{q})$:
\begin{equation}
\chi_g(\vec{q},\tau) = e^{-H\tau}\chi_g(\vec{q},0) \quad \underset{\tau \rightarrow \infty}{\longrightarrow} \quad \chi_0(\vec{q})
\label{relax}
\end{equation}
In improved relaxation runs, the propagation of the SPF expansion coefficients
$c_{i_\kappa j_\kappa}^{(\kappa)}(\tau)$ of Eq.(\ref{MCTDH3}) 
is performed via Eq.(\ref{relax}), but
the configuration coefficients $A_{j_1...j_f}(t)$ are obtained anew
at each time-step via a Davidson diagonalization.

For the two-state $^2B_2$-$^2A_1$ calculations, the wavefunction
is represented in the diabatic basis, each component of which has
an expansion of the form of Eq.(\ref{MCTDH2}), and different sets
of time-dependent single-particle functions 
$\varphi_{j_\kappa}^{(\kappa)}(q_\kappa,t)$.  Since the adiabatic-to-diabatic
transformation angle is not constant with geometry, the single-particle 
functions which represent the initial state will be different in the
diabatic basis than in the adiabatic basis (in which they would
be identical to the single-particle functions of the relaxation run).
For this reason, an iterative technique~\cite{Becketal} is employed
to minimize the error between the diabatic representation and
its adiabatic representation.

\section{Calculated Cross Sections for Dissociative Electron Attachment to Water } \label{resultssect}

\begin{figure*}
\begin{center}
\begin{tabular}{cc}
\includegraphics*[width=0.45\textwidth]{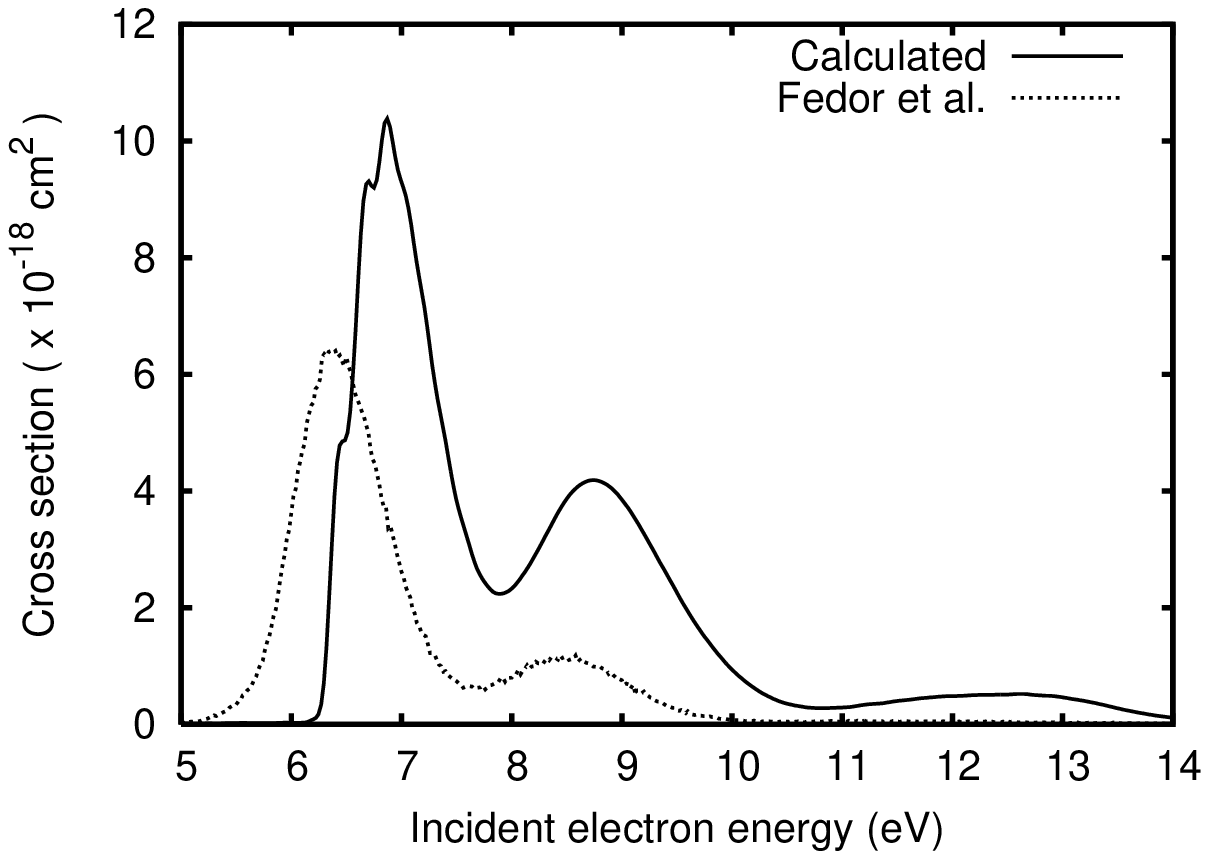} &
\includegraphics*[width=0.45\textwidth]{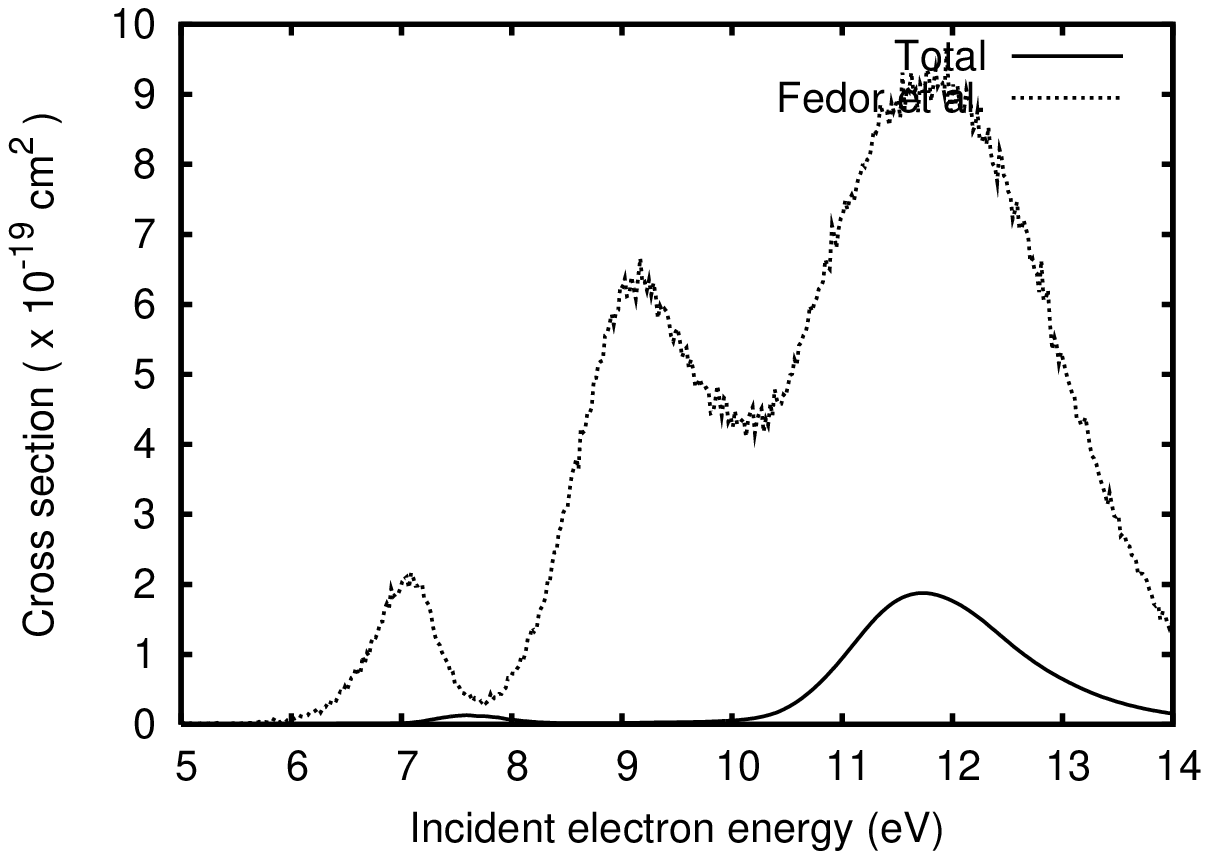} \\
\includegraphics*[width=0.45\textwidth]{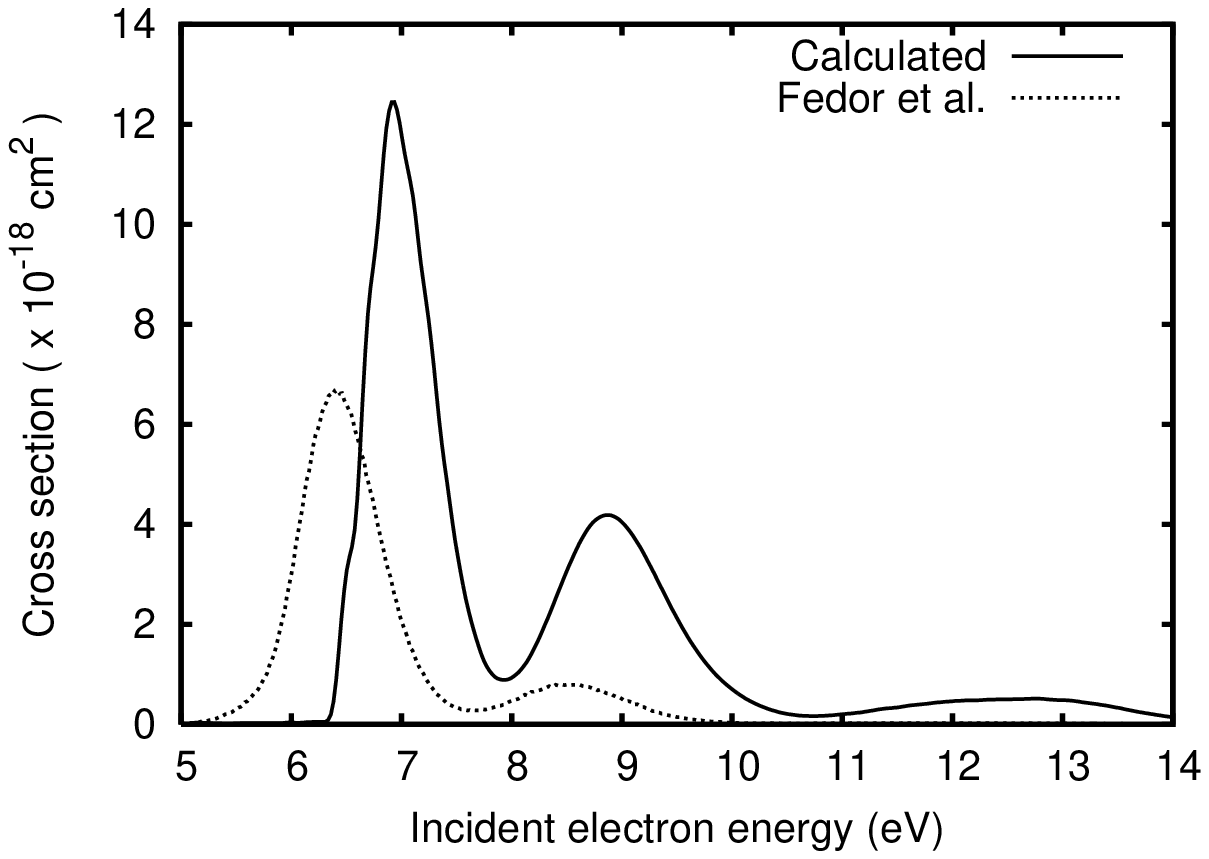} &
\includegraphics*[width=0.45\textwidth]{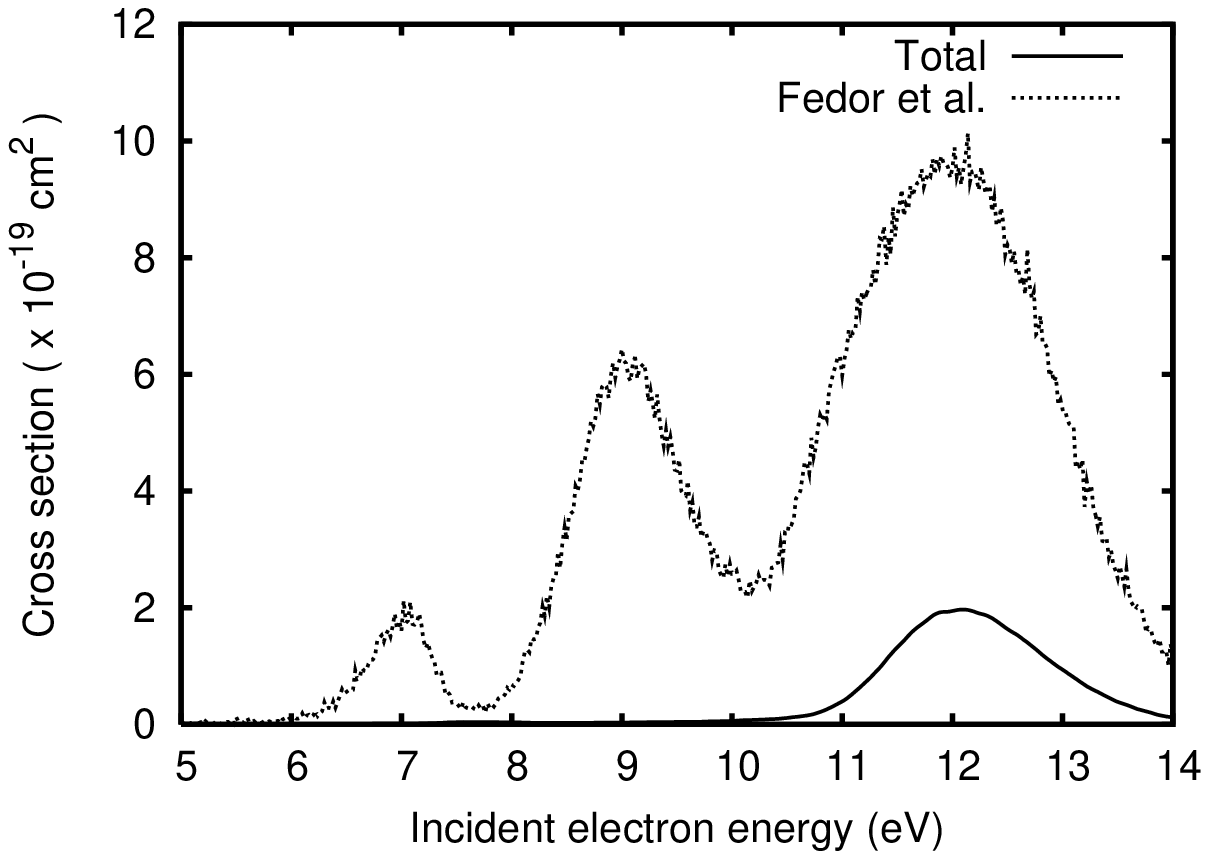} 
\end{tabular}
\end{center}
\caption[Total cross sections]{Cross sections for production of 
H$^-$ or D$^-$ (left) and O$^-$ (right) as a function
of incident electron energy, summed from different MCTDH calculations.  
Top, cross sections from H$_2$O; bottom, cross sections from D$_2$O.
Experimental results
of Fedor \textit{et al.}\cite{fedor} included for comparison.  The experimental
data, which do not have absolute normalization, are normalized to
agree with Compton and Christophoreau's~\cite{compton} H$^-$+OH
peak height for DEA via the $^2B_1$ resonance at 6.5$\times$
10$^{-18}$ cm$^2$.}
\label{xsects_total}
\end{figure*}

We will present cross sections for dissociative
electron attachment to water into the three different
atom-diatom arrangements that are present as asymptotes
of the $^2B_1$, $^2A_1$, and $^2B_2$  Feshbach resonances, 
\begin{equation}
\mathrm{H}_2\mathrm{O} + e^- \to 
\begin{cases} 
\mathrm{H}_2 + \mathrm{O}^- & 3.56 \ \mathrm{eV}  \nonumber \\
\mathrm{H}^- + \mathrm{OH} \quad (X \ ^2\Pi) & 4.35 \ \mathrm{eV}\nonumber \\
\mathrm{H}^- + \mathrm{OH}^* \quad (^2\Sigma) & 8.38 \ \mathrm{eV} ,\nonumber 
\end{cases}
\label{asymptotes2}
\end{equation}
resolved into the final rovibrational states of the diatomic fragment,
as a function of incident electron
energy.  The final state resolution allows us to determine the
kinetic energy of the diatomic fragment.  Therefore, we are able
to calculate cross sections as a function of both the incident
electron energy and the kinetic energy of the recoil, which
data we may easily compare with experiment.
These two dimensional data provide a comprehensive view of 
dissociative attachment via each of the resonances.  We calculate
the degree of rotational and vibrational excitation, and show
how these quantities change with the incident electron energies.

We have obtained converged cross sections for all channels 
considered, with two exceptions.  For the Renner-Teller
coupled $^2A_1$ (1 $^2A'$) and $^2B_1$ states, we have been
unable to obtain a non-zero result for the minor H$_2$+O$^-$ channel, as
well as its deuterated counterpart.  For the production of
H$^-$ from the $^2B_2$ state, coupled to the $^2A_1$ state
via the conical intersection, our calculations are not fully converged,
although we do obtain total cross sections
and, for the OH ($^2\Sigma$) fragment, final state resolution.

In general, our results are in qualitative, though not quantitative, agreement
with the experimentally measured cross sections
for the \textit{major} product arrangments observed in DEA
via each resonance state:
H$^-$+OH from the $^2B_1$ and $^2A_1$ resonances, and 
H$_2$+O$^-$ from the highest-energy $^2B_2$ resonance.
We have had difficulty obtaining results for the minor channels,
failing to reproduce the experimental result for the magnitude
of the cross section for production of H$_2$+O$^-$ via the
first two resonances, and not
being able to fully converge the calculation for production of
H$^-$+OH ($^2\Pi$ and $^2\Sigma$) via the $^2B_2$ resonance.

Total cross sections calculated for
the two anion-diatom arrangements, and for both D$_2$O and
H$_2$O, are presented in Fig.~\ref{xsects_total}, along with
recent experimental results from Fedor \textit{et al.}\cite{fedor}.
Calculated and experimental peak heights and locations are
collected in Table \ref{resultstable}.
In Fig.~\ref{xsects_total}, 
the experimental data are internormalized but not absolutely
normalized; therefore, for the purposes of comparison, we normalize
the experimental peak for the production of H$^-$+OH via the 
lowest-energy $^2B_1$ resonance to Compton and
Christophoreau's\cite{compton} result of 6.5$\times$10$^{-19}$ cm$^2$.
The calculated curves are obtained by summing the cross sections
into the individual rovibrational states $\chi_{j \nu}$ of the ion
+ diatom arrangement.  Therefore, the three-body cross sections
are neglected.  The recent experimental results of Ref. ~\cite{fedor}
do not resolve the kinetic energy of the atom-diatom recoil and therefore
do not distinguish between the two- and three-body DEA cross section.
Thus, to the degree that three body breakup is important, our calculated
results cannot be compared directly with these experiments, 
for incident electron energies which exceed the
three-body dissociation thresholds of either 8.04eV (H+H+O$^-$)
or 8.75eV (H+H$^-$+O).  

We can draw the following conclusions from Fig.~\ref{xsects_total}
and Table \ref{resultstable}.
First, it is clear that the entrance amplitude for the $^2B_1$ resonance
has been overestimated by our present study, because the magnitude of 
the DEA cross section via this resonance is entirely controlled by 
its entrance amplitude, and we have overestimated the experimental 
peak height by
nearly 60\%.  Also, similar to the result of our previous 
study\cite{haxton1,haxton2}, we overestimate the energy at which
the H$^-$ cross section via the $^2B_1$ state peaks by about 0.4eV.
The calculated peak location, 6.87eV, is larger than the vertical transition
energy for the $^2B_1$ resonance as defined by our potential-energy
surface constructed in paper I, which is 6.63eV.  This value is obtained through a
configuration-interaction treatment of the resonance; using 
complex Kohn scattering calculations, we obtained a value of 6.09eV.
The comparison between the calculated and experimental peak
locations indicates that
the physical value of the vertical transition energy for the $^2B_1$
Feshbach resonance is probably about 6.2eV, nearer to the complex
Kohn result.

Similar observations apply to the comparison between the calculated
and experimental results for DEA via the $^2A_1$ resonance
to produce H$^-$+OH; the calculated peak height is too large and
located at a higher energy than is experimentally observed.  
Therefore, it is possible that we have overestimated the entrance
amplitude and the vertical transition energy  for this resonance as well.
The transition energy as defined by our potential-energy 
surface is 9.01eV; from complex Kohn calculations we obtained
a lower result of 8.41eV, which is probably closer to the physical
value.
However, as we explain further below, the disagreement in
magnitude and location between the calculated and observed results may indicate
a breakdown of the local complex model for DEA via the $^2A_1$
resonance.

\begin{table}[!t]
\caption[Calculated peak cross sections]{Peak cross sections ($\sigma$, in units of 10$^{-19}$ cm$^2$ )
and peak locations ($E$, in eV) calculated for DEA 
via the three resonances, .  Experimental peak locations
are taken from the data of Fedor \textit{et al.}, except where noted. }
\label{resultstable} 
\begin{tabular}{|c|cccc|cccc|}
\hline
  & \multicolumn{4}{c|}{Calculated} & \multicolumn{4}{c|}{Experiment} \\
  & &   $^2B_1$            &   $^2A_1$    &    $^2B_2$  &   $^2B_1$            &   $^2A_1$    &    $^2B_2$ &  \\
\hline
OH + H$^-$  & $\sigma$ &  103.7      &   41.4   &    2.61\footnotemark[1]   &  65\footnotemark[2]   &  13\footnotemark[2]  &  &  \\
 (X $^2\Pi$) & $E$ & 6.87  &   8.74     & 11.54    & 6.4   &  8.4  &  &  \\
\hline
OH + H$^-$  & $\sigma$ &       &        &  3.67\footnotemark[1]    &    &    &  & \\
($^2\Sigma$) & $E$ &    &    &  12.68    &    &    &   &  \\
\hline
H$^-$  & $\sigma$&      &        &  5.21\footnotemark[1]    &    &    &  ? & \\
(total, $^2B_2$) & $E$ &     &    &  12.61    &    &    &   11.8\footnotemark[3] & \\
\hline
H$_2$ + O$^-$     &  $\sigma$ &   0.121     &   \textbf{0}    &   1.87    &  1.3\footnotemark[4]  & 3.2\footnotemark[4]   & 5.7\footnotemark[4] &  \\
&  $E$ &  7.62   &    &  11.75    &  7.1   &  9.0  &  11.8 &  \\
\hline
\hline
OD + D$^-$  & $\sigma$&  124.4     &    41.6      &   1.45\footnotemark[1]   & 52\footnotemark[2]   &  6\footnotemark[2]  &  &  \\
(X $^2\Pi$)&   $E$ & 6.93   &  8.87   &  11.93   &   6.4  &  8.5   & &   \\
\hline
OD + D$^-$  & $\sigma$&       &          &      1.96\footnotemark[1]      &    &    &  &  \\
($^2\Sigma$) &$E$ &     &    &    12.86  &    &    &  &   \\
\hline
D$^-$ & $\sigma$ &     &        &  2.60\footnotemark[1]    &   &   & ?  & \\
 (total, $^2B_2$)&$E$&     &    &  12.57    &    &    &  &    \\
\hline
D$_2$ + O$^-$   & $\sigma$     &  0.0242     &    \textbf{0}      &     1.97       &   ?  &  ?   & ?  &  \\
&  $E$ & 7.63   &    &  12.10    &  7.1   &  9.0    & 12.0 &   \\
\hline
 \end{tabular}
 \footnotetext[1]{ Calculation not converged for H$^-$+OH arrangement via $^2B_2$ resonance.}
\footnotetext[2]{ Compton and Christophoreau, Ref.~\cite{compton}}
\footnotetext[3]{ Jungen, Ref.~\cite{Jungen}}
\footnotetext[4]{Melton, Ref.~\cite{Melton}}

\end{table}

The data in Fig. \ref{xsects_total} and Table \ref{resultstable} indicate that, while
the calculations overestimate the cross sections for H$^-$+OH production, they
evidently underestimate those for H$_2$+O$^-$ production.  
 H$_2$+O$^-$ is the major channel for DEA via the $^2B_2$ resonance.
 As explained in Ref.~\cite{haxton3},
the presence of this channel is entirely due to nonadiabatic coupling
between the upper $^2B_2$ (2 $^2A'$) resonance and the lower $^2A_1$
(1 $^2A'$) resonance via their conical intersection.
As we will show, the magnitude of this cross
section is determined by active competition between different
product arrangements, the dynamic effects of both the real and
imaginary components of the surface, as well as the conical
intersection dynamics.  We regard the agreement with experiment
that we have obtained to be quite good, considering the complexity
of the system.  We note that the location of the calculated
peak for O$^-$ production which we have calculated (11.75eV) 
agrees well with the experimental value (11.8eV).  The peak
location may be contrasted with the vertical excitation energy,
which was calculated in paper I to be
12.83eV.  The peak maximum is a full 1 eV below the vertical
transition energy, which difference
reflects the influence of autodetachment
upon the nuclear dynamics.  The large autodetachment probability
weights those components of the propagating wave packet which
are closer to the product arrangement, i.e., lower on the
potential-energy surface, and results in a breakdown of the
multidimensional reflection principle.

The product channel H$_2$+O$^-$ is the minor one for DEA via
the first two resonances, $^2B_1$ and $^2A_1$.  We have failed
to reproduce the corresponding experimental results; our calculations
produce a very small cross section for DEA via the $^2B_1$ resonance
to produce this channel, and zero cross section for $^2A_1$.
We regard nonlocal effects to be a prime candidate for the physical
origin of this channel for the $^2B_1$ state; for the $^2A_1$
state, we suspect that three-body dissociation into H+H+O$^-$, 
which is not treated in the present study, may play a significant role
in this channel.

The probability of a dissociative attachment event is neatly divided into
distinct probabilities for attachment and survival by the local complex
potential model.
The norm of the driving term in the driven Schrodinger equation
of the LCP model, Eq.(\ref{eq:boomerang}), corresponds to 
the probability per for electron attachment, weighted by the envelope
of the initial vibrational state.  We define this quantity as the
attachment width $\Gamma_A$,
\begin{equation}
\label{gammaA}
\Gamma_A =2\pi\left\langle \phi_{\nu_i} \vert \phi_{\nu_i} \right \rangle,
\end{equation}
and list the values of $\Gamma_A$ for each of the resonances 
in Table \ref{survivaltable}.

Once the electron has attached, the loss of flux via the 
imaginary component of the complex-valued surfaces
determines the survival probability of the anion state.
The survival probability may be calculated by integrating the flux $F_{j \nu}(E)$
over energy:
\begin{equation}
\label{psurvdef}
P_{surv} = \frac{\sum_{j \nu} \int dE \ F_{j \nu}(E)}{\langle \phi_{\nu_i} \vert \phi_{\nu_i} \rangle}.
\end{equation}
The calculated survival probabilities
are also listed in Table \ref{survivaltable}.  
The survival probability for the lowest
$^2B_1$ resonance state is near 1, and therefore for this resonance the magnitude of the cross section
is controlled by the attachment probability.  For DEA via 
the $^2A_1$ state, the cross section is lowered by the effect of autodetachment,
though once attached the electron is more likely to survive to dissociation.
For DEA via the upper $^2B_2$ resonance, the large majority of the attached
wave packet is lost to autodetachment; therefore, the variation
of the lifetime of this state with nuclear geometry 
plays a large role in the dynamics.

\begin{table}[t]
\caption[Calculated attachment widths and survival probabilities]{Attachment
  widths and survival probabilities calculated for the three resonances using
  Eqs.(\ref{gammaA})-(\ref{psurv}). }
\label{survivaltable} 
\begin{center}
\begin{tabular}{|c|ccc|ccc|}
\multicolumn{1}{c}{ }   & \multicolumn{3}{c}{Attachment width}  & \multicolumn{3}{c}{Survival} \\
\multicolumn{1}{c}{ }   & \multicolumn{3}{c}{( $\times$10$^{-4}$ a.u. )}  & \multicolumn{3}{c}{probability} \\
\hline
          &   $^2B_1$     &   $^2A_1$    &    $^2B_2$  &   $^2B_1$     &   $^2A_1$    &    $^2B_2$ \\
\hline
H$_2$O       &   3.47       &   4.33     &    4.74     &   93.8\%    &    65.1\%    &    21.5\%     \\
D$_2$O       &   3.54       &   4.19     &    4.84     &   91.6\%     &   57.2\%    &   13.1\%     \\
\hline
\end{tabular}
\end{center}
\end{table}

We calculate the average
degree of rotational and vibrational excitation, as well as the average kinetic
energy of the anion recoil, for each of the final channels
by weighting the survival probability by the quantity of interest,
\begin{equation}
\left\langle \nu \right\rangle  = \frac{\sum_{j \nu} \nu \int dE \ F_{j \nu}(E)}{P_{surv}\langle \phi_{\nu_i} \vert \phi_{\nu_i} \rangle} \nonumber \\
\end{equation}
\begin{equation}
\left\langle j^2 \right\rangle  = \frac{\sum_{j \nu} j(j+1) \int dE \ F_{j \nu}(E)}{P_{surv}\langle \phi_{\nu_i} \vert \phi_{\nu_i} \rangle} \nonumber \\
\end{equation}
\begin{equation}
\label{psurv}
\left\langle E_{kin} \right\rangle  = \frac{\sum_{j \nu} \frac{M_{diatom}}{M_{total}}(E_{inc} - E_{j \nu}) \int dE \ F_{j \nu}(E)}{P_{surv}\langle \phi_{\nu_i} \vert \phi_{\nu_i} \rangle}. \\
\end{equation}
In the third line of Eq.(\ref{psurv}), $E_{inc}$ is the incident electron energy, $E_{j \nu}$ is
the energy of the final state relative to the ground vibrational state of H$_2$O, and $M_{total}$ and
$M_{diatom}$ are the masses of the original triatom and the diatomic fragment, respectively, so
that the resulting quantity is the kinetic energy of the anion recoil in the laboratory frame.
We present these results in Table \ref{erottable}, along with experimental
data on the anion recoil kinetic energy from Fedor \textit{et al.}~\cite{fedor}.

Our calculated average values of the
anion recoil kinetic energy, $\langle E_{kin} \rangle$, agree to varying degrees with the
results of Ref.~\cite{fedor}.  Our calculated values for average kinetic
energy release for the production of H$^-$ from the $^2B_1$ and $^2A_1$ resonances
are much larger than observed by these authors, but closer to the values measured by Belic {\em et al.}~\cite{belic}.
 As described in paper I,
the potential-energy surfaces which we have constructed for these resonance states reproduce
the energetics of the two-body asymptotes very well (to within 0.08eV for the ground vibrational
state).  Errors in the present results may therefore only come from errors in the vertical transition
energies, or a misrepresentation of the dynamics prior to breakup. 
As discussed above, it is likely that our vertical transition energies for the
$^2B_1$ and $^2A_1$ resonances are too high, perhaps by as much as 0.4eV
relative to the proper physical values.  Most of this excess energy
may be transmitted to the kinetic energy of the H$^-$ recoil, due to the
small mass of hydrogen relative to the H$_2$O molecule.  While our
calculated results exceed the experimental result of Ref.~\cite{fedor}
by more than 0.4eV, they are within 0.4eV of the Belic {\em et al.} value for the $^2B_1$ state and 0.85 eV for the $^2A_1$ state.  Fedor \textit{et al.} comment that the values  obtained
for the kinetic energy release of H$^-$ via the $^2B_1$ and $^2A_1$
resonances 
by Belic {\em et al.}~\cite{belic} ``may be considered as more accurate.''
The discrepancy with  Belic {\em et al.} for $^2B_1$ result supports our recommendation
that the physical transition energy for the $^2B_1$ state be taken to be
approximately 6.2eV.  The maximum kinetic energy release at our
calculated peak locations of 6.87 and 8.74eV for the $^2B_1$ and
$^2A_1$ states, is, respectively, 2.38 and 4.15eV.  Our results
are therefore very near the maximum values and reflect the small
average degree of vibrational excitation which we calculate.

\begin{table}[t]
\caption[Degree of excitation of diatomic fragments]{Expectation values
of final vibrational quantum number, $\langle \nu \rangle$, and that of angular momentum quantum
number, $\langle j^2 \rangle$, of diatomic fragment, as well as the expectation value of 
the kinetic energy of the anion recoil, $\langle E_{kin} \rangle$, for each resonance, as calculated with Eq.(\ref{psurv}).
Average kinetic energy determined by the experimental method of Ref. \cite{fedor}, final column. }
\label{erottable}

\begin{tabular}{c|cccc}
\hline
\hline
   Diatomic fragment  &   $\langle \nu \rangle$  &  $\langle j^2 \rangle$  & $\langle E_{kin} \rangle$  & $\langle E_{kin} \rangle$, exp't.\footnotemark[1] \\
\hline
\multicolumn{1}{l|}{$^2B_1$} &\\
 H$^-$+OH      &  1.28  &  107      &   2.04eV   &  0.96eV \\
       &    &        &      &  (1.5eV)\footnotemark[2]  \\
   H$_2$ + O$-$                   &    3.52  &  412   &   0.154  &   0.12  \\
   D$^-$ + OD                     &   1.75   &  240   &   1.89  &  0.70   \\
   D$_2$ + O$-$                   &   3.34  &  821   &     0.125   &   0.14   \\
\hline
\multicolumn{1}{l|}{$^2A_1$ (1 $^2A'$)} &\\
 H$^-$ + OH  ($K$=0)               &  2.11  &  121  &   3.35    &   1.55  \\
                &    &    &       &   (2.5 eV)\footnotemark[2]\\  
  D$^-$ + OD  ($K$=0)            &    2.98   &  221 &  3.16   &   1.20   \\
\hline
\multicolumn{1}{l|}{$^2B_2$ (2 $^2A'$)} &\\
 H$^-$ + OH ($^2\Pi$)       &  \footnotemark[3]  &  \footnotemark[3] & \footnotemark[3]   &       \\
   H$^-$ + OH ($^2\Sigma$)  &  4.69\footnotemark[3]  &  439\footnotemark[3] & 3.45\footnotemark[3]   &      \\
   H$_2$ + O$-$             &  7.75         &  405        &  0.413 &  0.57   \\ 
   D$^-$ + OD ($^2\Pi$)     &  \footnotemark[3]  &  \footnotemark[3] & \footnotemark[3]   &     \\
   D$^-$ + OD ($^2\Sigma$)  &  6.63\footnotemark[3]  &  819\footnotemark[3] & 3.35\footnotemark[3]   &      \\
   D$_2$ + O$-$             & 13.0          &  725        & 0.684  &  0.79    \\
\hline
\hline
\end{tabular}
 \footnotetext[1]{Data from Ref.\cite{fedor}, except where noted.}
 \footnotetext[2]{Ref. \cite{belic}}
 \footnotetext[3]{The calculation for H$^-$ production via $^2B_2$ is not converged.}

\end{table}

Our values for the average kinetic energy release of the major O$^-$
fragment from the $^2B_2$ resonance agree much better with the
results of Fedor \textit{et al.}; we again underestimate the
experimental result, but only by 28\% and 13\%, respectively, for the
nondeuterated and deuterated target.  For this channel,
the degree of excitation of the H$_2$ (D$_2$) fragment is large,
and therefore the kinetic energy of the atom-diatom recoil is
less than its maximum allowed value.  The maximum kinetic energy
release at our calculated peak (11.75eV) is 0.91eV, more than
twice our result for the average value.  Therefore, more energy
goes into the rovibrational excitation of the H$_2$ fragment than
into the kinetic energy of the recoil.

We do not calculate the three-body dissociative electron
attachment cross section, i.e.,
\begin{equation}
\mathrm{H}_2\mathrm{O} + e^- \to 
\begin{cases} 
\mathrm{H} + \mathrm{H} + \mathrm{O}^- & 8.04 \ \mathrm{eV}\\
\mathrm{H}^- + \mathrm{H} + \mathrm{O}  & 8.75 \ \mathrm{eV} .
\end{cases}
\label{asymptotes3}
\end{equation}
The complex absorbing potential flux formalism~\cite{jae96:6778,Becketal,mey03:251} 
which
is employed within the MCTDH implementation~\cite{mctdh82:i}
is not appropriate for the three-body breakup channel, at
least when used in conjunction with 
the Jacobi coordinate systems used here. We do, however, produce
rigorous results for the two-body channels by projecting upon
the bound rovibrational final states as in Eq.(\ref{projector})
and summing.  

Our surfaces, as described in paper I,
are not designed to reproduce the dynamics
leading to three-body dissociation either.
Due to our neglect of the shape/Feshbach resonance intersection
on the $^2B_2$ manifold, which is a true characteristic of the
physical system and which leads to the double-valuedness of the
physical surface, we cannot accurately represent the dynamics
leading to the three-body dissociation channels with our single $^2B_2$
surface.  The $^2B_2$ manifold is coupled to the $^2A_1$ state
in the three-body region by the conical intersection, and therefore
it is possible that this omission affects the dynamics via the
$^2A_1$ state as well.

It is clear that we may only rigorously compare our results
with experiment for the two-body channels.  The comparison is complicated
by the fact that the experimental results are sometimes not
final-state resolved, such as those presented in Fig. \ref{xsects_total},
and in any case always incorporate a finite
resolution in determining the energy of the incident electron energy 
and the kinetic energy of the recoil.  The energetics of
the asymptotes of the physical surfaces, which are mirrored very well by our constructed
potential-energy curves, dictate that for the lowest $^2B_1$ state
the three-body channels are closed, but that for DEA via the other
two resonances, at least one three-body channel is open.

In the following subsections, we give the principal findings of the
nuclear dynamics calculations for each channel that was studied.
Further details are given in the EPAPS archive\cite{epaps}.  Most
calculations were carried out for the ground vibrational state and for total
rotational angular momentum $J=0$ (or, in the case of the Renner-Teller
coupled 1 $^2A'$-$^2B_1$ calculations, $R_{z'}=0$).  Rovibrationally
excited initial states were also examined, and these are listed
in the descriptions of the individual calculations which follow.

\subsection{Dissociative electron attachment via the $^2B_1$ state}

We have previously~\cite{haxton1, haxton2} performed a calculation on the $^2B_1$
state, which is superseded by the present treatment.  
We perform the calculation using the one uncoupled $^2B_1$ potential-energy surface.
We have confirmed that Renner-Teller coupling to the $^2A_1$ state at linear 
geometry has a negligible
effect on the dynamics, at least for DEA via the ground rovibrational state of the target.

The treatment in our previous study~\cite{haxton1, haxton2} was not able to resolve the cross
section in the minor O$^-$+H$_2$ channel ($\frac{1}{40}^{\mathrm{th}}$ of the major channel), due to
deficiencies in the potential-energy surface.  While the present study
does obtain converged O$^-$+H$_2$ cross sections, they are two orders of magnitude below the observed cross sections, and peak at
energies well above the experimental peaks; it is
therefore clear that we have not represented the dynamics leading to this minor channel
accurately. The failure in this regard may be due to 
small deficiencies in the potential-energy surface, or to the presence
of significant nonlocal effects in this minor channel.

In the previous study, we 
reproduced the magnitude of the OH+H$^-$ cross section to within a few percent.
We continue to regard that close agreement to be essentially fortuitous.
Apart from an overall scale factor, the present calculations for  production of H$^-$ 
reproduce the shape and energy dependence of  the experimental
results very well.  It appears that the current value calculated for the width of the $^2B_1$
resonance at the equilibrium geometry of the neutral, 10.31meV, is larger the physical value, 
the latter being closer to our previously calculated value of 
6.0meV.  It is likely that a more accurate description of electron correlation
than we could include in the complex Kohn calculations of paper I
is required to reproduce the resonance wavefunction of the $^2B_1$ state.

\begin{figure}
\begin{center}
\begin{tabular}{cc}
\includegraphics*[width=0.45\textwidth]{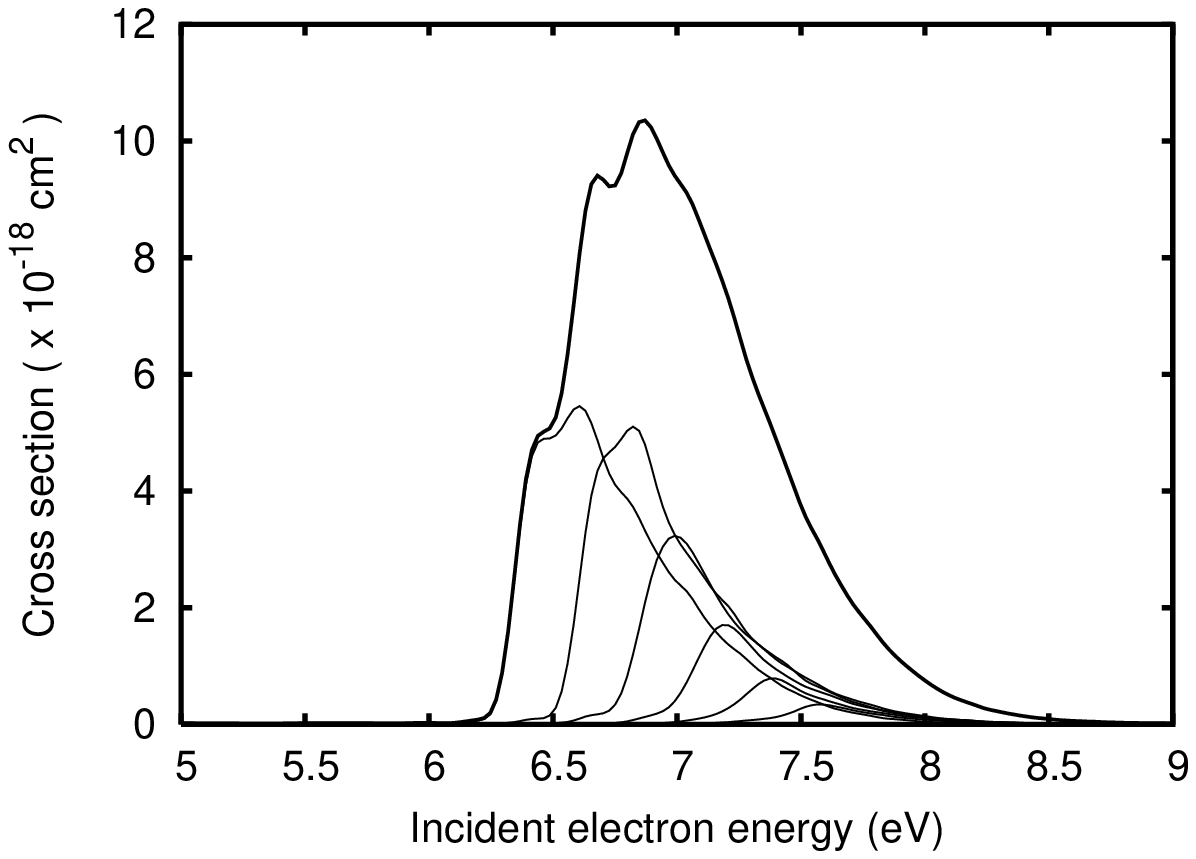} \\
\includegraphics*[width=0.45\textwidth]{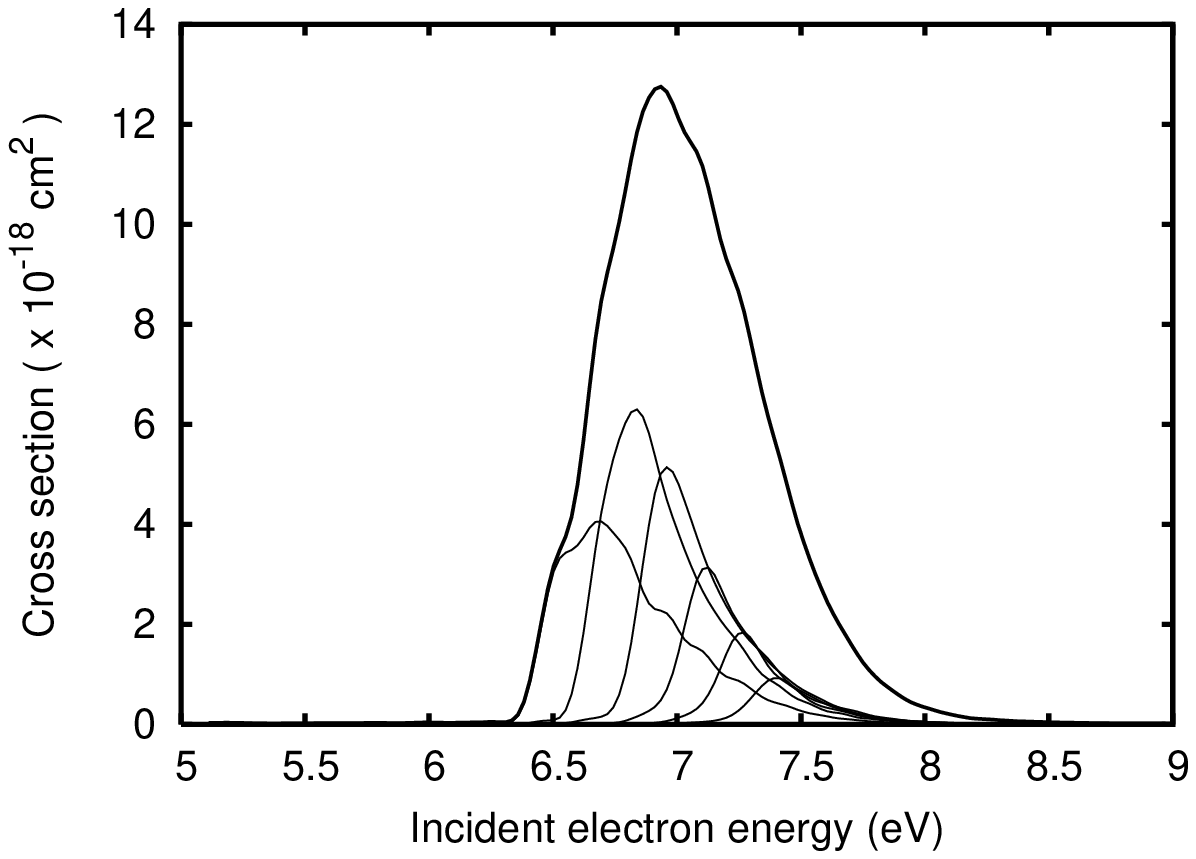}
\end{tabular}
\end{center}
\caption[Cross sections, H$^-$+OH from $^2B_1$]{Cross sections for production of H$^-$+OH ($\nu$), top, or D$^-$+OD ($\nu$), bottom, from $^2B_1$ state as a function
of incident electron energy.  Total, thick line; vibrational states 0 (ground)
through 5, thinner lines.}
\label{OH_H_2b1}
\end{figure}

\begin{figure}
\begin{center}
\begin{tabular}{c}
\includegraphics*[width=0.45\textwidth]{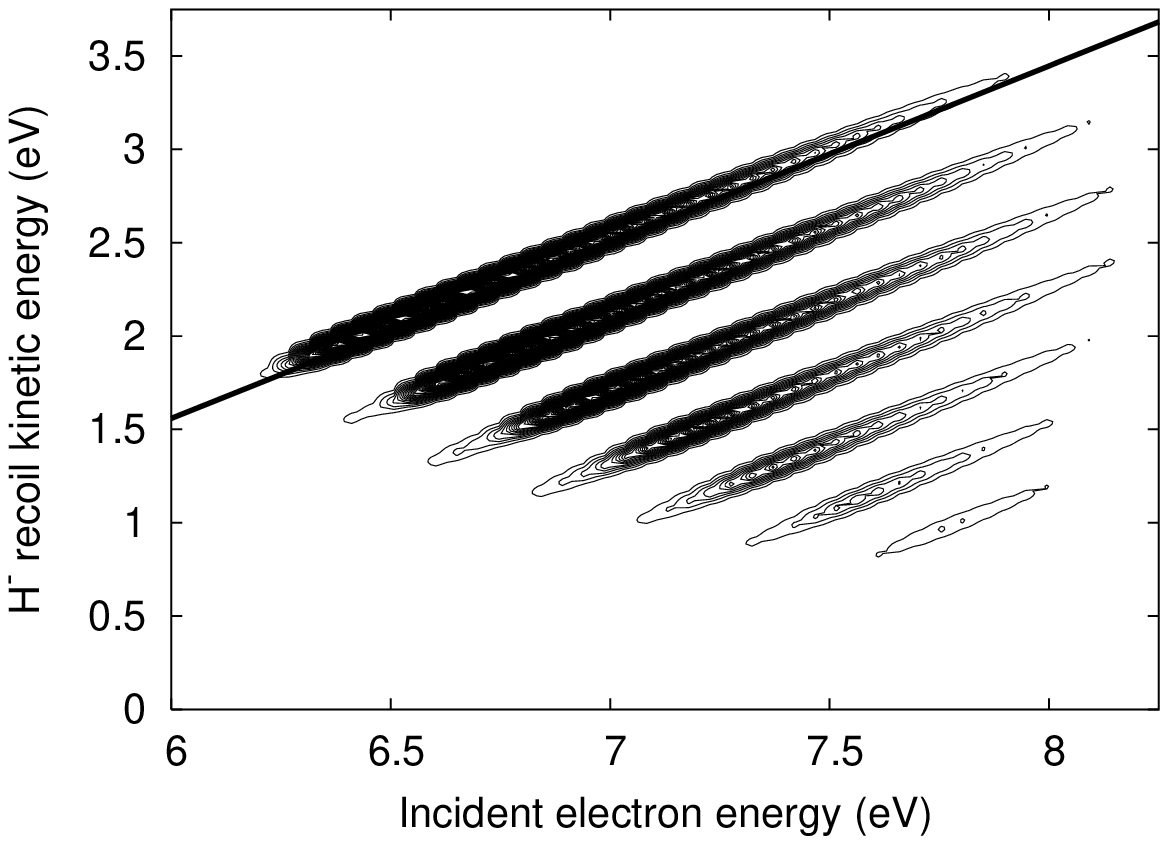} \\
\includegraphics*[width=0.45\textwidth]{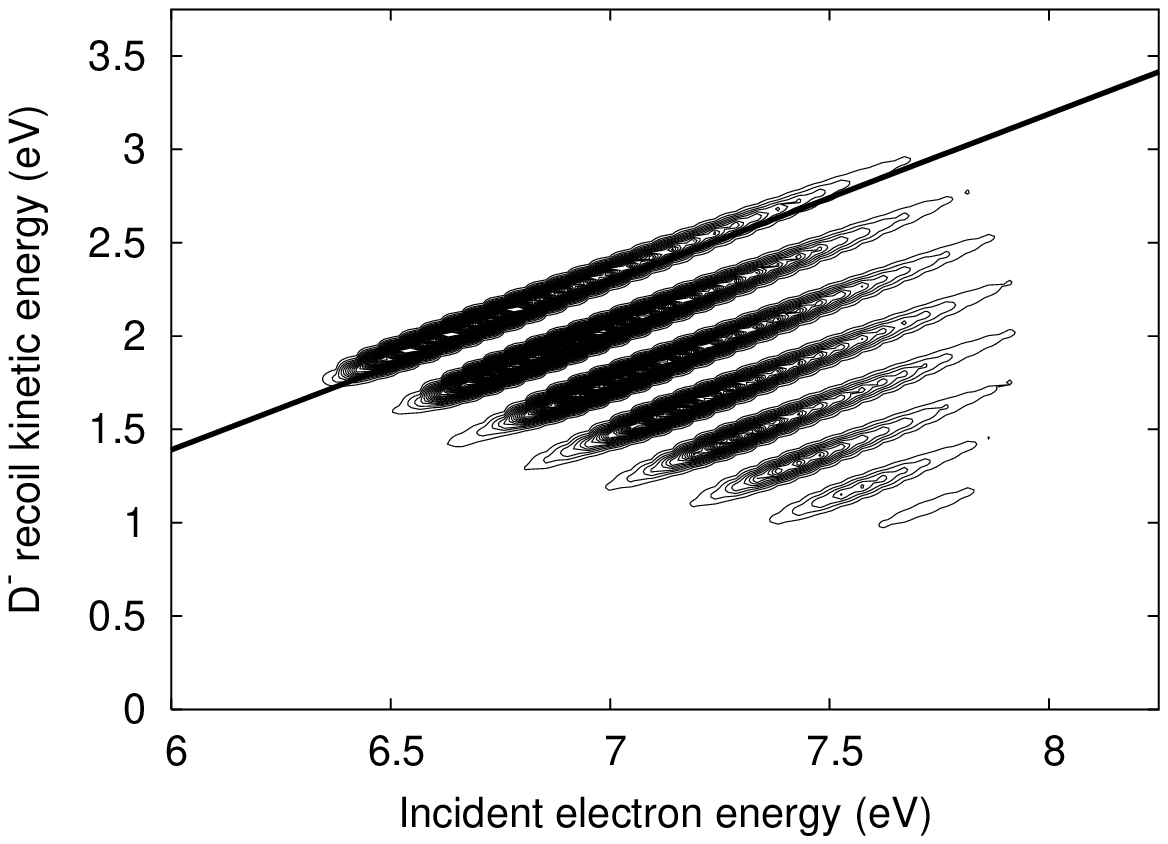} 
\end{tabular}
\end{center}
\caption[H$^-$ / D$^-$ kinetic energy distributions for $^2B_1$, 2D]{Top, 
cross section for production of 
H$^-$+OH from $^2B_1$ resonance as a function of incident electron energy and
H$^-$ fragment kinetic energy, unshifted, with
the physical value of the maximum kinetic energy available plotted as bold line;
bottom, deuterated.  The physical value for the maximum kinetic energy 
is slightly lower than the value
corresponding to our calculated surfaces.  
Contours every 2 $\times$ 10$^{-17}$ cm$^2$ eV$^{-1}$. }
\label{keplot_2b1J1}
\end{figure}

\begin{figure}
\begin{center}
\begin{tabular}{c}
\includegraphics*[width=0.45\textwidth]{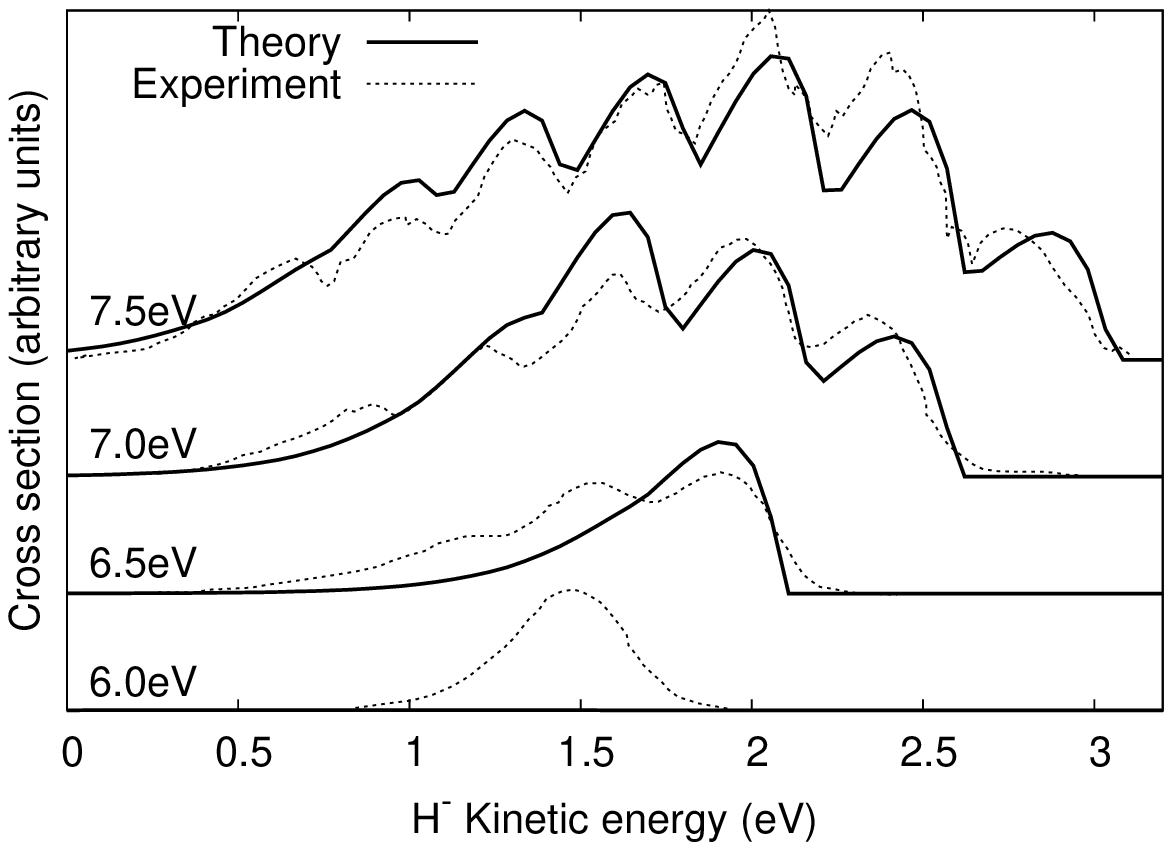} \\
\includegraphics*[width=0.45\textwidth]{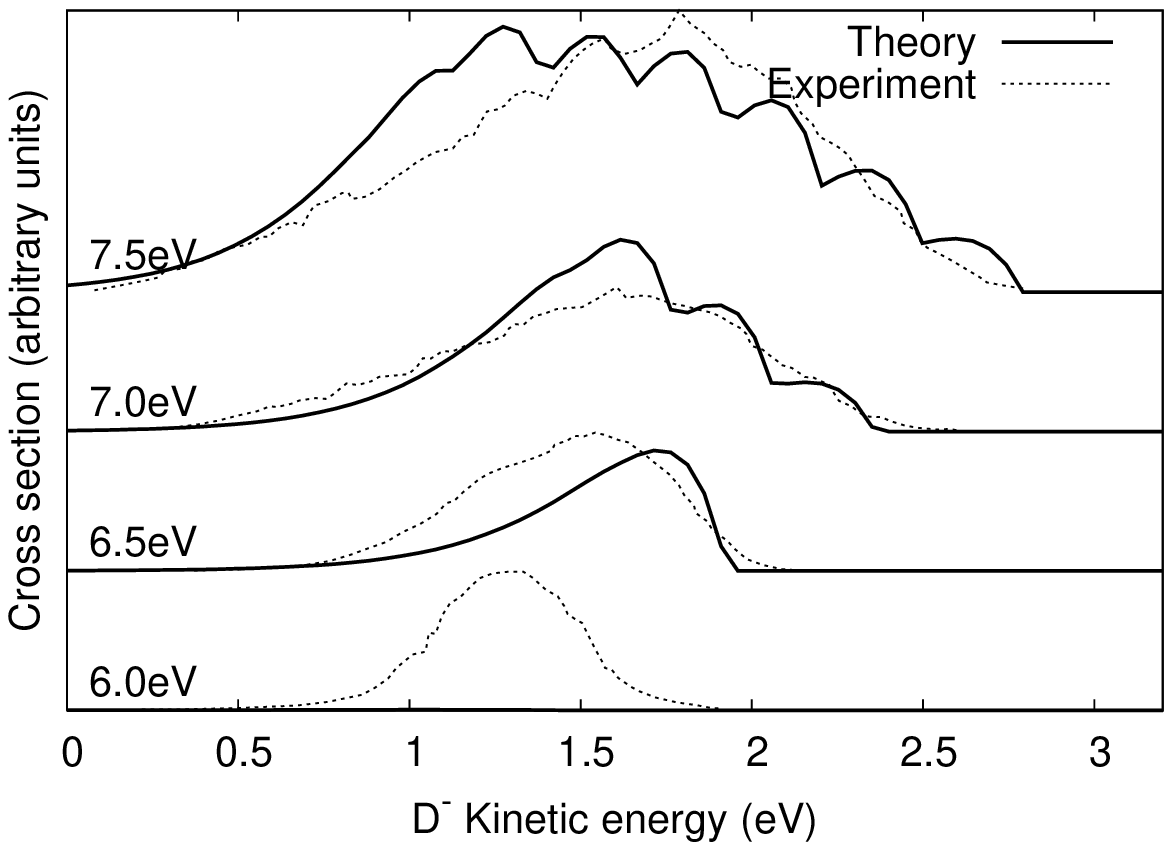} 
\end{tabular}
\end{center}
\caption[H$^-$ / D$^-$ kinetic energy distributions for $^2B_1$]{Top, production of 
H$^-$+OH via the $^2B_1$ resonance at different incident electron energies, as a function of
H$^-$ fragment kinetic energy, unshifted, on arbitrary and different
scales for each incident electron energy.  Bottom, D$^-$+OD.
Calculated results have been broadened using a 150meV linewidth, consistent
with the plotted experimental results from Belic, 
Landau and Hall~\cite{belic}.}
\label{belfig}
\end{figure}

\subsubsection{Production of OH (X $^2\Pi$) + H$^-$ via $^2B_1$ state}

This is the dominant channel for DEA to H$_2$O, having a peak cross section
of approximately 6$\times$10$^{-18}$ cm$^2$.  
Cross sections as a function of incident electron energy are shown in
Fig.~\ref{OH_H_2b1}.  We calculate a peak cross section of 10.35 $\times$ 10$^{-18}$ cm$^2$
at 6.87eV.  The magnitude of this cross section is larger than the experimental value
(6.6 $\times$ 10$^{-18}$ cm$^2$), and the location of the peak is displaced upward by 0.4eV.
For this resonance, autodetachment is nearly negligible.  Therefore,
the excess in the magnitude which we calculated reflects the fact that the  calculated width values, and hence the entrance amplitudes, are too large.

The cross sections  in Fig.~\ref{OH_H_2b1} 
are very similar in shape to those
produced previously~\cite{haxton1,haxton2}, though they are larger in magnitude.
At low incident electron energies, the first vibrational state is produced exclusively,
and subsequent vibrational states have sharp onsets.  As the degree of vibrational
excitation increases beyond the first few quanta, the magnitude of the cross
section decreases.  Despite the fact that the first five excited vibrational
states are clearly visible in Fig.~\ref{OH_H_2b1}, the average number of quanta
excited is only 1.28.  The average kinetic energy release, therefore, is near
its maximum value of 2.38eV.  
The degree of rotational
excitation calculated for this state is relatively low ($\left\langle j^2 \right\rangle$=107 for
H$_2$O).

A two-dimensional view of the data is provided in
Fig.~\ref{keplot_2b1J1}, where the kinetic
energy of the anion recoil, as determined by a full final-state
resolution of the products, is plotted versus incident electron energy; the contour lines indicate the magnitude of the cross section.  
The kinetic energy of the anion recoil in the laboratory frame is
\begin{equation}
E_{kin/anion} = \frac{M_{diatom}}{M_{total}} (E_{inc} - E_{j \nu}),
\end{equation}
as in Eq.(\ref{psurv}).
This figure shows that the degree of rotational excitation for production
of both H$^-$+OH and D$^-$+OD is small compared
to the vibrational spacing of the OH fragment, because there are separate lobes
corresponding to each vibrational state.  For the deuterated case, the lobes are thicker and closer together.
The thick line in this figure corresponds to the maximum kinetic energy
available, as determined by the physical energetics of the system; the
maximum kinetic energy as determined by the energetics of the constructed
surface is slightly higher.

In Fig.~\ref{belfig}, we plot the cross section as a function of H$^-$ kinetic energy for several values of
incident electron energy. To compare with the experimental results of 
Belic, Landau and Hall~\cite{belic}, which reflect the finite resolution of the
kinetic energy of the anion recoil, we incorporate the
experimental resolution of 150meV in the ion kinetic energy direction.
Such resolution effectively smears each vibrational peak into the next, and there are no hard
zeroes visible in the data of Fig.~\ref{belfig}.  A key result of
our calculation is that with better experimental resolution,
the individual vibrational peaks should be able to be resolved, not
only for H$_2$O, but also for D$_2$O, and that these peaks should be
fully separated.  The experimental resolution of
Ref.~\cite{belic} was insufficient to delineate the separate vibrational
peaks for D$_2$O.  We doubt that these authors have resolved the rotational
structure for H$^-$+OH production at 7.5eV incident electron energy,
as they claim.

The isotope effect observed for this channel has been a matter of some
interest.  Compton and Christophoreau~\cite{compton} observed the
ratio of peak heights for the deuterated (D$_2$O) to the  nondeuterated
(H$_2$O) species to be 0.75, and the ratio of the
energy-integrated cross sections, which approximate the ratio of
survival probabilities $P_{surv}$ calculated with Eq.(\ref{psurv}),
to be 0.60.  In contrast, we observe a larger peak for the deuterated
species, and similar survival probabilities $P_{surv}$, both near 1.

The recent results of Fedor \textit{et al.}~\cite{fedor} resolve this
discrepancy.  The peak heights which they obtain for H$^-$+OH
production versus D$^-$+OD production via the $^2B_1$ resonance
indicate a \textit{larger} peak for D$^-$+OD, reversing the
prior experimental evidence, and putting experiment and theory
on qualitatively similar ground.  It is clear that the ratio of
peak heights obtained by Fedor \textit{et al.}, while not 
explicitly calculated by these authors, is nearer to 1 than
the present theoretical results, but it is reassuring that the
trend for both experiment and theory is in the same direction.
The combination of the results of Ref.~\cite{fedor} and the
present results indicate that the survival probability for the
physical $^2B_1$ state is indeed near 1, and that minimal flux
is lost via the autodetachment for DEA via this resonance.

Plots of the propagated wave packet are shown in the EPAPS archive\cite{epaps}.

\subsubsection{Production of H$_2$+O$^-$ via the $^2B_1$ state}

This channel is by far the minor channel for DEA via the $^2B_1$ resonance.
The peak of the H$_2$+O$^-$ cross section is approximately $1/40^{\mathrm{th}}$
the height of the peak for the major H$^-$+OH channel \cite{Melton}.  Being
such a minor channel, it presents a more difficult challenge for theoretical
methods such as MCTDH, and a greater test for the local complex potential model.  
We were able to obtain converged cross
sections with the present treatment, although the magnitudes of our calculated
values are far below the experimental results.  Therefore, it is clear that we
have not represented the dynamics into this channel accurately.  It is possible
that small discrepancies in our calculated surface are to blame, or that the 
LCP model is inadequate.

\begin{figure}[!t]
\begin{center}
\begin{tabular}{c}
\includegraphics*[width=0.45\textwidth]{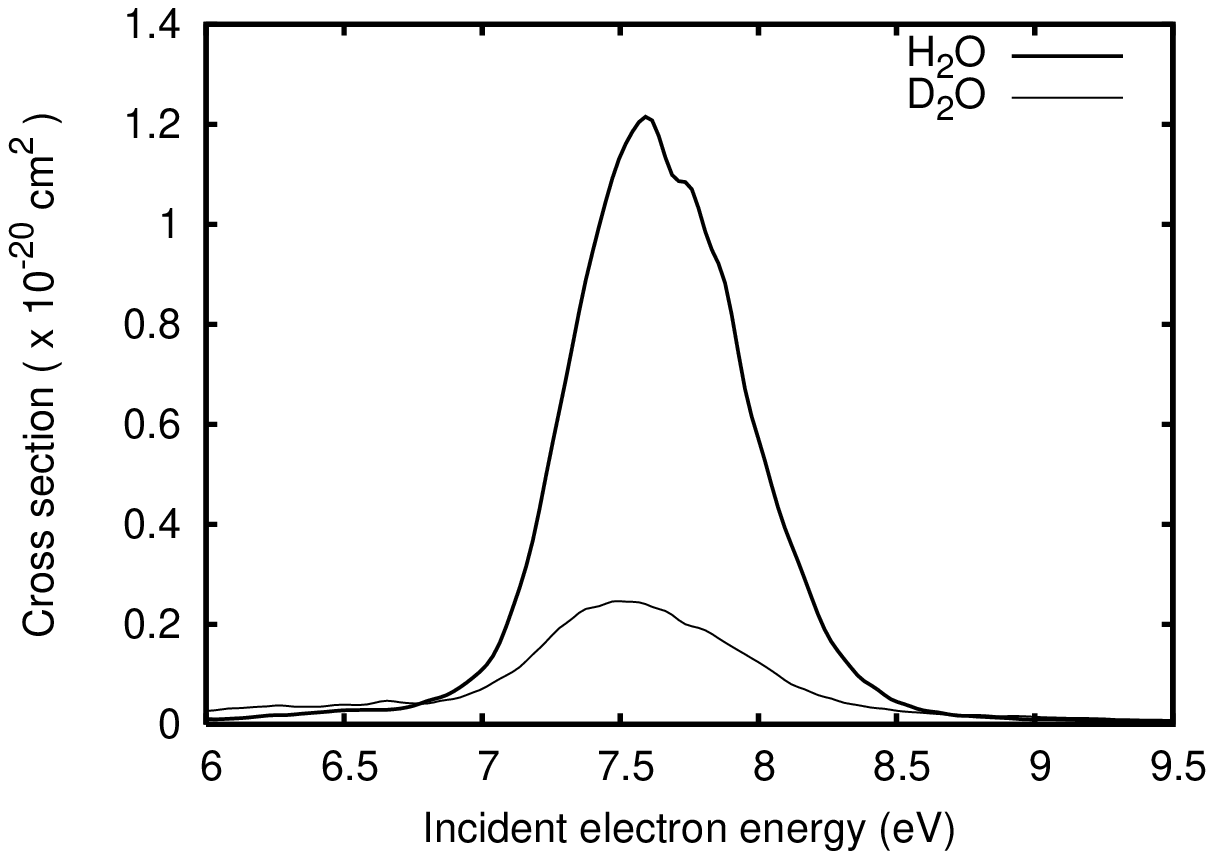} \\
\includegraphics*[width=0.45\textwidth]{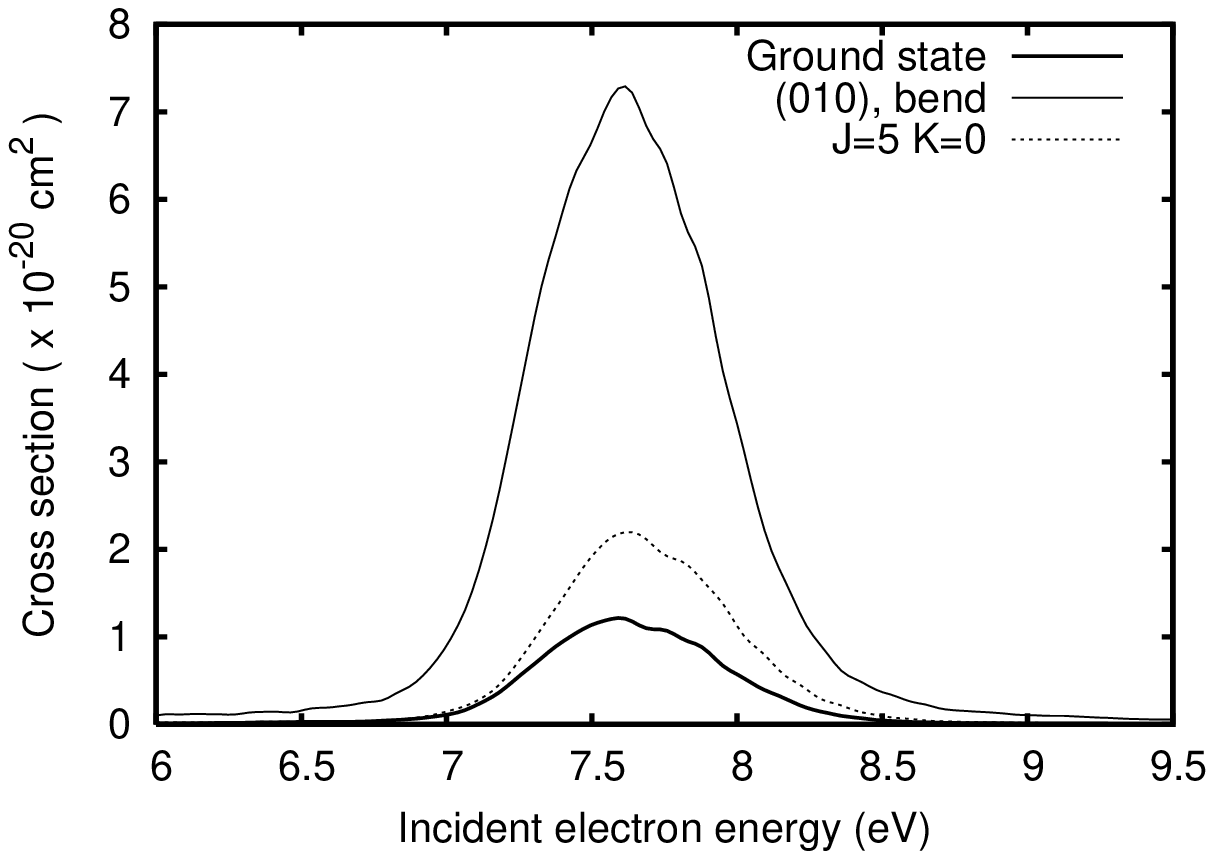}
\end{tabular}
\end{center}
\caption[Cross sections, H$_2$+O$^-$ from $^2B_1$]{Cross sections calculated for production of H$_2$+O$^-$ via the $^2B_1$ state 
as a function of incident electron energy.
Top, calculated isotope effect: comparison of H$_2$O and D$_2$O. Bottom, effect of target excitation: ground initial state result is compared to result
from (010) target state with one quantum of bend and to result from 
$J=5$, $K=0$.}
\label{H2_O_2b1}
\end{figure}

We present the cross sections calculated for H$_2$+O$^-$ production 
as a function of incident electron energy in Fig. \ref{H2_O_2b1}.
We compare the total cross section for H$_2$ production from H$_2$O with that for
D$_2$ production from D$_2$O in the top panel of this figure.
The cross sections peak at 7.6eV, 0.5eV above  the experimental peak at 7.1eV,
and are far smaller than the experimental result.
Although our representation of the nuclear dynamics leading to this channel
is clearly lacking, we performed additional calculations in which the target state
of H$_2$O was rovibrationally excited.  We performed two calculations for total
angular momentum $J=5$, employing the centrifugal sudden (CS) approximation with $K=0$,
in the $R$-embedding coordinate system, as well as a calculation with $J=0$ but one quantum
of bend, the (010) state.  The total cross sections for production of H$_2$ from these
excited states are compared to the ground initial state result in the top panel
of Fig. \ref{H2_O_2b1}.

As is clear from these results, initial excitation of the target may play a large
role in determining the magnitude of the DEA cross section for H$_2$+O$^-$ production
via the $^2B_1$ state, but is insufficient to explain the discrepancy between
the theoretical and experimental results.  The effect of bending excitation
increases the cross section dramatically; rotational excitation to $J=5$
also enhances the cross section by approximately a factor of 2.  The
excitation energy of the bending mode is approximately 0.2eV; that of the
$J=5$, $K=0$ state is approximately 0.056eV.  These quantities may be
compared to the value of $kT$ at 373.15$^\circ$ K, which is 0.032eV.
This comparison indicates that the bending state is not significantly
populated in typical experimental setups and should not be responsible
for the magnitude of the observed cross sections.  Comparison of the
rotational energy to $kT$ indicates that the degree of rotational
excitation of the target may determine the precise value of the peak
cross section observed in experiment.  However, rotational excitation
of the sample is insufficient to explain the discrepancy between our results
and the experimental ones.  For the $R$-embedding coordinate system,
the $K=0$ projection of angular momentum is the most likely to enhance
the DEA cross section for production of H$_2$, because that projection
minimizes the centrifugal potential in the $r_{HH}$ coordinate [see Eq.(\ref{hamiltonian})].
The calculated enhancement is due to the effect of the centrifugal
potential in $R$, which ``pushes'' the wave packet toward large $R$,
where the O$^-$+H$_2$ potential well lies.

Because the width $\Gamma$ of the $^2B_1$ resonance is small for all nuclear
geometries, one might expect any nonlocal effects in the resonant nuclear dynamics
to be small as well.  However,  we are here  considering a minor channel,
which is only barely accessible with LCP dynamics on the constructed potential-energy surface.  
If nonlocal effects were to open a new dynamical pathway, or otherwise effectively lower the
dynamical barrier to the H$_2$+O$^-$ well, the magnitude of such effects
would not have to be great in order to produce a noticeable enhancement of
such a small cross section.  Therefore, we regard nonlocal effects to be
a strong candidate for the source of the experimentally observed 
cross section for production of H$_2$+O$^-$ via the $^2B_1$ resonance.

\subsection{Dissociative electron attachment via the $^2A_1$ (1 $^2A'$) state, Renner-Teller coupled to the $^2B_1$ state}

These calculations are performed in the diabatic ($l_{z'}=\pm 1$) basis, which diagonalizes
the nuclear kinetic energy operator with the Renner-Teller effect, 
as per the discussion in Sec. \ref{triatomicsect},
employing the centrifugal sudden Hamiltonian of Eq.(\ref{hamiltonian}).
The initial state is the adiabatic $^2A_1$ (1 $^2A'$) state,
comprised of equal parts $l_{z'}=\pm 1$.
Like the other calculations which we present
that incorporate rotational motion, they are parametrized by the body-fixed angular
momentum quantum number $K$, which is the projection of the total angular momentum
onto the embedding axis.  However, for these Renner-Teller calculations $K$ is 
interpreted as the eigenvalue of the projection of the nuclear rotational angular
momentum $R_{z'}$, not the total angular momentum $J_{z'}$, upon the embedding axis,
and therefore the diabatic basis $l_{z'}=\pm 1$ corresponds to $J_{z'}=K\pm 1$.~\cite{co2_2}

We have obtained cross sections for the major H$^-$+OH
(X $^2\Pi$) channel of DEA via the $^2A_1$ resonance.  However, for the minor H$_2$+O$^-$
channel, we have not been able to obtain converged, nonzero cross sections.
The mechanism for DEA via the $^2A_1$ resonance to produce H$_2$+O$^-$
remains unknown. It is possible that three-body breakup, which we have not treated, is important here.

The considerations of Ref.~\cite{haxton3} indicate
that the nuclear dynamics of DEA via the $^2A_1$ resonance may
hold some surprises, and that the LCP model
may be insufficient for a full description thereof.  
In particular, as demonstrated in paper I,
the width of the $^2A_1$ resonance becomes
large as the nuclear geometry
moves toward the H$^-$+OH product arrangment.  We have
calculated width values as high as 0.15eV for this resonance state
for such stretched geometries, despite the fact that the resonance
state lies only slightly above the neutral at these geometries
and ultimately becomes bound as the atom-diatom distance
is further increased.  The explanation for this
behavior is that the electronic structures of the neutral and
anion become highly correlated and different from each other 
at such stretched geometries, and
as a result, there is considerable shape resonance character mixed
into the $^2A_1$ Feshbach resonance.  The radically peaked behavior
of the width of the $^2A_1$ state may portend a breakdown of the
LCP model, which relies on the implicit assumption that the 
background-resonance coupling is a relatively smooth
function of nuclear geometry.  Also, the fact that the $^2A_1$
resonance is coupled to the neutral target by an $s$-wave matrix
element indicates that virtual state effects may play a large role
as the resonance becomes bound.  Such virtual state effects cannot be properly
described by the LCP model, but have been taken
into account in other systems using effective range theory  as, for example, 
in Refs.~\cite{wim1, wim2}.
The fact that we have overestimated the magnitude of the experimental cross section
for H$^-$+OH production via the $^2A_1$ resonance indicates that a breakdown of the
LCP model may be responsible for the loss of flux via autodetachment.

\subsubsection{Production of H$^-$ + OH via the Renner-Teller coupled $^2A''$ ($^2B_1$) \& 1 $^2A'$ ($^2A_1$) states}

The cross sections for total H$^-$+OH production 
were easy to converge, as the channel involved is
the main channel, the dissociation direct, and the dynamics apparently
reasonably separable in the $r_{OH}$ Jacobi internal coordinate system. 
The relatively small size of the single particle function (SPF) expansions
required to converge the calculation (see the EPAPS archive\cite{epaps}) support this conclusion.  
We found the Renner-Teller coupling to have a negligible effect on both the magnitude
of the total cross section and its breakdown into rotational and vibrational states, so
we only report results for $K=0$.  The cross sections (for $K=0$) into final vibrational
and rotational levels of the diatomic fragment, for both the deuterated and nondeuterated
cases, are given in the EPAPS archive\cite{epaps}.

Our theoretical treatment overestimates the cross section for DEA into
this channel via the $^2A_1$ resonance.  Our peak heights, 4.14 and 4.16 $\times$10$^{-18}$
cm$^2$ for the nondeuterated and deuterated target, respectively, are approximately three times larger
than Melton's observed peak height of 1.3 $\times$10$^{-18}$ cm$^2$.  Although the recent results
of Fedor \textit{et al.}~\cite{fedor} do resolve this peak better than previous experiments,
and indicate that Melton's peak height value may be too low, there is a clear 
discrepancy between theory and experiment here.  We attribute the discrepancy to
virtual state effects, as discussed in Ref.~\cite{haxton3}, which may
lead to significant autodetachment as the $^2A_1$ (1 $^2A'$) state becomes 
bound.  There is also the possibility that we have overestimated the
entrance amplitude for this state, as we have done for the $^2B_1$ state.

\begin{figure}
\begin{center}
\begin{tabular}{c}
\includegraphics*[width=0.45\textwidth]{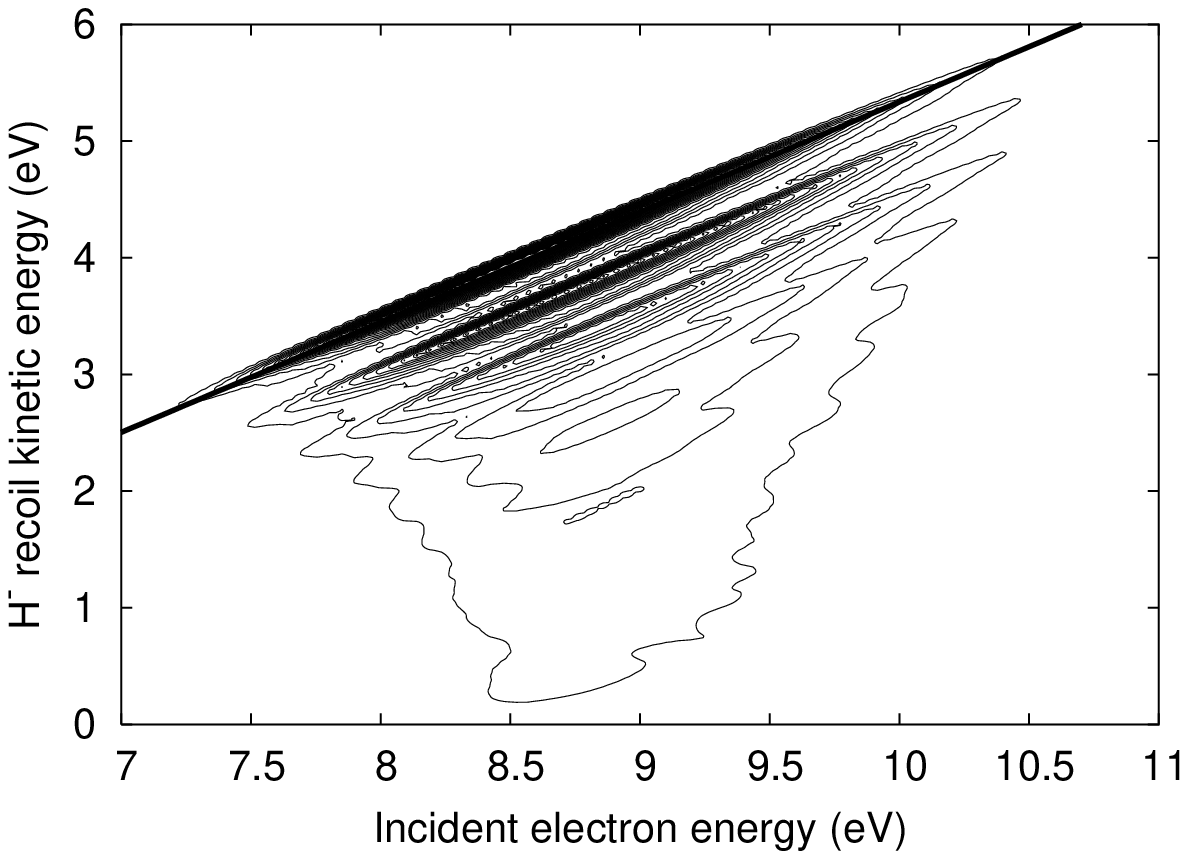} \\
\includegraphics*[width=0.45\textwidth]{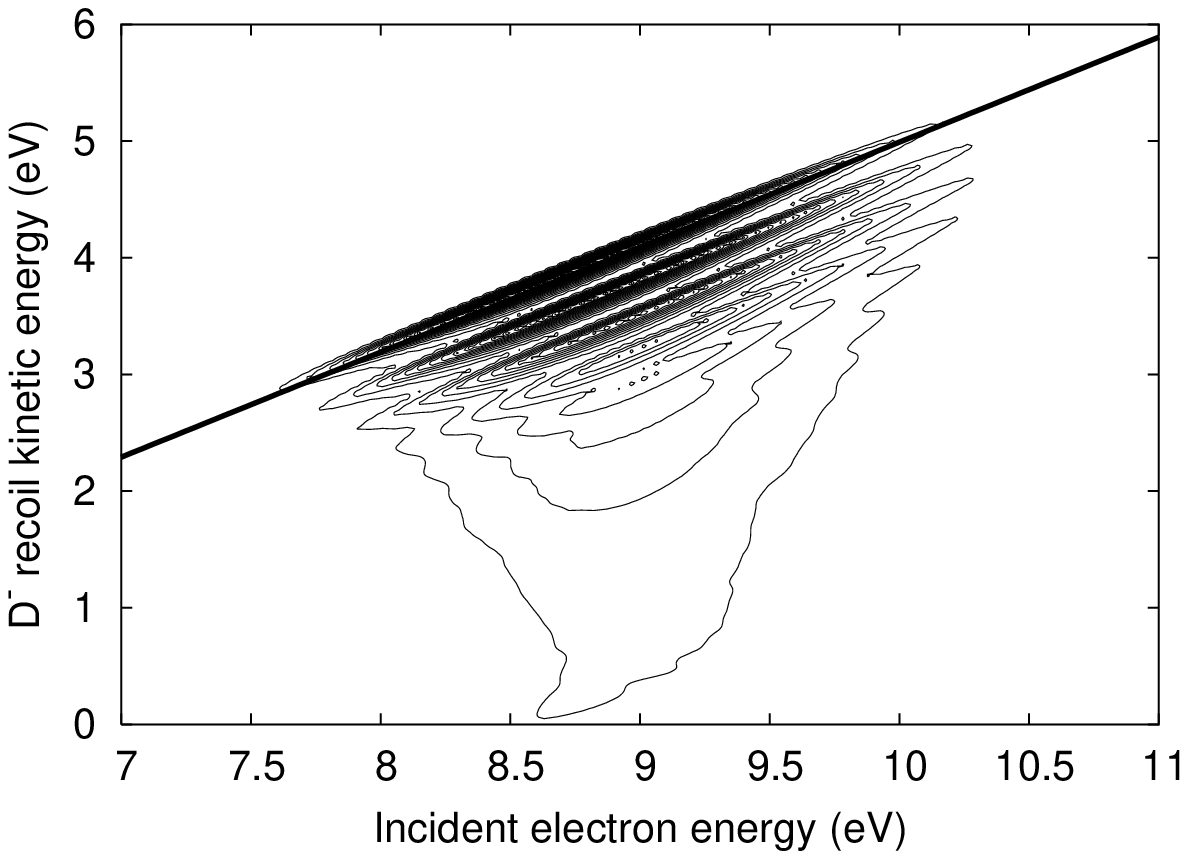} 
\end{tabular}
\end{center}
\caption[H$^-$ / D$^-$ kinetic energy distributions for $^2A_1$, 2D]{Top, production of 
H$^-$+OH from the $^2A_1$ state, as a function of incident electron energy and
H$^-$ fragment kinetic energy, as in previous plots;
bottom, deuterated.
 The maximum kinetic energy available, as determined
from the physical energetics, is plotted with a bold line.
Contours every 6 $\times$ 10$^{-18}$ cm$^2$ eV$^{-1}$. }
\label{keplot_a1b1J1}
\end{figure}

The degree of vibrational excitation is higher for this resonance than for the
$^2B_1$ resonance: the values of $\langle \nu \rangle$ 
calculated from Eq.(\ref{psurv}) are 2.03 and 1.28 for the $^2A_1$ and $^2B_1$ resonances,
respectively.
This difference is most likely due to the gradient of the
potential-energy surface in the symmetric stretch direction, which is larger
at the equilibrium geometry of the neutral for the
$^2A_1$ surface than for the $^2B_1$ surface.  
The behavior of the propagated wave packet, which is plotted in the EPAPS
archive\cite{epaps}, is similar to that found for the $^2B_1$ resonance: the wave packet 
experiences an initial impulse in the symmetric stretch
direction, but then is bifurcated by the developing potential wall in this direction,
and reflected into either H$^-$+OH channel.  The vibrational excitation is
the result of the wave packet oscillating in the $r$ direction as it
passes down the OH potential well.

The degree of rotational excitation within the OH fragment
is also higher for the $^2A_1$ state than for the $^2B_1$ state.  Using Eq.(\ref{psurv}), we 
calculate an average degree of rotational excitation $\langle j^2 \rangle$=119
for this resonance, compared to 107 for the $^2B_1$ resonance.
This results from the larger
gradient of the potential-energy surface in the bend direction for the $^2A_1$
surface compared to  the $^2B_1$ surface.  The $^2A_1$ wave packet
is given an impulse in the bend direction, which corresponds to excitation
of rotational quanta $j$.  This excitation persists in the final state, as
demonstrated by these calculations.

Fig.~\ref{keplot_a1b1J1} presents two-dimensional plots of the cross section as a function of both
incident electron energy and final anion recoil kinetic energy.  Fig. \ref{keplot_a1b1J1}
displays a clear difference from Fig. \ref{keplot_2b1J1}.  This figure shows
that for DEA via the $^2A_1$ state, the degree of rotational excitation of the
diatomic fragment is high enough that the 
different vibrational states are distinguishable,
but not completely separated.

\begin{figure}
\begin{center}
\includegraphics*[width=0.45\textwidth]{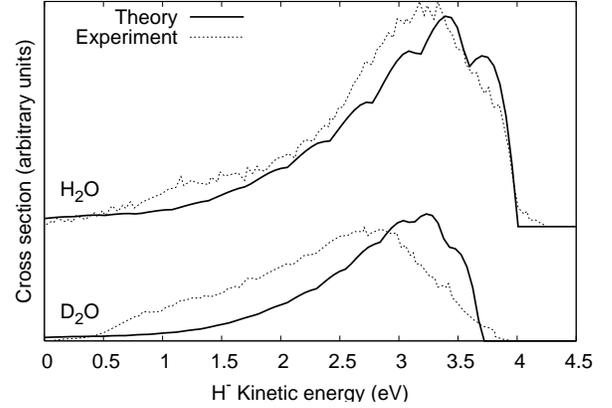} 
\end{center}
\caption[H$^-$ and $D^-$ kinetic energy distributions via $^2A_1$ resonance, 
with experimental result]
{Production of 
H$^-$+OH ($^2\Pi$) and D$^-$+OD ($^2\Pi$) via the $^2A_1$ state 
at 8.5eV incident electron energy,
as a function of
fragment kinetic energy, as in previous plots.
Calculated results have been broadened using a 150meV linewidth. 
to compare with 
the experimental results from
Belic, Landau and Hall~\cite{belic}, also plotted.}
\label{belfig7}
\end{figure}

\begin{figure}[!b]
\begin{center}
\includegraphics*[width=0.45\textwidth]{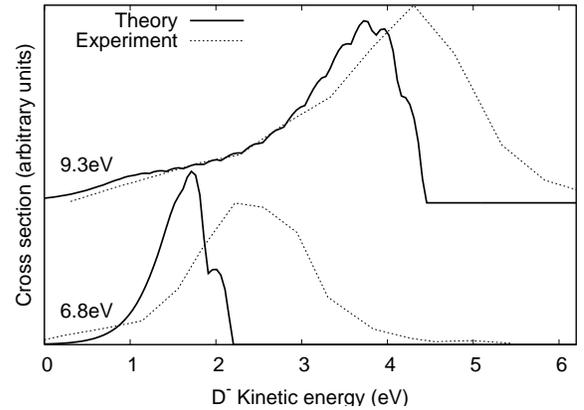} 
\end{center}
\caption[D$^-$ kinetic energy distributions for $^2B_1$ and $^2A_1$ at 6.8 and 9.3eV]{
Production of 
D$^-$+OD ($^2\Pi$) at 
9.3eV ($^2A_1$) and 6.8eV ($^2B_1$), as a function of
D$^-$ fragment kinetic energy, as in previous figures,
with
experimental results from Curtis and Walker~\cite{curtiswalker}.  }
\label{CWfig3a}
\end{figure}

We compare our results for the laboratory-frame, anion 
recoil kinetic energy distribution at an incident
electron energy of 8.5eV with the corresponding
results of Belic, Landau and Hall\cite{belic}, in Fig. \ref{belfig7}. 
(Two-dimensional plots of the cross section as a function of both incident
electron energy and final anion recoil kinetic energy can be found
in the EPAPS archive\cite{epaps}.)  In contrast to the case of H$^-$ production
from the $^2$B$_1$ state, here the degree of rotational excitation of the diatomic
fragment is high enough that the different vibrational states are distinguishable,
but not completely separated.
The experimental resolution of Ref.~\cite{belic} was insufficient to delineate
the different peaks for various OH ($\nu$), although Fig. \ref{keplot_a1b1J1}
demonstrates that with sufficient resolution, the vibrational structure should
be apparent.
In Fig. \ref{CWfig3a}, we compare to the experimental results of Curtis and
Walker\cite{curtiswalker}.

\subsubsection{Failure to calculate production of H$_2$ + O$^-$ via
  dissociative attachment to the $^2A_1$ state}

We have been unable to obtain a nonzero cross section for DEA via the $^2A_1$ (1 $^2A'$) state,
Renner-Teller coupled to the $^2B_1$ state, leading to H$_2$+O$^-$.  We have
attempted calculations for $K$=0 (uncoupled), 1, 2, 3, and 4.  Within the MCTDH calculations,
we employed single-particle function (SPF) expansions of up to 24$\times$36$\times$30, 
with no success.  With this large SPF expansion, and propagation times of up to 100~fs,
we regard the representation of the LCP dynamics within the MCTDH ansatz to be accurate. We suspect that O$^-$ production from $^2A_1$ may be dominated by three-body breakup into
H+H+O$^-$, which we have not treated.

\subsection{Dissociative electron attachment via the $^2B_2$ state,
involving the conical intersection with the $^2A_1$ state}

As described in Ref.\cite{haxton3} and paper I,
dissociative electron attachment to H$_2$O via the highest-energy $^2B_2$ state
must involve the conical intersection that this state exhibits with the
$^2A_1$ state.  The gradient of the potential-energy surface leads directly
toward this conical intersection from the equilibrium geometry of the
neutral.  The conical intersection forms a line in the three-dimensional
space of nuclear geometries, and occurs within $C_{2v}$ symmetry, where
the OH bond lengths are equal.  

We performed a diabatization on the results of configuration-interaction
calculations on $^2A_1$ and $^2B_2$ (1 and 2 $^2A'$) resonances, as described in
paper I,
to produce a set of diabatic $^2A_1$ and $^2B_2$ surfaces
along with a coupling surface.  These diabatic surfaces are employed in all
of the following calculations.

Before describing the individual calculations, a few preliminary remarks about
the experimental observations are in order.  
Although absolute cross sections for H$^-$ production via the $^2B_2$ resonance
are not available,
the experimental evidence~\cite{Melton, fedor} 
indicates that for both D$_2$O and H$_2$O target states, the branching ratio
between H$^-$ production and O$^-$ production highly favors O$^-$. 
Therefore, the dynamics of DEA beginning in the $^2B_2$ state are much
different from those for the lower-energy $^2A_1$ and $^2B_1$ resonances,
which yield far more H$^-$ than O$^-$.

This observation is not surprising, in light of the potential-energy surfaces 
which we have calculated and shown in paper I; the upper 
2 $^2A'$ surface was demonstrated to be quite different from those of the
lower resonances.  In particular, the dynamics beginning on the $^2B_2$ (2 $^2A'$)
surface will begin with a decrease of the H-O-H bond angle $\theta_{HOH}$, which
motion will favor the H$_2$+O$^-$ product arrangement.  However, as we will
show, there appears to be active competition between the two product arrangements,
and the branching ratio observed in experiment is likely the product of both
the shape of the real-valued component to the 2 $^2A'$ surface and the behavior
of its imaginary component.

As is clear from Table \ref{resultstable}, the cross sections we calculate for this 
channel  are smaller than the observed cross sections.  However, the comparision 
is complicated by the fact that the three-body
dissociation channels are open for incident electron energies sufficient to reach
the $^2B_2$ resonance, and we produce cross sections only for the two-body dissociation
channel; the disagreement may therefore be due to a large contribution of the
three-body breakup channel to the dominant production of O$^-$.
However, the locations of the calculated and experimental peak maxima for production
of O$^-$ from the $^2B_2$ resonance agree very well: both cross sections peak at
about 11.8eV.  Although the presence of the three-body dissociation channel may shift
the peak, this comparison indicates that we have probably accurately represented the
vertical transition energy for the $^2B_2$ resonance.  The vertical transition
energy as represented by our configuration-interaction surface is 12.83eV,
and therefore we recommend a value of approximately 12.8eV for the appropriate
physical transition energy.  This value is above the value of 11.97eV given by the complex Kohn calculations of paper I.

The calculated branching ratio between  H$_2$+O$^-$ (D$_2$+O$^-$) and  
OH+H$^-$ (OD+D$^-$) production
is near unity, but   the experimental ratio (for the undeuterated product)
exceeds 1 by a large factor.
  As we will show, the dynamics within the
first few femtoseconds after attachment are controlled by both real 
and  imaginary components of the potential-energy surface, the latter
consuming most of the propagated wave packet within the first six femtosectonds.

We have examined the effect of rotational excitation upon the cross section for production
of H$_2$+O$^-$ from DEA via the $^2B_2$ state, and find it to be negligible.

\subsubsection{Production of H$^-$ + OH ($^2\Pi$ and $^2\Sigma$) via the upper $^2B_2$ state}

The calculation for the production of OH ($^2\Pi$ and $^2\Sigma$)+H$^-$ via the $^2B_2$ Feshbach resonance,
which is coupled to the $^2A_1$ resonance via their conical intersection, proved
difficult to converge. This is evidenced by   raggedness in the OH ($^2\Pi$) channel cross sections.
A final state resolution in this channel was not possible , although we
were able to resolve the final states of OH ($^2\Sigma$).

\begin{figure}
\begin{center}
\begin{tabular}{c}
\includegraphics*[width=0.45\textwidth]{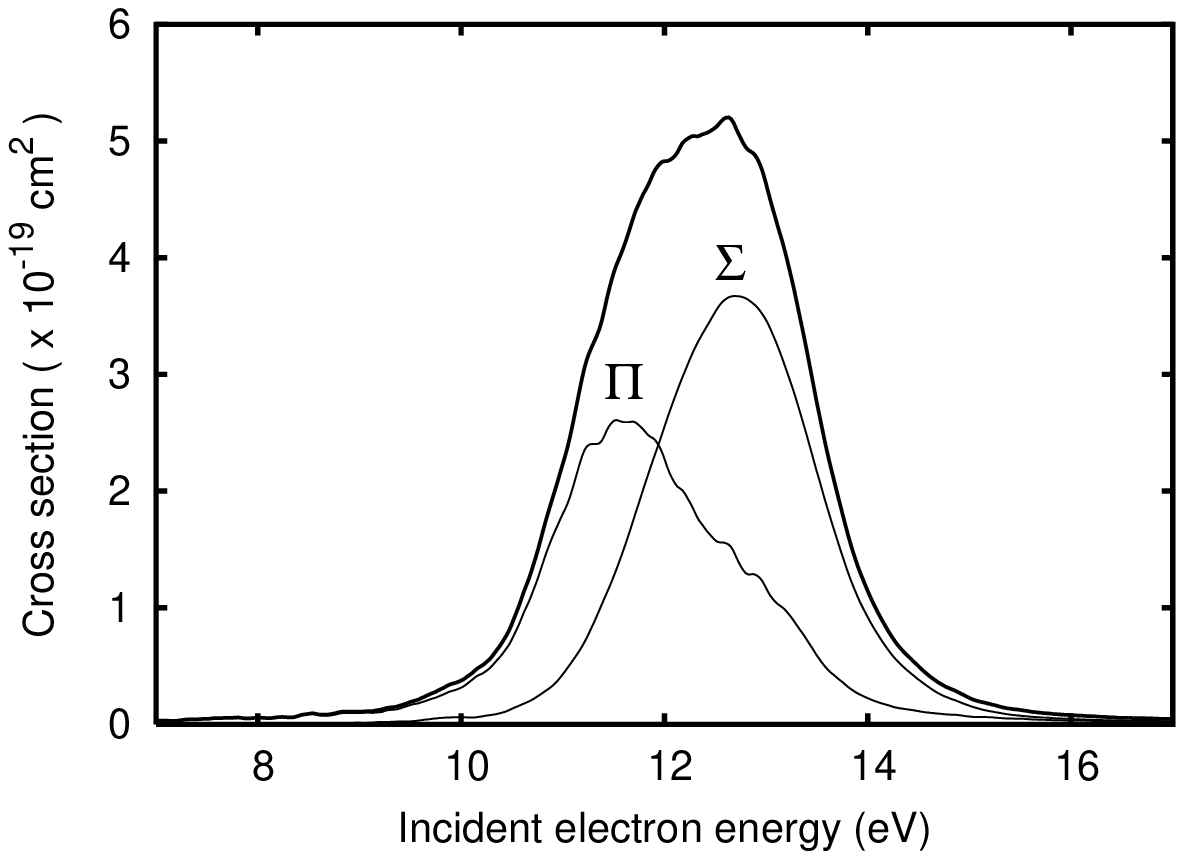} \\ 
\includegraphics*[width=0.45\textwidth]{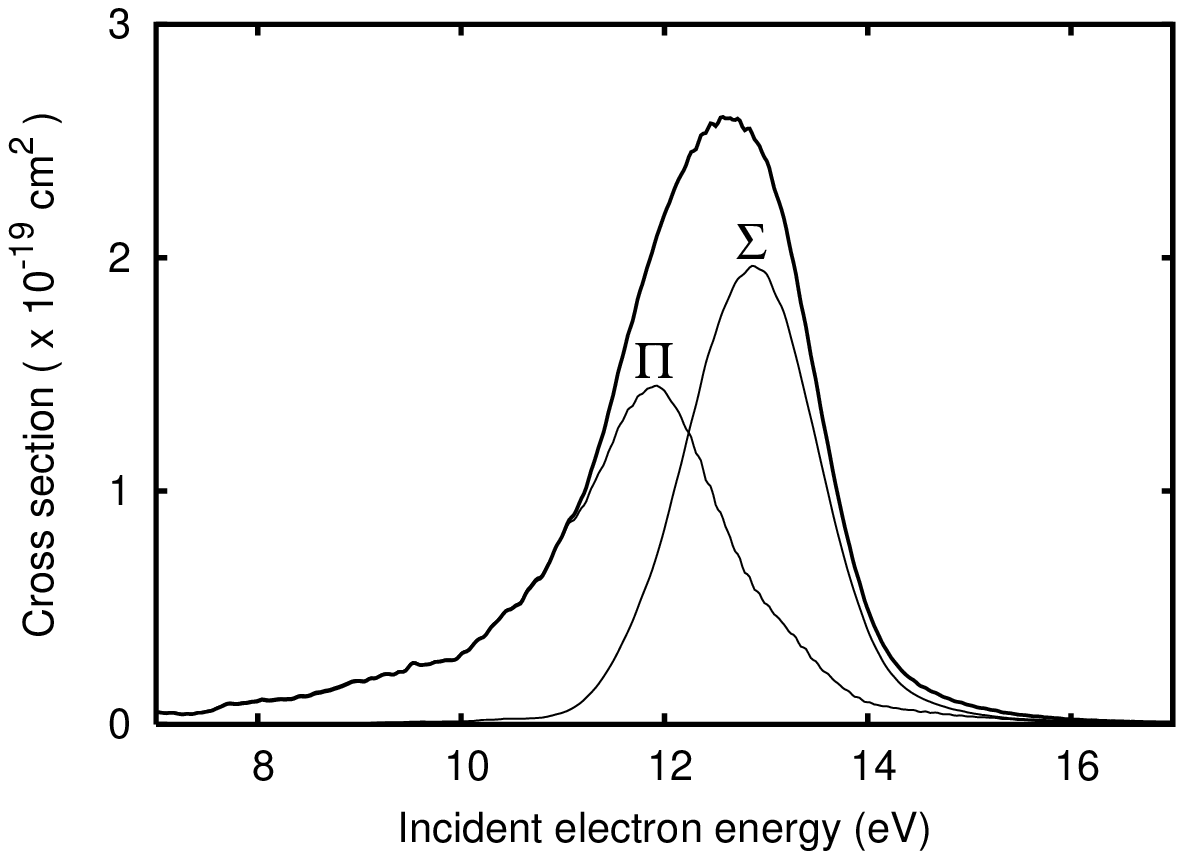} 
\end{tabular}
\end{center}
\caption[Cross sections, H$^-$ from $^2B_2$ calculation]{Top, total cross section 
calculated for production of H$^-$+OH ($^2\Pi$) versus H$^-$+OH ($^2\Sigma$) 
from $^2B_2$ state as a function
of incident electron energy.  Bottom, deuterated version.}
\label{OH_H_b2a1}
\end{figure}

\begin{figure*}
\begin{center}
\begin{tabular}{ccc}
\resizebox{0.3\textwidth}{!}{\includegraphics*[0.55in,0.6in][3.95in,3.06in]{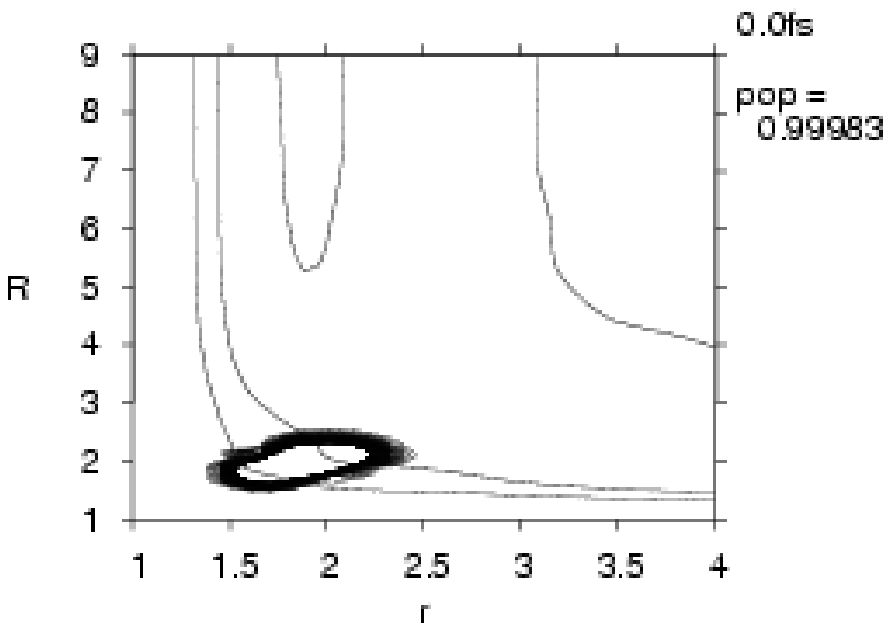}} &
\resizebox{0.3\textwidth}{!}{\includegraphics*[0.55in,0.6in][3.95in,3.06in]{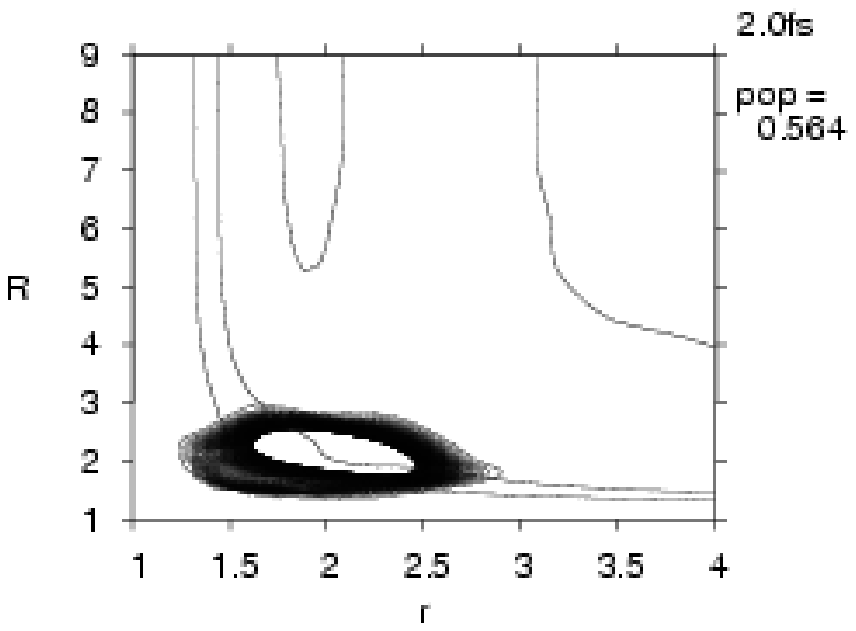}} &
\resizebox{0.3\textwidth}{!}{\includegraphics*[0.55in,0.6in][3.95in,3.06in]{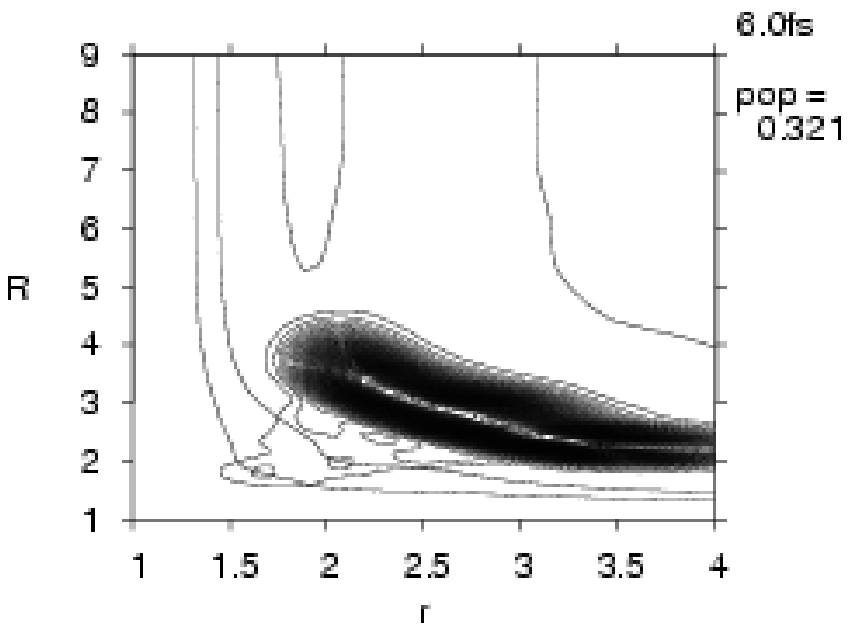}} \\
\resizebox{0.3\textwidth}{!}{\includegraphics*[0.55in,0.6in][3.95in,3.06in]{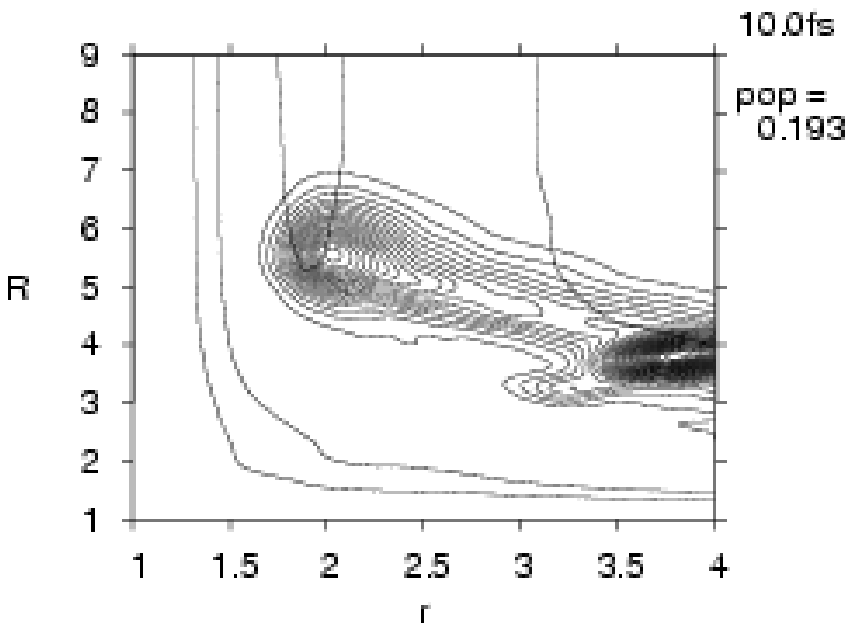}} &
\resizebox{0.3\textwidth}{!}{\includegraphics*[1.25in,1.1in][4.65in,3.6in]{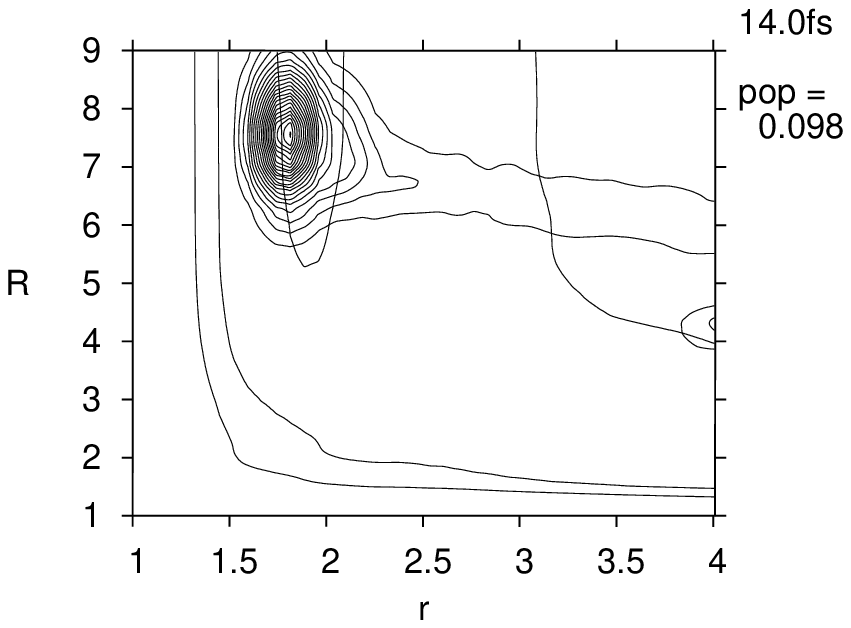}} &
\resizebox{0.3\textwidth}{!}{\includegraphics*[1.25in,1.1in][4.65in,3.6in]{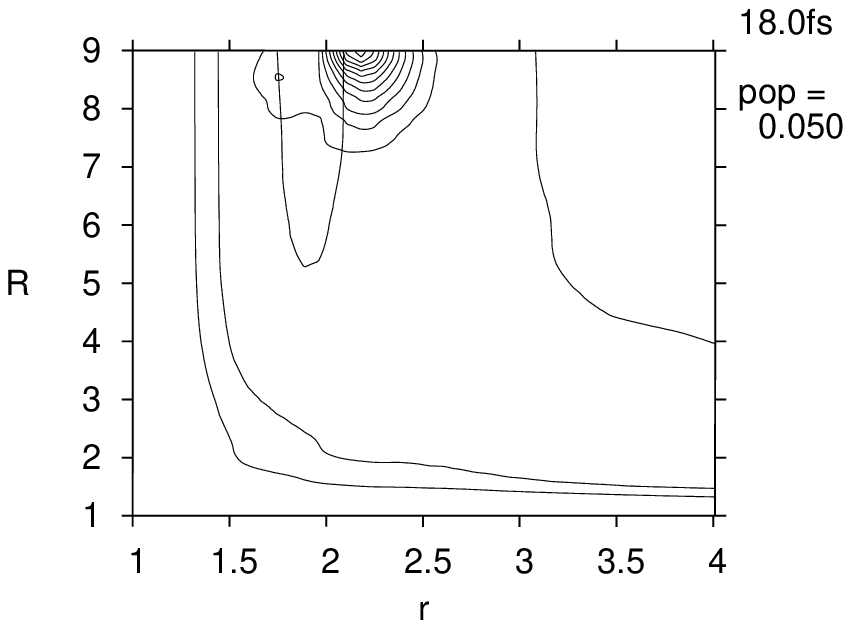}} 
\end{tabular}
\end{center}
\caption[$^2B_2$ propagation, 1 $^2A'$ component]{Propagation of wave packet on coupled $^2B_2$ and $^2A_1$ surfaces for H$^-$+OH ($^2\Pi$ / $^2\Sigma$) channels,
adiabatic 2 $^2A'$ ($\rightarrow$ $^2\Sigma$) component,
with real part of 2 $^2A'$ potential-energy surface at $\gamma$=90$^\circ$.
Bond lengths, units of bohr.  Density is integrated over $\gamma$.}
\label{b2a1J1prop2}
\end{figure*}

\begin{figure*}
\begin{center}
\begin{tabular}{ccc}
\resizebox{0.3\textwidth}{!}{\includegraphics*[1.25in,1.1in][4.65in,3.6in]{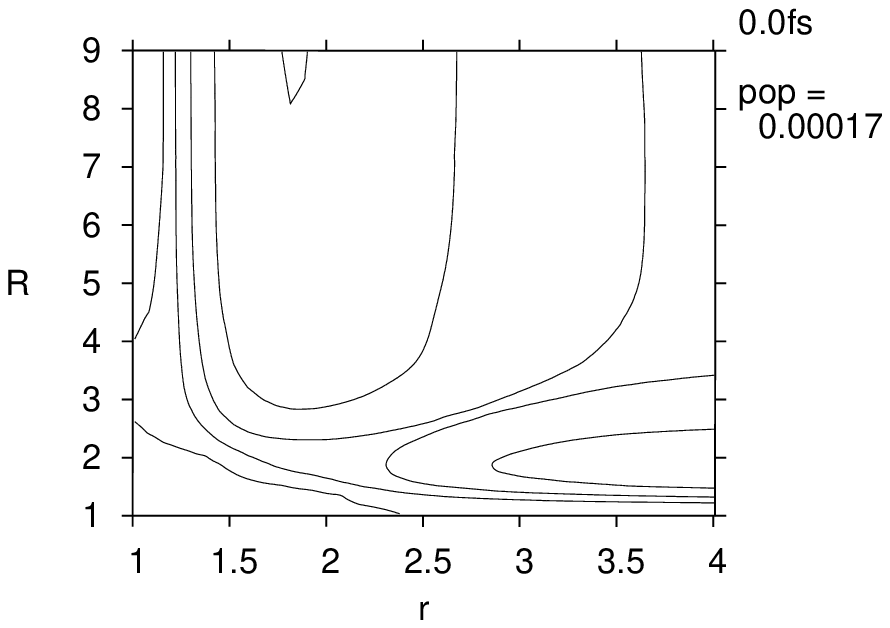}} &
\resizebox{0.3\textwidth}{!}{\includegraphics*[1.25in,1.1in][4.65in,3.6in]{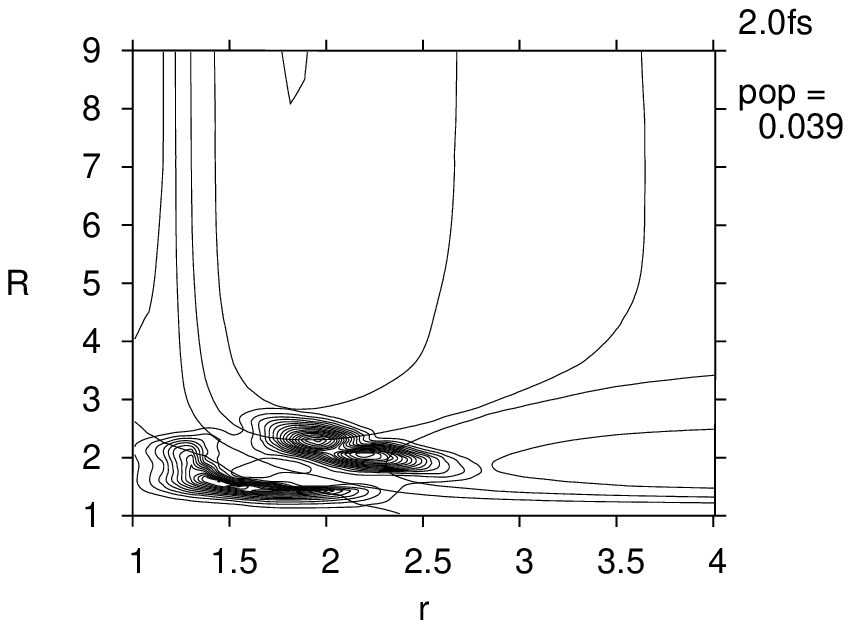}} &
\resizebox{0.3\textwidth}{!}{\includegraphics*[1.25in,1.1in][4.65in,3.6in]{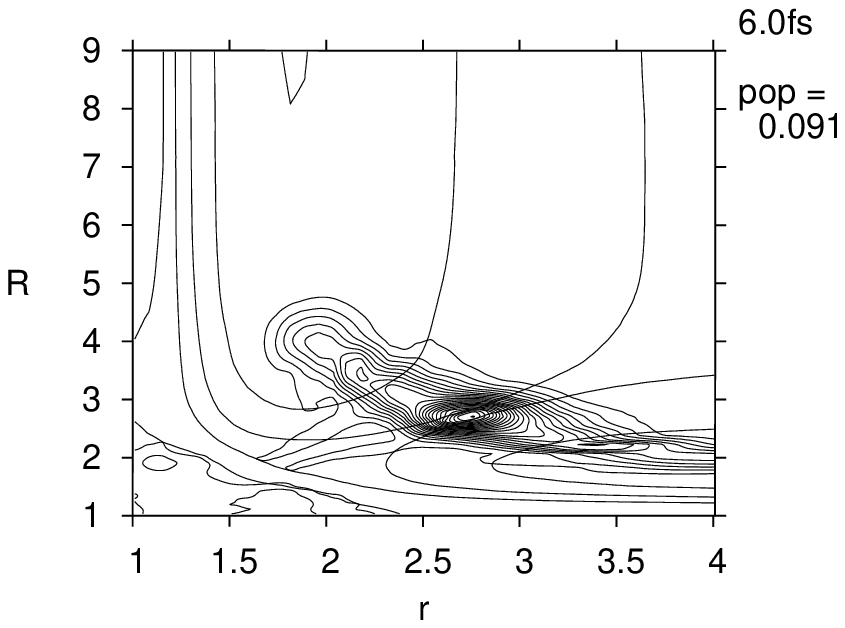}} \\
\resizebox{0.3\textwidth}{!}{\includegraphics*[1.25in,1.1in][4.65in,3.6in]{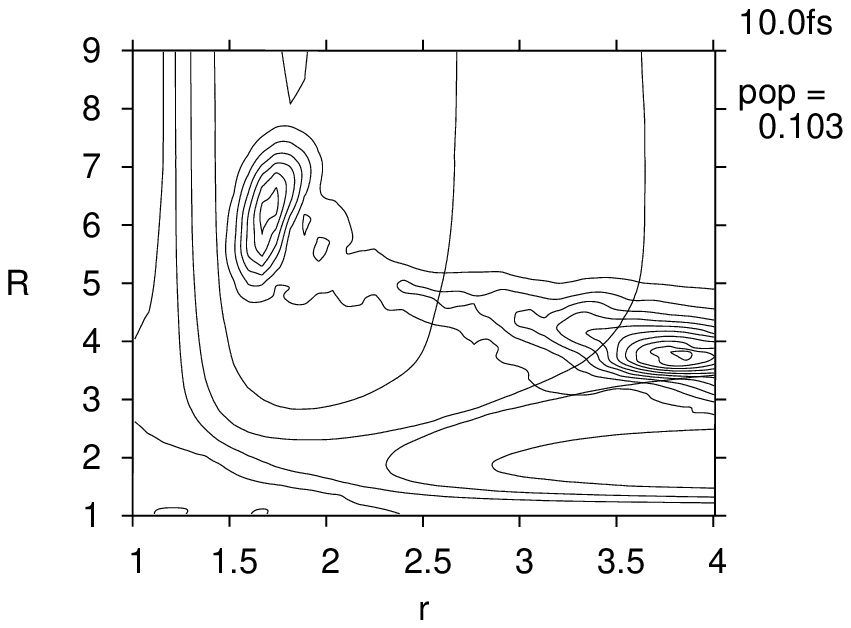}} &
\resizebox{0.3\textwidth}{!}{\includegraphics*[1.25in,1.1in][4.65in,3.6in]{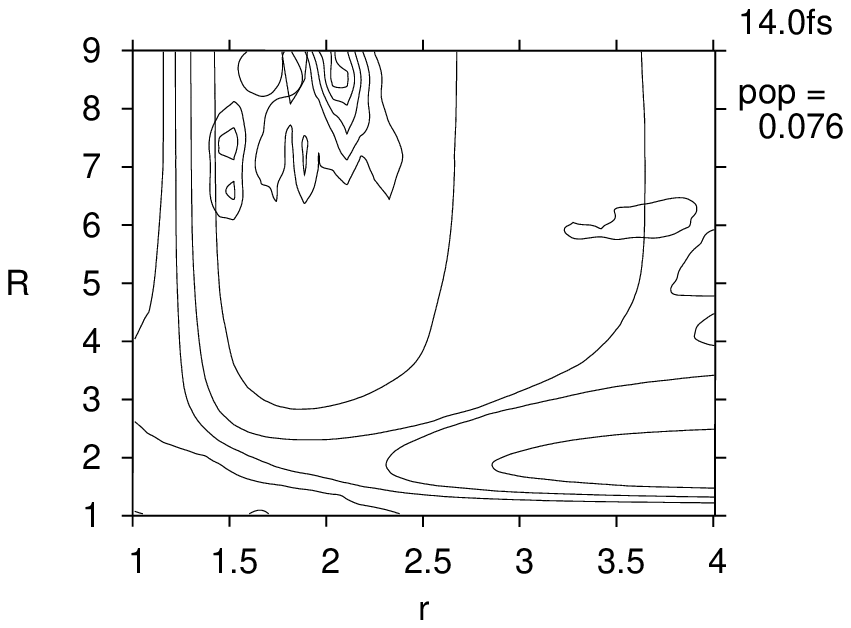}} &
\resizebox{0.3\textwidth}{!}{\includegraphics*[1.25in,1.1in][4.65in,3.6in]{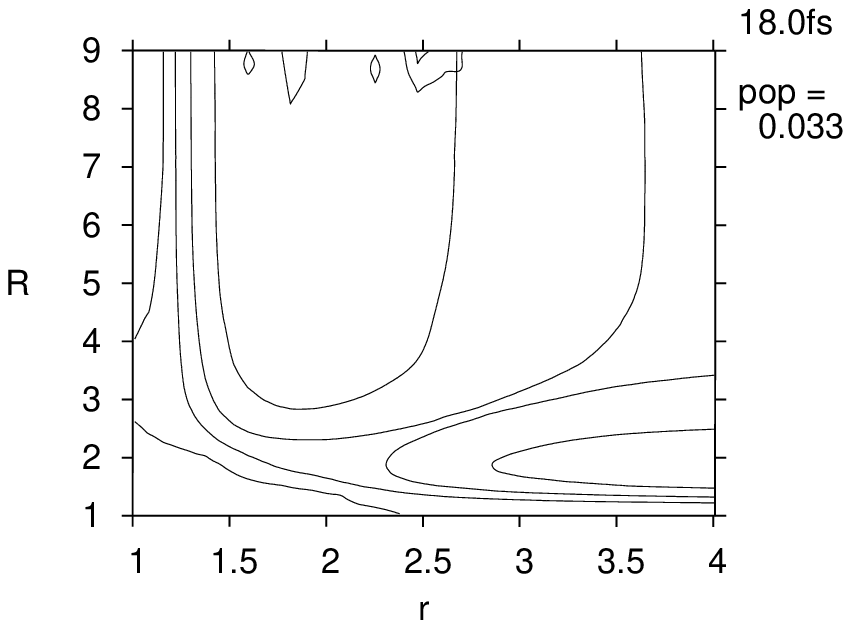}} 
\end{tabular}
\end{center}
\caption[$^2B_2$ propagation, 1 $^2A'$ component]{Propagation of wave packet on coupled $^2B_2$ and $^2A_1$ surfaces for H$^-$+OH ($^2\Pi$ / $^2\Sigma$) channels,
adiabatic 1 $^2A'$ ($\rightarrow$ $^2\Pi$) component,
with real part of 1 $^2A'$ potential-energy surface at $\gamma$=90$^\circ$.
Bond lengths, units of bohr.  Density is integrated over $\gamma$.}
\label{b2a1J1prop1}
\end{figure*}

With reference to the discussion in Ref.~\cite{haxton3} and paper I,
the constructed
diabatic $^2B_2$
surface correlates to the species H$^-$+OH ($^2\Sigma$), as does the adiabatic
$^2B_2$ (2 $^2A'$) surface.  The diabatic $^2A_1$ surface correlates to H$^-$
plus ground state OH ($^2\Pi$), as does the adiabatic 1 $^2A'$ state.  Therefore,
dynamics beginning on the $^2B_2$ surface that leads to production of
the ground-state H$^-$+OH ($^2\Pi$) species must proceed via the
off-diagonal coupling to the $^2A_1$ surface.  From the viewpoint of the
diabatic basis, the off-diagonal coupling must in this case lead to
a transition between the diabatic $^2B_2$ and $^2A_1$ surfaces;
from the viewpoint of the adiabatic basis, the dynamics must proceed
through the conical intersection via the singular derivative couplings
inherent in that basis.

We present the calculated total cross sections for production of either
H$^-$+OH or D$^-$+OD in Fig. \ref{OH_H_b2a1}.  The  results are similar 
in shape, but the magnitude of the cross sections for the deuterated case are approximately
half those  of the nondeuterated case.
Differences in the reduced masses in the dissociative
direction result in a relatively longer time during which the deuterated
species may undergo autodetachment and, consequently, a smaller survival probability for the deuterated anion,  As shown in Table \ref{survivaltable},
the survival probability for the nondeuterated $^2B_2$ state is
21.5\%, whereas for the deuterated species it is only 13.1\%.
Unfortunately, there is  no experimental
data for comparison that measures the relative magnitude of the H$^-$ and D$^-$ peaks
for the highest-energy $^2B_2$ resonance.

An obvious feature of the results presented in Fig. \ref{OH_H_b2a1}
is that the branching ratio of OH ($^2\Pi$) to OH ($^2\Sigma$) production
depends on the incident energy of the electron.
This ratio varies from 100\% (only $^2\Pi$ is produced) at onset to zero (only 
$^2\Sigma$ produced) at higher energy.  At low energy
the observed cross section is the result of dynamics in which the
wave packet makes a nonadiabatic transition from the upper $^2B_2$ (2 $^2A'$)
surface to the lower $^2A_1$ (1 $^2A'$) surface, whereas at high energy,
the observed cross section is due to dynamics in which there is no transition.
Thus, the nuclear dynamics via the $^2B_2$ state  involve the conical
intersection to produce a branching ratio that varies with incident energy
in an interesting way.

We have only been able to achieve final-state resolution
for the OH ($^2\Sigma$) fragment.  
Tow-dimensional views of the cross sections for H$^-$+OH($^2\Sigma$) production,
as a function of both the incident electron energy and the kinetic energy of the
H$^-$ fragment, are given in the EPAPS archive\cite{epaps}, along with comparisons
of our calculated results with previous experiment.
Our calculations reproduce the approximate level of excitation
within the diatomic fragment, as the theoretical and experimental results are both
centered near the same kinetic energy, $\sim$2.75eV for H$^-$ from H$_2$O, and
$\sim$1.5eV for D$^-$ from D$_2$O.  We cannot make a more quantitative comparison, because there
are no experimental values for the average kinetic energy release in this channel.

The wave packet dynamics for DEA leading to H$^-$ production via the 
$^2B_2$ state coupled to the $^2A_1$ state are shown in Figs.~\ref{b2a1J1prop2}
and \ref{b2a1J1prop1}.  The former shows the reduced density on the adiabatic
2 $^2A'$ surface, where it starts; the latter shows that on the adiabatic 1 $^2A'$ surface.  These
plots were obtained by transforming the propagated wave packets from the 
diabatic basis to the adiabatic basis.  The wave packet
initially has no magnitude on the lower 1~$^2A'$ surface.  Nonadiabatic coupling
changes this situation as the wave packet is propagated.  The norm of the
propagated wave packet on the 1 $^2A'$ surface reaches a maximum of 0.112 at 
$t$=9.1~fs by which time
a portion of the wave packet has reached the dissociative H$^-$+OH ($^2\Pi$)
well of the 1~$^2A'$ surface.  The portion of the wave packet within this
well (see the bottom-left panel of Fig. \ref{b2a1J1prop1})
lies beyond $R$=4.5$a_0$ where the resonance becomes bound,
and so it continues toward dissociation with negligible loss of
flux.  The subsequent decrease of the norm of the wave packet on the 1 $^2A'$ surface
is therefore due to the consumption of other parts of the wave packet by the
imaginary component of this surface, and to its absorption by the complex
absorbing potentials.

As described in paper I,
the magnitude of the width for the
upper 2 $^2A'$ surface is generally large, though it decreases slowly
as the H$^-$+OH ($^2\Sigma$) well is approached, and abruptly as 
the H$_2$+O$^-$ well is approached.  As a result, the wave packet which
begins upon the upper 2 $^2A'$ surface is rapidly consumed, and its norm
decreases from exactly 1 to 0.321 within six femtoseconds.  At this
time, the combined norm on both surfaces is 0.402.  The calculated total
survival probability for this resonance, $P_{surv}$, calculated with Eq.(\ref{psurvdef}),
is 21.5\% (see Table \ref{survivaltable}).  From this comparison 
we can see that the
majority of the autodetachment for this resonance occurs within the first
six femptoseconds; its survival probability is 40.2\% within this initial time period,
and 21.5/40.2 = 53.5\% thereafter.

\begin{figure}[t]
\begin{center}
\includegraphics*[width=0.45\textwidth]{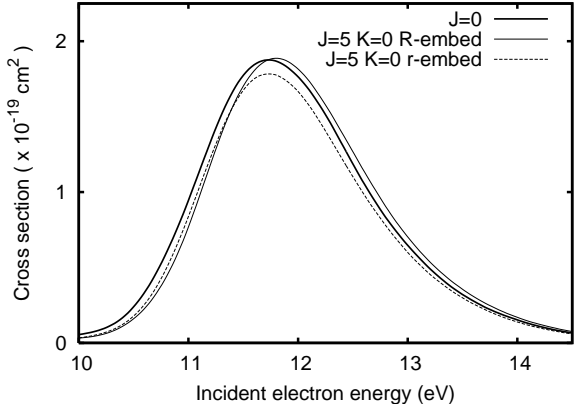} 
\end{center}
\caption[Cross sections, H$_2$+O$^-$ from $^2B_2$, rotationally excited]{Cross sections for production
  of H$_2$ ($\nu$)+O$^-$ from $^2B_2$ state, with rotational 
excitation of the target, as a function
of incident electron energy.  Bold line, total cross sections
for ground rotational state ($J=0$); thin line, calculations for
$J=5$, $K=0$.}
\label{H2_O_b2a1.excite}
\end{figure} 

\begin{figure}
\begin{center}
\begin{tabular}{c}
\includegraphics*[width=0.45\textwidth]{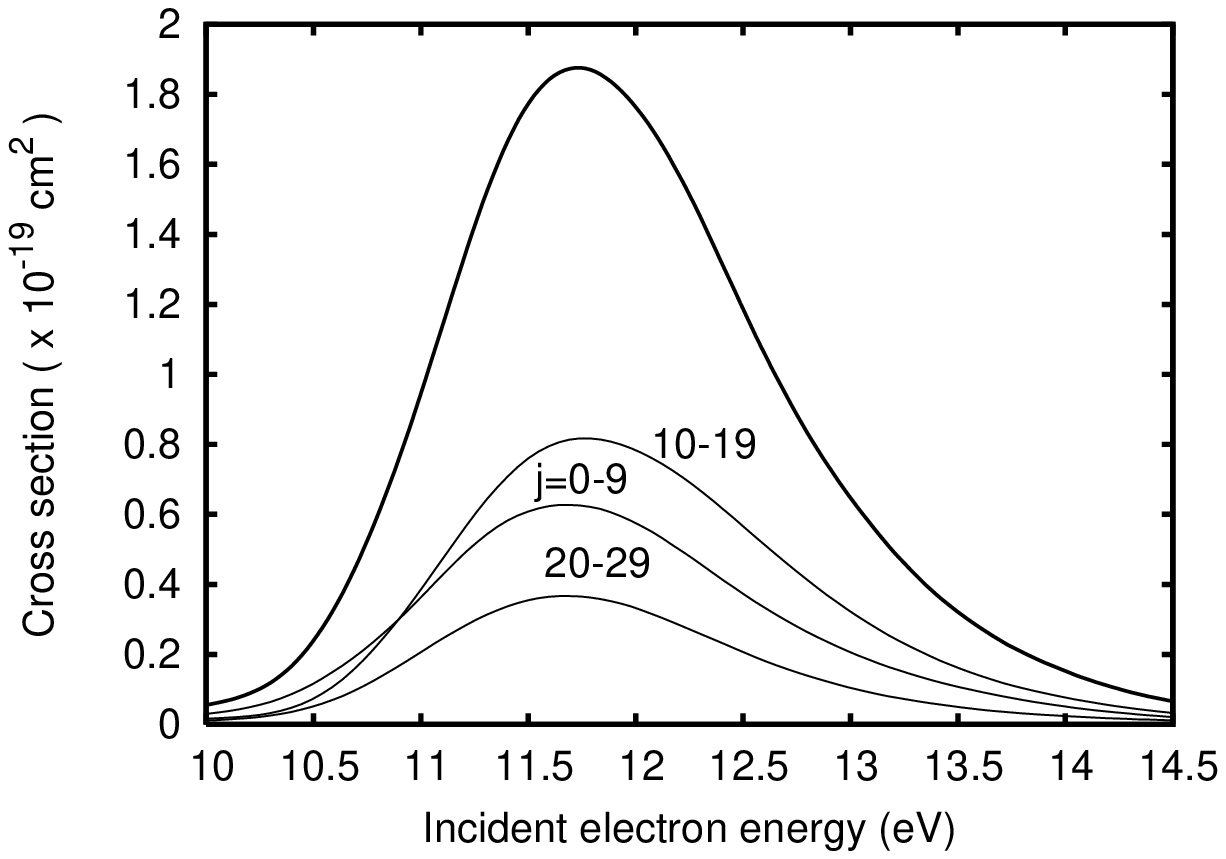} \\
\includegraphics*[width=0.45\textwidth]{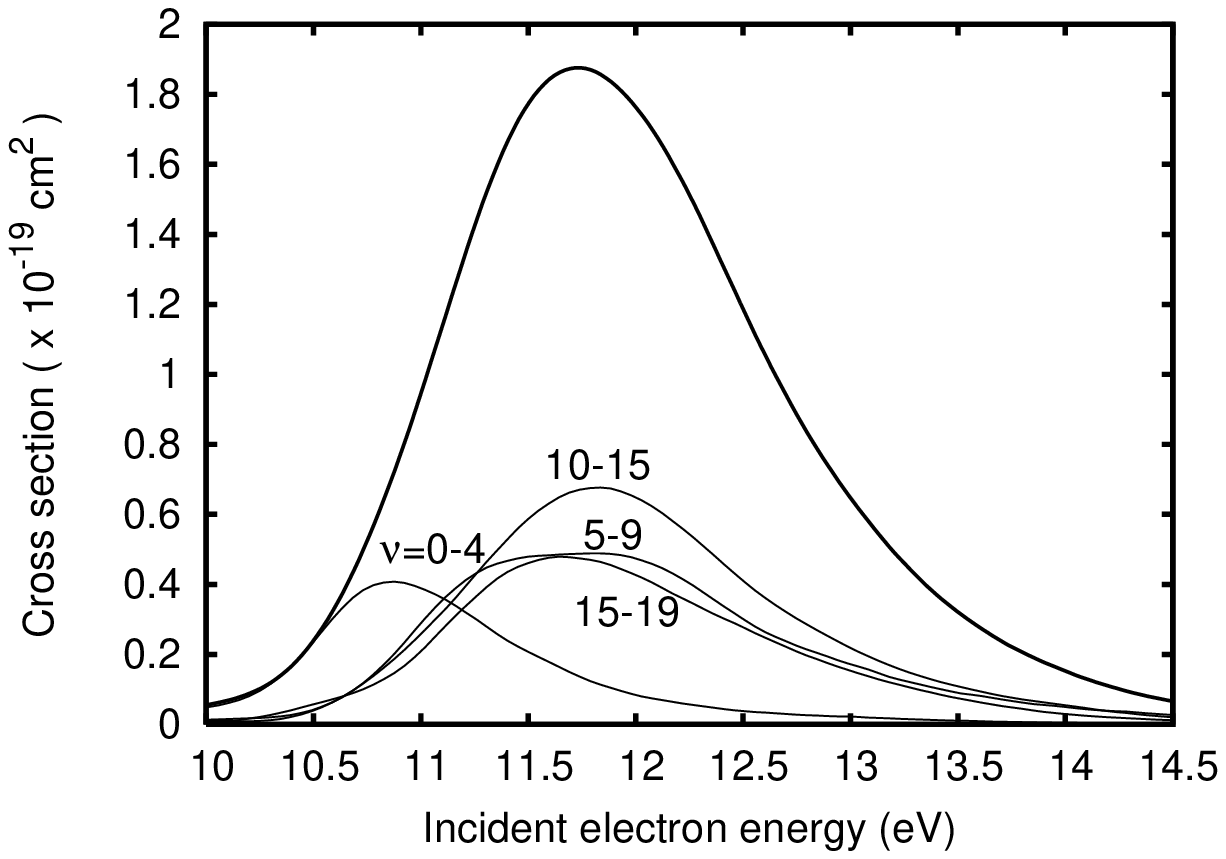}
\end{tabular}
\end{center}
\caption[Cross sections, H$_2$+O$^-$ from $^2B_2$]{Cross sections for production
  of H$_2$ ($\nu$)+O$^-$ from $^2B_2$ state as a function
of incident electron energy, showing degree of vibrational and rotational
excitation of the H$_2$ fragment.  Total cross section, thick line.  
Top, cross sections
summed over vibrational quantum number $\nu$ and binned by rotational
quantum number $j$; bottom, $\nu$ and $j$ reversed.}
\label{H2_O_b2a1}
\end{figure}

\begin{figure*}
\begin{center}
\begin{tabular}{ccc}
\resizebox{0.3\textwidth}{!}{\includegraphics*[0.55in,0.5in][3.9in,3.65in]{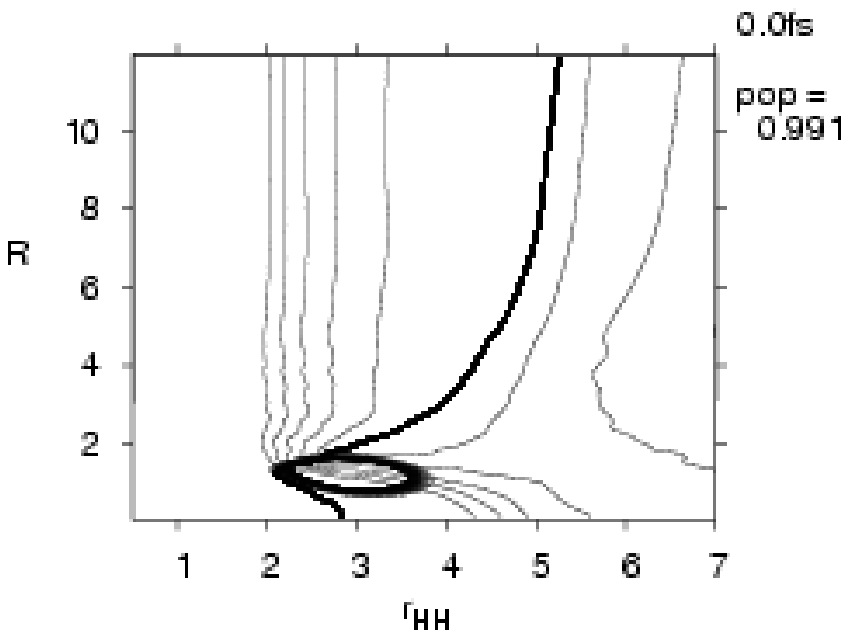}} &
\resizebox{0.3\textwidth}{!}{\includegraphics*[0.55in,0.5in][3.9in,3.65in]{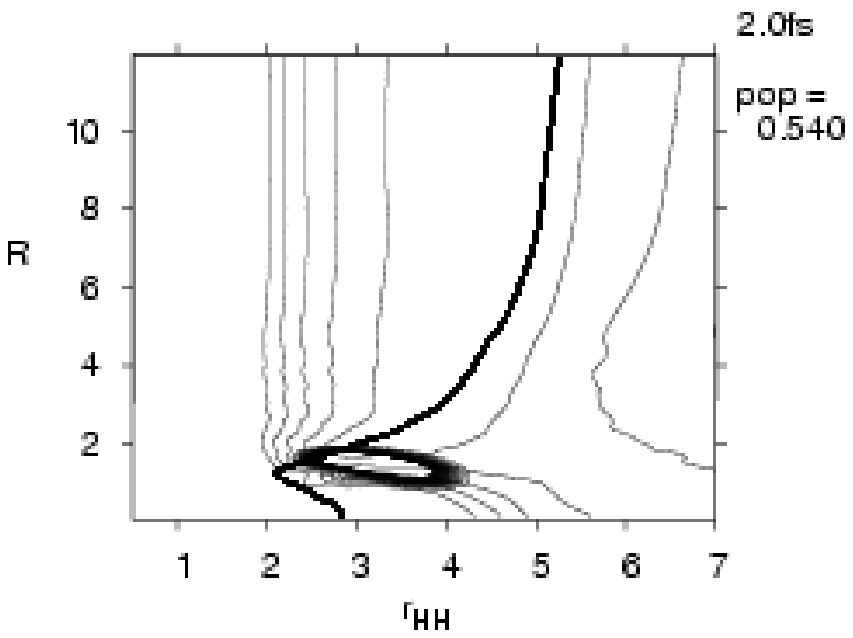}} &
\resizebox{0.3\textwidth}{!}{\includegraphics*[0.55in,0.5in][3.9in,3.65in]{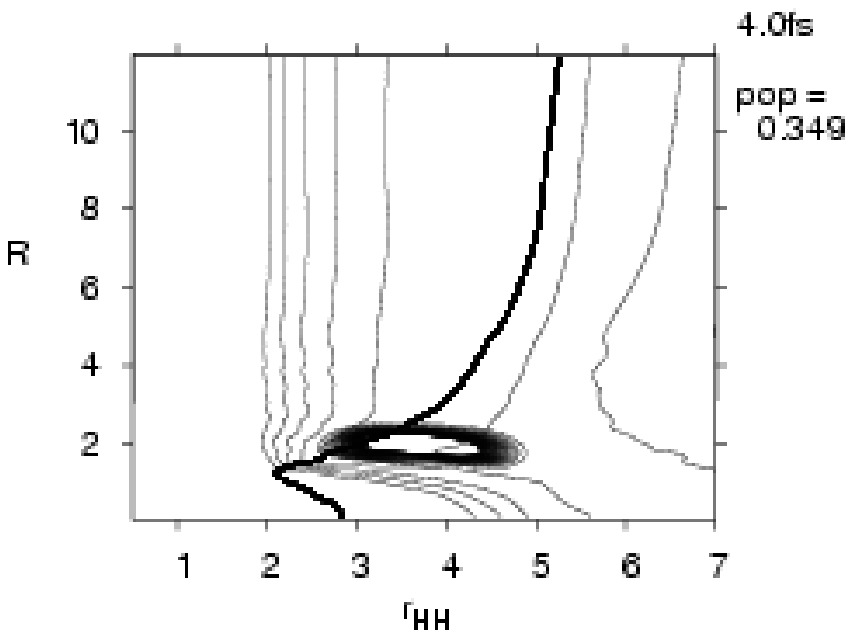}} \\
\resizebox{0.3\textwidth}{!}{\includegraphics*[0.55in,0.5in][3.9in,3.65in]{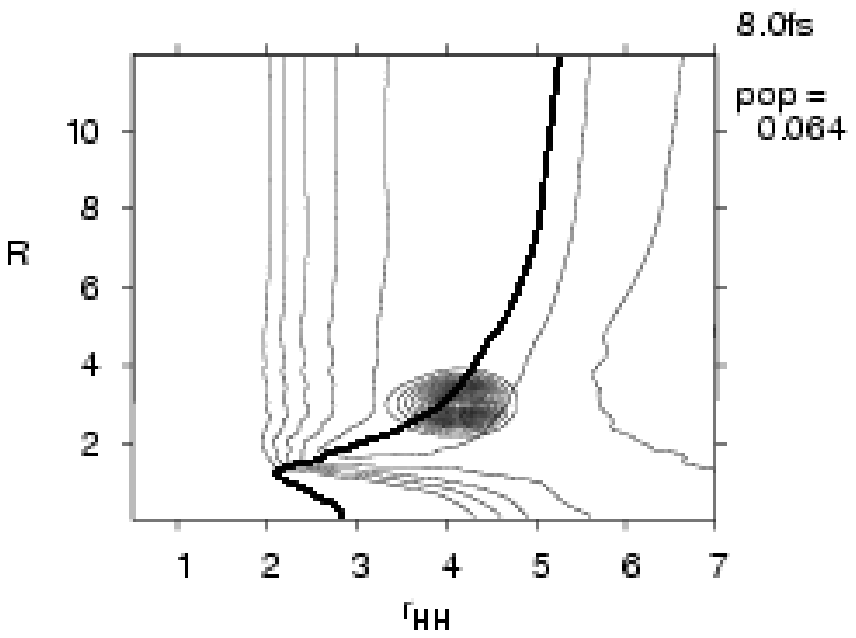}} &
\resizebox{0.3\textwidth}{!}{\includegraphics*[0.55in,0.5in][3.9in,3.65in]{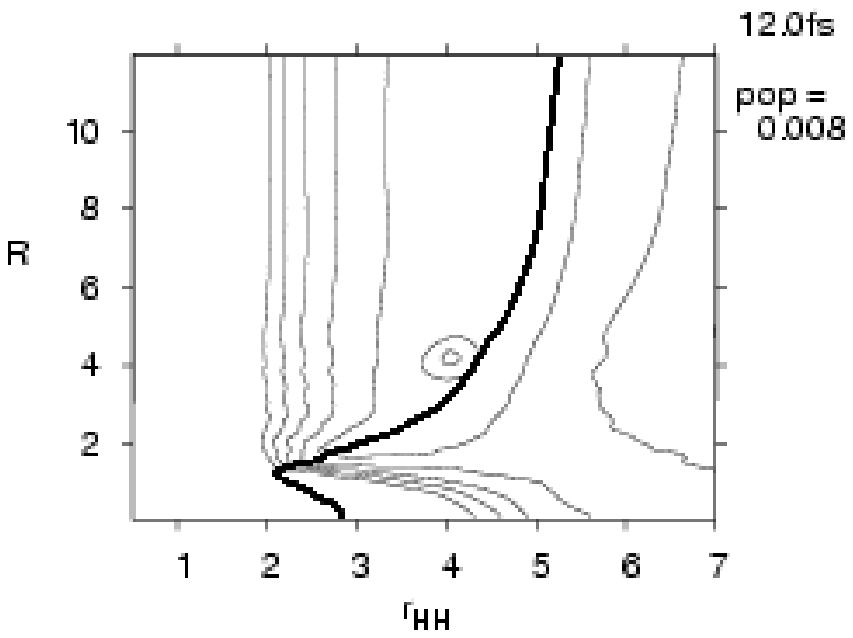}} & 
\resizebox{0.3\textwidth}{!}{\includegraphics*[0.55in,0.5in][3.9in,3.65in]{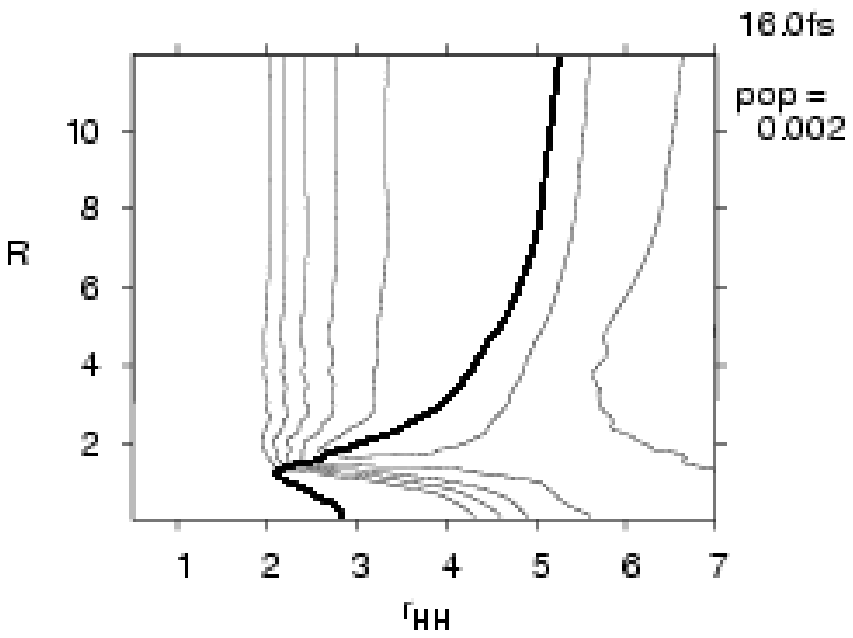}}  
\end{tabular}
\end{center}
\caption[$^2B_2$ propagation, 1 $^2A'$ component]{Propagation on coupled $^2B_2$ and $^2A_1$ surfaces for H$_2$+O$^-$ channel,
adiabatic 2 $^2A'$ component.
The reduced density (integrated over $\gamma$) of the 
adiabatic 2 $^2A'$ component of the propagated wave packet is plotted 
with the real part of the 2 $^2A'$ potential-energy surface at $\gamma$ =
90$^\circ$ ($C_{2v}$ geometry).  The location of the conical intersection is
marked with a bold line. Bond lengths, units of bohr.}
\label{b2a1J2prop2}
\end{figure*}
\begin{figure*}
\begin{center}
\begin{tabular}{ccc}
\resizebox{0.3\textwidth}{!}{\includegraphics*[0.55in,0.5in][3.9in,3.6in]{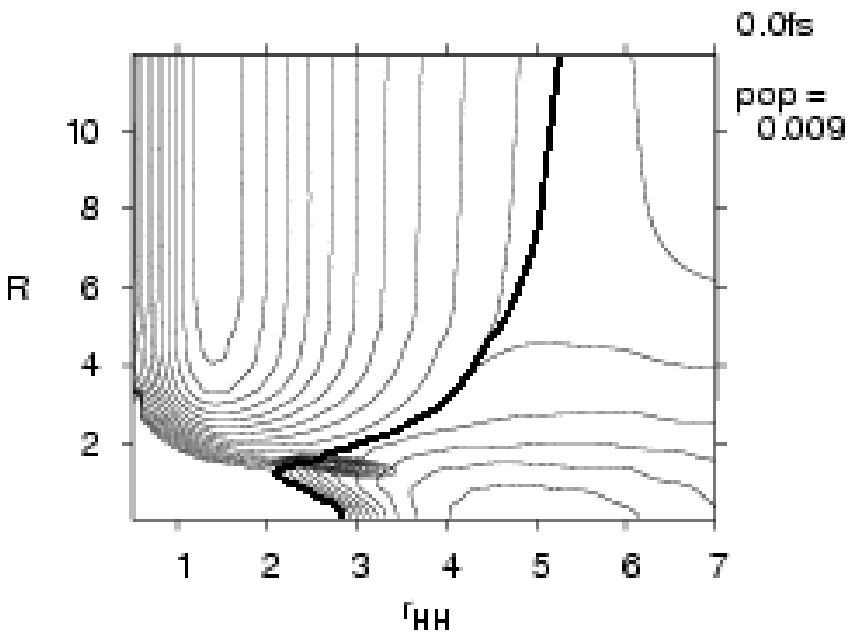}} &
\resizebox{0.3\textwidth}{!}{\includegraphics*[0.55in,0.5in][3.9in,3.6in]{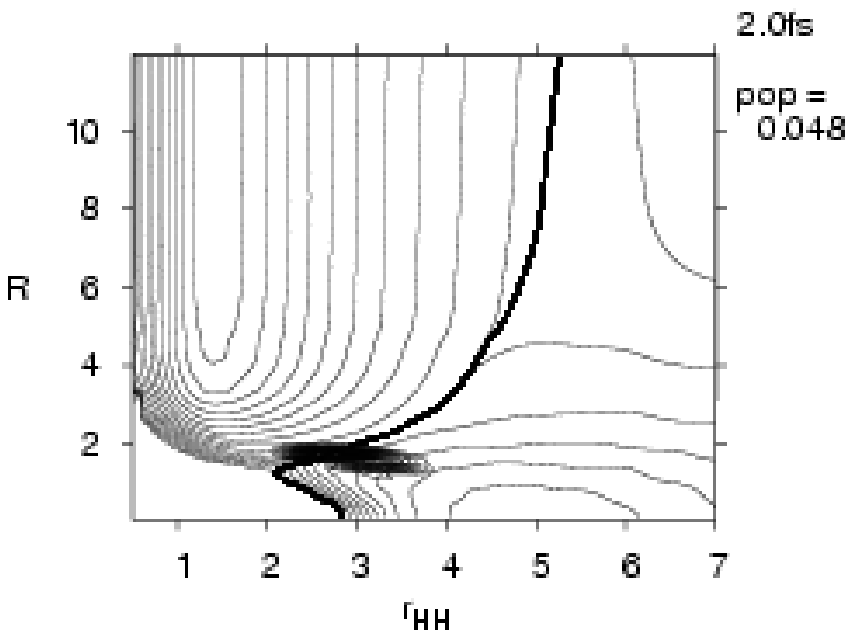}} &
\resizebox{0.3\textwidth}{!}{\includegraphics*[0.55in,0.5in][3.9in,3.6in]{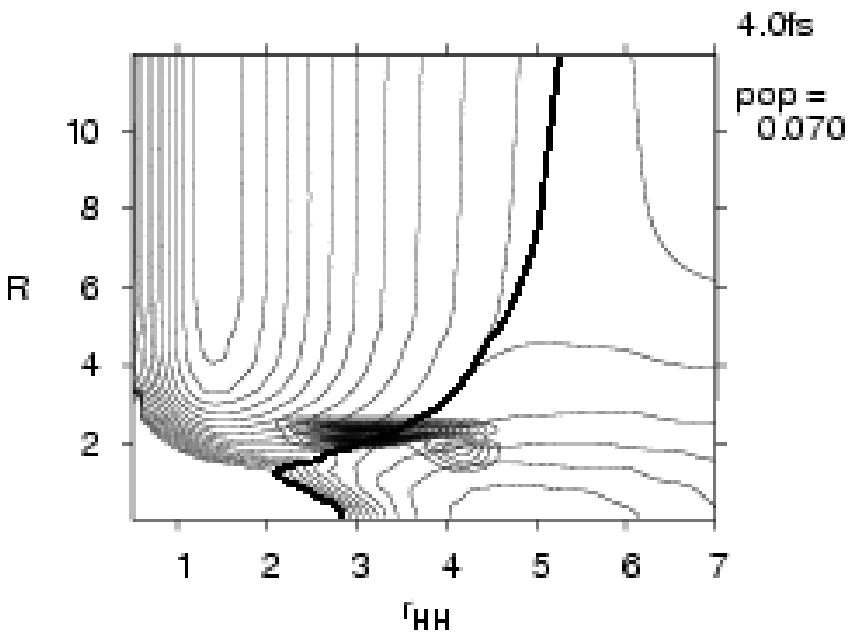}} \\
\resizebox{0.3\textwidth}{!}{\includegraphics*[0.55in,0.5in][3.9in,3.6in]{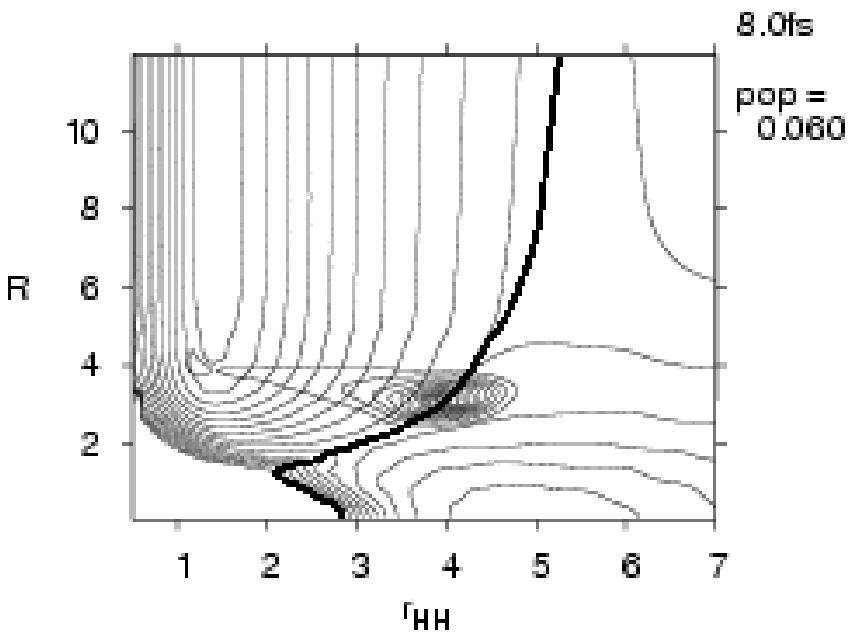}} &
\resizebox{0.3\textwidth}{!}{\includegraphics*[0.55in,0.5in][3.9in,3.6in]{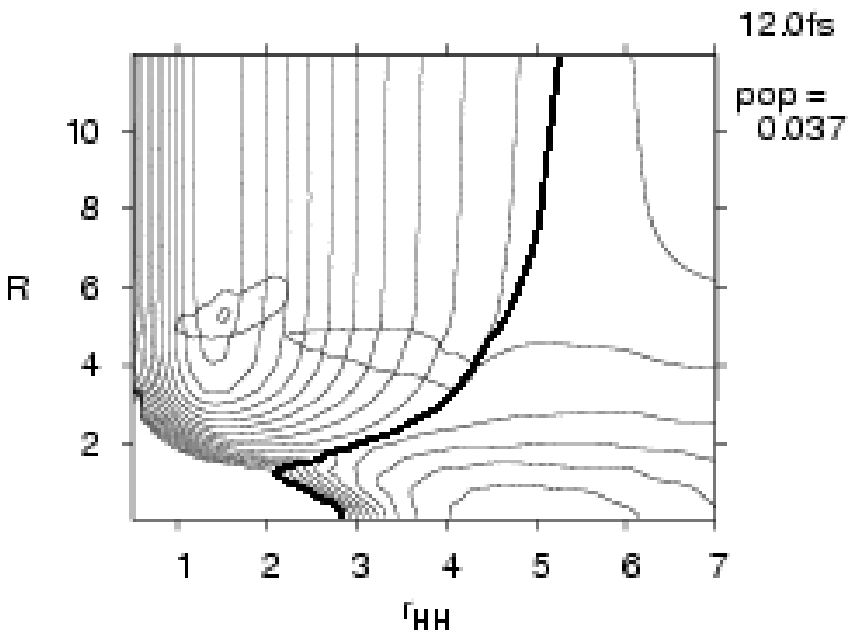}} &
\resizebox{0.3\textwidth}{!}{\includegraphics*[0.55in,0.5in][3.9in,3.6in]{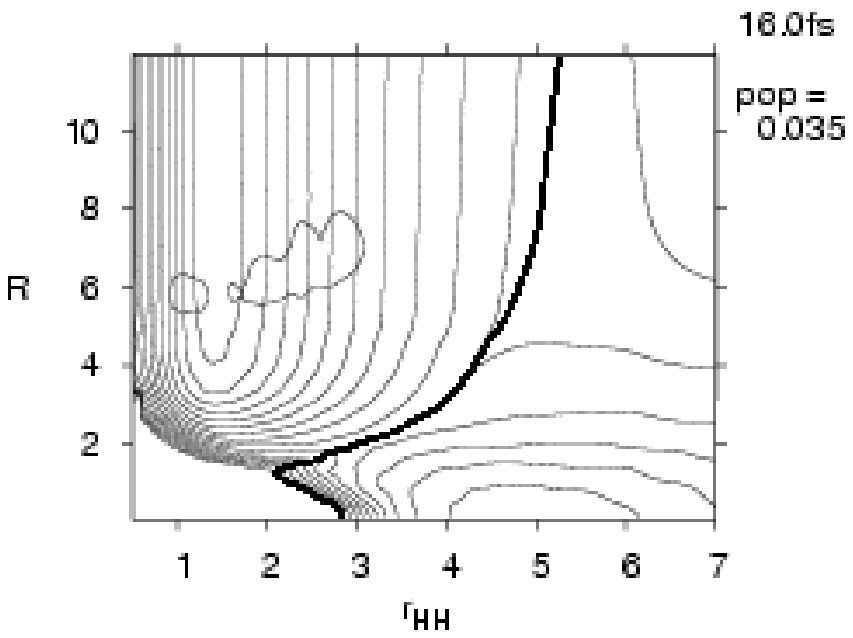}} 
\end{tabular}
\end{center}
\caption[$^2B_2$ propagation, 1 $^2A'$ component]{Propagation on coupled $^2B_2$ and $^2A_1$ surfaces for H$_2$+O$^-$ channel, adiabatic 1 $^2A'$ component.
The reduced density (integrated over $\gamma$) of the 
adiabatic 1 $^2A'$ component of the propagated wave packet is plotted 
with the real part of the 1 $^2A'$ potential-energy surface at $\gamma$=90$^\circ$ ($C_{2v}$ geometry).  The location of the conical intersection is marked with a bold line.}
\label{b2a1J2prop1}
\end{figure*}

The high degree of vibrational excitation ($\langle \nu \rangle$) in both
the OH ($\Pi$) and OH ($\Sigma$) channels 
is apparent in the oscillations of the dissociating
wave packet within each potential well, visible in the lower panels of
Figs.~\ref{b2a1J1prop2} and \ref{b2a1J1prop1}.  Fig.~\ref{b2a1J1prop1}
shows that in this case  there is additional
structure to the dissociating wave packet  on the lower 1~$^2A'$ surface;
however, this  structure
is most likely due to the calculation not being fully converged.

\subsubsection{Production of H$_2$ + O$^-$ via the upper $^2B_2$ state}

The channel H$_2$+O$^-$ is the dominant channel
observed in experiment for dissociative attachment to water via the
highest-energy $^2B_2$ resonance.  As discussed at length in 
Ref.~\cite{haxton3} and paper I, 
this channel is not present as an asymptote on the 
$^2B_2$ (2 $^2A'$) surface, and therefore, the system must undergo a nonadiabatic
transition via the conical intersection to the lower 1 $^2A'$ surface in order
to reach this product channel.  In the context of the representation
which we constructed in paper I, the system must follow the
\textit{diabatic} $^2B_2$ surface past its crossing with the $^2A_1$ diabatic
surface.  The H$_2$+O$^-$ channel is present as an asymptote of the
diabatic $^2B_2$ surface.   As described in Ref.~\cite{haxton3} and paper I, 
the adiabatic $^2A_1$ surface does not have
a bound asymptote in this arrangement; it correlates to O$^-$+H$_2$ ($\sigma_g^1 \sigma_u^1$) instead.

The calculated peak cross section  for this channel, 1.87$\times$10$^{-19}$cm$^2$, 
is smaller than the experimental value, 5.7$\times$10$^{-19}$cm$^2$,  reported by 
Melton~\cite{Melton} .  The comparison with
experiment is again complicated by the fact that we calculate only the
two-body DEA cross section, while the available experimental data do not 
discriminate between production of O$^-$+H+H and  O$^-$+H$_2$.
A possible explanation for the discrepancy between our calculated
results and experiment is the presence of a large three-body component to
O$^-$ production via the $^2B_2$ resonance. Rotational excitation of the 
target H$_2$O molecule, on the other hand, cannot
account for this discrepancy.  We have performed several calculations
in which rotational excitation of the target is included.  These include
calculations for total angular momentum $J=5$, projection $K=0$.  We find that
the effect of such rotational excitation is minimal, as  Fig.~\ref{H2_O_b2a1.excite} shows.

We calculate a very high degree of rotational and vibrational excitation in the 
H$_2$ or D$_2$ fragment.  The average degree of vibrational excitation $\langle \nu \rangle$
calculated from Eq.(\ref{psurv} is 7.75 for the H$_2$ fragment and 13.0 for the D$_2$
fragment.  The corresponding values for $\langle j^2 \rangle$ are 405 and 725, respectively.
Figure \ref{H2_O_b2a1} shows the total cross sections, as well as the cross sections
into either rotational or vibrational states, summed over the opposite quantum number.
The degree of vibrational excitation evidently decreases with incident electron energy,
while the degree of rotational excitation shows little correlation with incident
electron energy.

The high degree of rotational and vibrational excitation of the diatomic fragment
reduces the kinetic energy of the atom-diatom recoil.  This is reflected in the
cross sections for production of both H$_2$ and D$_2$ via the $^2B_2$ resonance, which have
the greatest magnitude nearere the lower range of recoil energy.  Two-dimensional plots
of these cross sections as functions of both incident electron energy and the kinetic 
energy of the recoil are shown in the EPAPS archive\cite{epaps}.

Plots of the propagated wave packet for DEA via the $^2B_2$ resonance 
are shown in Figs.~\ref{b2a1J2prop2} and \ref{b2a1J2prop1}.  
The first of these shows the magnitude-squared of the 1 $^2A'$ component to the propagated wave packet, integrated over $\gamma$, and the latter shows the 2 $^2A'$ component.  
The corresponding potential-energy surfaces,
evaluated at $\gamma$=90$^\circ$, are also plotted, along with the location
of the conical intersection seam which appears as a bold line.

The wave packet begins on the upper surface and proceeds to the lower surface only via nonadiabatic coupling near the conical intersection.
As described in paper I, the gradient of the  upper 2 $^2A'$ resonance surface leads downhill toward its conical intersection
with the 1 $^2A'$ resonance, leading the propagated wave packet
toward the seam.  This behavior is clearly visible in Fig.~\ref{b2a1J2prop2}
and \ref{b2a1J2prop1}.  The 2 $^2A'$ wave packet follows the conical intersection
seam in Fig.~\ref{b2a1J2prop2}, until it is consumed by the large imaginary
component to that potential-energy surface and by nonadiabatic coupling to the
1 $^2A'$ state along the intersection. 
The wave packet appears
on the 1 $^2A'$ surface in Fig.\ref{b2a1J2prop1} along the conical intersection,
and a small portion of it is able to reach the H$_2$+O$^-$ well of that
state.

The magnitude of the cross section for production of H$_2$+O$^-$ from the
$^2B_2$ resonance is therefore controlled by several competing effects.
The shape of the real part of the potential-energy surface determines the dynamically
accessible pathways and favors localization of the 2 $^2A'$ wave packet
near the conical intersection.  At the same time, the large imaginary
component to this surface consumes the wave packet and decreases the
amount of flux available to enter the conical intersection.  
On the lower 1 $^2A'$ surface, the amount of flux that enters the H$_2$
potential well is determined by the shape of that potential-energy surface,
since the conical intersection is outside the potential well and only a
fraction of the wave packet is propagated into the well.

\section{Conclusion}

We have presented the results of a fully \textit{ab initio} study of 
dissociative electron attachment to H$_2$O that includes
the full dimensionality of nuclear motion. We have attempted to 
calculate the cross sections for all the major and minor two-body channels  which 
are present as asymptotes of the Born-Oppenheimer,
$^2B_1$, $^2A_1$, and $^2B_2$ adiabatic electronic Feshbach resonances.
While we have qualitatively described the principal features that have been
experimentally observed, it is clear that a  fully quantitative description 
of this process has yet to be achieved.

The nuclear dynamics calculations were carried out using the MCTDH method
within the framework of the local complex potential model. For the major channel DEA,
H$^-$+OH (X $^2\Pi$) production through the lowest $^2B_1$ resonance,
the underlying assumptions of the model are well satisfied and we have obtained 
reasonably good agreement  with the experimental observations. 
Another notable feature of the present study is the quantification of the mechanism in
the major channel that leads to production of O$^-$ through the $^2B_2$ resonance.
Our earlier speculation~\cite{haxton3} that a conical intersection between the $^2A_1$
and $^2B_2$ states would play the key role in this process has been confirmed by the present study.

The present treatment has been limited to a consideration of DEA only into the final-state two-body channels. This limitation undoubtedly explains our inability to produce a non-zero
cross section for O$^-$ production via the $^2A_1$ resonance, which is likely to be
dominated by three-body breakup. Three-body breakup may also play a role in O$^-$ production
via the $^2B_2$ resonance, and its neglect here could explain why our calculated cross sections 
are smaller than the experimental results, which did not differentiate two- and three-body channels.

Physics beyond the local complex potential model may be at
work in some of the minor channels.  Dissociative
electron attachment via the $^2A_1$ Feshbach resonance may involve an
even greater variety of complicated resonant as well as non-resonant phenomena not described
by the LCP model. A variety of effects that go beyond the LCP model could be at play in the 
production of H$^-$ via  $^2A_1$ Feshbach resonance, including coupling to a broader shape resonance and even non-resonant virtual state effects. The neglect of such effects could well
explain our overestimation of the cross section
for production of H$^-$+OH via the $^2A_1$ state. Even for DEA via the lowest-energy  
$^2B_1$ state, nonlocal physics may be important in the minor channel, which leads to
H$_2$+O$^-$. 

We have achieved considerable success in describing the mean features of DEA to water,
clarified the mechanisms for the two-body breakup channels, and found evidence to suggest
that three-body breakup to produce O$^-$ might be important.  Nonetheless, many challenges
remain before a complete and quantitative understanding of this fundamental, but complicated,
system will be realized.

\begin{acknowledgments}

This work was performed under the auspices of the US Department of Energy
by the University of California Lawrence Berkeley National Laboratory
under Contract DE-AC02-05CH11231 and
was supported by the U.S. DOE Office of Basic Energy
Sciences, Division of Chemical Sciences.  The authors acknowledge many helpful
discussions with H.-Dieter Meyer concerning various aspects of the MCTDH method.

\end{acknowledgments}

\end{document}